\documentclass[%
 reprint,
 amsmath,amssymb,
 aps,
nofootinbib]{revtex4-2}

\usepackage{amsmath,latexsym,amssymb,amsfonts}
\usepackage{booktabs}   
\usepackage{siunitx}    
\usepackage{multirow}   

\newcolumntype{P}{S[table-format=1.0]} 
\newcolumntype{C}{S}                  
\newcolumntype{R}{S}   

\usepackage{dcolumn}
\usepackage{appendix}
\usepackage{mathtools}
\usepackage{graphicx}
\usepackage{rotating}

\usepackage{parallel,enumitem}
\usepackage[usenames,dvipsnames]{color}

\usepackage{dcolumn}
\usepackage{bm}
\usepackage{hyperref}
\usepackage{mathrsfs}
\hypersetup{
    colorlinks=true,
    linkcolor=blue,
    filecolor=magenta,      
    urlcolor=cyan,
    pdftitle={Overleaf Example},
    pdfpagemode=FullScreen,
    }
\usepackage{tensor}
\usepackage{hyperref} 

\begin{document}	
\title{Modified Teukolsky formalism: Null testing and numerical benchmarking}

\author{Fawzi Aly${}^{1}$
}
\thanks{Corresponding author}
\email{mabbasal[AT]buffalo.edu}
\author{ Mahmoud A. Mansour${}^2$
}
\email{ mansour[AT]iis.u-tokyo.ac.jp}

\author{ Luis Lehner${}^3$
}
\email{llehner[AT]perimeterinstitute.ca}

\author{ Dejan Stojkovic${}^1$
}
\email{ds77[AT]buffalo.edu}

\author{Dongjun Li${}^{4}$ }
\email{dongjun[AT]illinois.edu}

\author{Pratik Wagle${}^{5}$}
\email{pratik.wagle[AT]aei.mpg.de}

\affiliation{${}^1$HEPCOS, Physics Department, SUNY at Buffalo, Buffalo, New York, USA,}

\affiliation{${}^2$Physics Department, The University of Tokyo, Kashiwanoha, Kashiwa, Chiba 277-8574, Japan,}
\affiliation{${}^3$Perimeter Institute for Theoretical Physics, 31 Caroline St., Waterloo, ON, N2L 2Y5, Canada,}

\affiliation{${}^4$Illinois Center for Advanced Studies of the Universe \& Department of Physics}

\affiliation{${}^5$Max Planck Institute for Gravitational Physics (Albert Einstein Institute), D-14476 Potsdam, Germany}
\date{\today}

\begin{abstract}
Next-generation gravitational-wave detectors will make black-hole ringdown an increasingly sensitive probe of small departures from General Relativity in the strong-field regime. This motivates obtaining high-precision 
predictions of gravitational effective field theory, as spectral shifts can be quite small. Here we perform a focused stress test of the modified-Teukolsky framework by designing two null diagnostics. First, we consider an action with redundant operators  that must produce zero first-order vacuum QNM shifts. Second, we exploit a Ricci-flat identity relating two physical cubic Riemann theories to test such a relation is satisfied by the ringdown spectra obtained. We compute the shifts using two independent numerical approaches: the eigenvalue-perturbation and generalized continued-fraction (Leaver-type) methods. Both null tests are passed across multiple multipoles and overtones, and the control-operator results agree in magnitude with the benchmark values reported in Ref.~\cite{deRham_EFT_GW_2020}. These validations support using the framework for obtaining accurate predictions for robust strong-field tests, with straightforward extensions to rotating backgrounds and coupling with matter fields.
\end{abstract}

\maketitle

\section{Introduction}\label{introduction}

With next-generation ground- and space-based detectors—Einstein Telescope, Cosmic Explorer, and LISA—on the horizon, sensitivity and bandwidth will expand substantially, enabling more precise probes of the nonlinear, dynamical regime in which any departures from General Relativity (GR), if present, may manifest \cite{ET_Punturo_2010,ET_ScienceCase_2020,CE_Reitze_2019,CE_Evans_2021,LISA_AmaroSeoane_2017}. This motivates effective-field-theory (EFT) extensions of gravity—one branch within the broader landscape of beyond-GR models—which encode high-scale physics through controlled higher-derivative and non-minimal couplings that remain consistent with weak-field bounds yet can yield clean strong-gravity signatures \cite{deRham_EFT_GW_2020,Cano_2024_HD_Kerr_QNM}. 

Black-hole ringdown is a standard testbed for beyond-GR imprints: EFT induced deformations can shift GR quasinormal-mode (QNM) spectra, break even/odd isospectrality, and introduce additional radiative degrees of freedom that alter excitation and damping \cite{Berti_Review_2009,Silva_2024_QNMs_beyondGR,Li_2024_Isospectrality_MTF,Cardoso_2009_dCS_Schw,Weller_2025_BumpyBHs_PRD}. These expectations can be modeled via perturbatively equivalent strategies: ($i$) derive and solve \emph{modified master equations} with deformed potentials (modified Teukolsky/Regge–Wheeler-Zerilli(RWZ)) \cite{Cano_2024_HD_Kerr_QNM,deRham_EFT_GW_2020}; or ($ii$) work perturbatively in the EFT couplings so the GR Teukolsky/RWZ operator acts on the corrections, sourced by the GR gravitational perturbation (up to first order in book-keeping parameter $\epsilon$) and any effective matter source \cite{Li_MTE_2023,Wagle_2024_dCS_SlowRot_Eqs,Li_2024_Isospectrality_MTF}. Moreover, QNM shifts could be computed via various techniques for instance  ($a$) the \emph{eigenvalue perturbation} (EVP) method  \cite{Li_2024_Isospectrality_MTF,Zimmerman_QNMBeyondKerr_2014,Hussain_SpectralShifts_2022}, ($b$) \emph{direct integration} with asymptotic matching \cite{WagleYunesSilva2022_dCS_DirectIntegration} and ($c$) \emph{generalized continued-fraction} (Leaver-type) schemes adapted to the modified equations \cite{Leaver_1985_CFM,Leaver1990_RN,Cano_2024_HD_Kerr_QNM}. 

Demonstrations span both parity-preserving and parity-violating sectors: in higher-derivative gravity (HDG), high-order slow-rotation expansions reproduce Kerr QNM shifts with good accuracy up to $a/M\!\sim\!0.7$ and agree with metric-perturbation results \cite{Cano_2024_HD_Kerr_QNM}; in dCS gravity-- in which parity is violated-- the slow-rotation master system for $\Psi_{0,4}$ plus the pseudoscalar was derived \cite{Wagle_2024_dCS_SlowRot_Eqs}, and the resulting QNM frequency shifts including overtones were computed with cross-gauge (IRG/ORG) and RWZ consistency, and are typically more sensitive to EFT corrections \cite{Li_2025_dCS_SlowRot_QNMs}. Moreover, when extra fields couple to curvature, \emph{frequency contamination} leaks additional tones into the gravitational sector \cite{Lestingi_Contamination_2025,Silva_2024_QNMs_beyondGR}.

Because EFT effects in astrophysical black holes are expected to be relatively ``tiny", QNM computations require high precision and stringent consistency checks. Given the various avenues for modeling and exploring such beyond-GR imprints (as some summarized above), it is essential to validate both the modified Teukolsky formalism (MTE) framework itself \emph{and} the numerical tools used within. In this work, we establish controlled consistency tests of this kind.

In EFT, several higher-derivative operators are {redundant} on Ricci-flat backgrounds: using local field redefinitions built from the leading equations of motion, terms proportional to Ricci scalar $R$, or Ricci tensor $R_{\mu\nu}$, and their derivatives can be removed \cite{deRham_EFT_GW_2020,Lu_Stelle_2015,Cano_LeadingKerr_2019}. In particular, operators with linear Ricci contributions to the EFT-Lagrangian: $R\,R_{\mu\nu\alpha\beta}R^{\mu\nu\alpha\beta}$ and $R^{\mu \nu} R_{\mu \alpha \beta \gamma} {R_{\nu}}^{\alpha \beta \gamma}$  do \emph{not} shift asymptotically flat, vacuum observables at first order in EFT perturbation parameter $\zeta$—including QNM spectra, excitation factors, scattering amplitudes, and cross sections (see, e.g., \cite{deRham_EFT_GW_2020,Cano_2024_HD_Kerr_QNM}). This furnishes a sharp \emph{null test}: we consider such redundant deformations into the QNM computation and verify that the spectrum is not affected up to $\mathcal{O}(\zeta)$, thereby validating the two-parameter bookkeeping $(\epsilon,\zeta)$ in the modified-Teukolsky program and exposing any spurious numerical contamination. We use this controlled setting to benchmark EVP, and generalized continued fractions with diagnostics focused on truncation systematics and the heightened sensitivity of overtones.

\noindent\textbf{Paper structure.}
Section~\ref{sec:setup} formulates the EFT-modified Teukolsky framework by specifying the Lagrangian, background geometry, and master operator, and discusses axial--polar isospectrality via a corrected superpotential $W(r)$. Section~\ref{subsec:numerics} introduces two independent QNM solvers and their associated diagnostics: the eigenvalue-perturbation (EVP) method (Sec.~\ref{subsec:evp_contour}) and a Leaver-type generalized continued-fraction pipeline (Sec.~\ref{subsec:leaver_pipeline}). Numerical results are presented in Sec.~\ref{numerical_results}, and we conclude with a summary and outlook in Sec.~\ref{conclusion}. Explicit EFT effective potentials are collected in Appendix~\ref{Effective potentials}, while additional implementation details and numerical checks are provided in Appendix~\ref{More_Details_On_Numerical_Methods}. Full even/odd Frobenius recurrence coefficients are compiled in a companion \textit{Mathematica} notebook and will be made publicly available upon publication.

\section{Setup}\label{sec:setup}

We consider the low-energy EFT of gravity in four
dimensions, including all parity-even curvature operators of canonical
dimension six. We work with signature $(-,+,+,+)$ and natural units
$\hbar=c=1$. Following the conventions of Ref.~\cite{deRham_EFT_GW_2020},
the dimension-6 part of the Lagrangian is
\begin{widetext}
\begin{equation}
\begin{gathered}
\mathcal{L}_{D6}
  = \frac{\sqrt{-g}}{M^{2}} \Big[
      d_{1}\, R \Box R
    + d_{2}\, R_{\mu\nu} \Box R^{\mu\nu}
    + d_{3}\, R^{3}
    + d_{4}\, R\, R_{\mu\nu} R^{\mu\nu}
    + d_{6}\, {R_{\mu}}^{\nu} {R_{\nu}}^{\rho} {R_{\rho}}^{\mu}
    + d_{7}\, R^{\mu\nu} R^{\alpha\beta} R_{\mu\alpha\nu\beta} \\[2pt]
  \quad
    + d_{5}\, R\, R_{\mu\nu\alpha\beta} R^{\mu\nu\alpha\beta}
    + d_{8}\, R^{\mu\nu} R_{\mu}{}^{\alpha\beta\gamma} R_{\nu\alpha\beta\gamma}
    + d_{9}\, {R_{\mu\nu}}^{\alpha\beta} {R_{\alpha\beta}}^{\gamma\sigma} {R_{\gamma\sigma}}^{\mu\nu}
    + d_{10}\, R_{\mu}{}^{\alpha}{}_{\nu}{}^{\beta}
             R_{\alpha}{}^{\gamma}{}_{\beta}{}^{\sigma}
             R_{\gamma}{}^{\mu}{}_{\sigma}{}^{\nu} \Big] \\
  \equiv \frac{\sqrt{-g}}{M^{2}} \sum_{i=1}^{10} d_i\,\mathcal{O}_i \, .
\end{gathered}
\end{equation}
\end{widetext}

where $\Box$ is the d'Alembertian operator, $g\equiv\det(g_{\mu\nu})$, $d_i$ are the Wilson coefficients and $R_{\mu \nu \alpha \beta}$ is the Riemann tensor. 

On Ricci-flat backgrounds ($R_{\mu\nu}=0=R$), operators at least quadratic in $R$ or $R_{\mu\nu}$
(i.e., $\mathcal{O}_i$ with $i=1,2,3,4,6,7$) do not contribute to the background nor to linearized
vacuum perturbations at the order we consider.
For later convenience we introduce the combination
\begin{equation}
d_{58} \equiv 4d_{5}+d_{8}\, .
\end{equation}

The remaining four operators split into two categories: those linear in the Ricci
scalar/tensor ($\mathcal{O}_5,\mathcal{O}_8$) and those purely cubic in the Riemann tensor
($\mathcal{O}_9,\mathcal{O}_{10}$). The former are removable by local field redefinitions and,
therefore, do not induce first-order QNM shifts in vacuum; in contrast, the latter
are genuinely dynamical and do correct the QNM spectrum at first order. Accordingly, we define two subsets:
\begin{itemize}
\item Two \emph{null-test operators},
\begin{equation}
\begin{aligned}
\mathcal{O}_{5} &= R\,R_{\mu\nu\alpha\beta}R^{\mu\nu\alpha\beta},\\
\mathcal{O}_{8} &= R^{\mu\nu}R_{\mu\alpha\beta\gamma}{R_\nu}^{\alpha\beta\gamma},
\end{aligned}
\end{equation}

\item Two genuinely \emph{dynamical cubic invariants},
\begin{equation}
\begin{aligned}
\mathcal{O}_{9} &= {R_{\mu\nu}}^{\alpha\beta}\,{R_{\alpha\beta}}^{\gamma\sigma}\,{R_{\gamma\sigma}}^{\mu\nu},\\
\mathcal{O}_{10} &= R_{\mu}{}^{\alpha}{}_{\nu}{}^{\beta}\,
                   R_{\alpha}{}^{\gamma}{}_{\beta}{}^{\sigma}\,
                   R_{\gamma}{}^{\mu}{}_{\sigma}{}^{\nu}.
\end{aligned}
\end{equation}
\end{itemize}

Moreover, in four dimensions, $\mathcal{O}_{10}$ is related to $\mathcal{O}_{9}$ via the algebraic identity \cite{Edgar_2002}:

\begin{widetext}
\begin{equation}\label{eq:d10_identity_ops}
\mathcal{O}_{10}
=
\frac{1}{2}\mathcal{O}_{9}
-\frac{3}{8}\mathcal{O}_{5}
-3\,R^{\mu\alpha}R^{\nu\beta}R_{\mu\nu\alpha\beta}
-4\,{R_{\mu}}^{\nu}{R_{\nu}}^{\rho}{R_{\rho}}^{\mu}
+\frac{9}{2}\,R\,R_{\mu\nu}R^{\mu\nu}
-\frac{5}{8}\,R^{3}.
\end{equation}
\end{widetext}
In particular, on Ricci-flat backgrounds this reduces to
$
\mathcal{O}_{10}=\frac12\,\mathcal{O}_{9}-\frac{3}{8}\mathcal{O}_{5}\, .
$
Thus, at $\mathcal{O}(\zeta)$ the QNM shift obtained from $\mathcal{O}_{9}$ must equal twice the shift from $\mathcal{O}_{10}$, up to contamination from the null-test operator $\mathcal{O}_{5}$.

\textit{Problem Statement(null tests).} Any nonzero QNM shift sourced by $\mathcal{O}_5$ or $\mathcal{O}_8$
at $\mathcal{O}(\zeta)$ would indicate a breakdown of our perturbative setup and/or numerical implementation
rather than a genuine EFT effect. Also, for equal
Wilson coefficients, the ratio of QNM shifts from $\mathcal{O}_9$ and $\mathcal{O}_{10}$ must be equal to $2$ on Ricci-flat backgrounds up to numerical contamination.

Finally, our control (even-parity) sector matches the even-parity cubic operator basis used in
Ref.~\cite{Cano_2023}. That reference also considers the (odd-parity) cubic invariant
\begin{equation}\label{eq:odd_operator}
\tilde{\mathcal{O}}_{11}
\equiv \frac{1}{2}\,\epsilon^{\mu\nu\alpha\beta}
{R_{\mu\nu}}^{\rho\sigma}\,{R_{\rho\sigma}}^{\delta\gamma}\,{R_{\delta\gamma\alpha\beta}},
\end{equation}
For simplicity, we will not consider $\tilde{\mathcal{O}}_{11}$ in this work.

\subsection{Background geometry and master perturbation equations}

We work on a static, spherically symmetric background. In Schwarzschild-like coordinates $(t,r,\theta,\phi)$, the metric takes the form
\begin{equation}
\mathrm{d}s^2 = -A(r)\,\mathrm{d}t^2 
+ \frac{\mathrm{d}r^2}{B(r)} 
+ r^2\,\mathrm{d}\Omega^2,
\quad
\frac{\mathrm{d}r}{\mathrm{d}\bar{r}_*}
=\sqrt{A(r)B(r)}\,,
\label{eq:tortoise}
\end{equation}
where $\bar{r}_*$ is the tortoise coordinate associated with the
EFT-deformed background and $\mathrm{d}\Omega^2$ is the line element on the unit $2$--sphere. We define
\begin{equation}
f(r) \equiv 1 - \frac{r_g}{r}\,,
\end{equation}
so that in the GR limit ($\zeta\to 0$) one has $A(r)=B(r)=f(r)$
and $\bar r_* \to r_*$, where $r_*$ is the standard Schwarzschild
tortoise coordinate.

The EFT deformation parameter is
\begin{equation}
\zeta \equiv M^{-2} M_{\mathrm{Pl}}^{-2} r_g^{-4}\,.
\end{equation}
At $\mathcal{O}(\zeta)$, the EFT
corrections to the background functions $A(r),B(r)$ and the
axial/polar potentials can be written in closed form; for completeness
we provide the full GR and EFT potentials for both parities in
Appendix~\ref{Effective potentials}.

For odd/even-parity perturbations we adopt the master-equation form of
Eq.~(2.23) in Ref.~\cite{deRham_EFT_GW_2020}:
\begin{equation}
\label{eq:master}
\begin{gathered}
\mathcal{L}^{(1)}_{o/e,\,n\ell}
\equiv
\frac{\mathrm{d}^2}{\mathrm{d}\bar{r}_*^2}
+ \omega^2_{n\ell} - \bar{V}^{o/e}_{n\ell}(r;\omega_{n \ell})\,,\\[4pt]
\mathcal{L}^{(1)}_{o/e,\,n\ell}\,\Psi^{o/e}_{n\ell} = 0\,,
\end{gathered}
\end{equation}
where the subscripts $(n,\ell)$ denote the overtone index and angular multipole, respectively. With effective potentials
\begin{equation}
\begin{aligned}
\bar{V}^{o/e}_{n\ell}(r;\omega_{n \ell})
&=\sqrt{A(r)B(r)}
\Big(V^{o/e,\,\mathrm{GR}}_{\ell}(r)
+\zeta\,V^{o/e}_{n\ell}(r)\Big)\\
&\quad-\;\zeta\,\Delta c(r)\,\omega^2_{n \ell}\,.
\end{aligned}
\end{equation}
Note that $V^{o/e,\,\mathrm{GR}}_{\ell}(r)$ is $\omega$-independent, whereas the EFT-deformed potential $\bar V^{o/e}_{\ell}(r;\omega)$ depends parametrically on $\omega$; evaluating it at $\omega=\omega_{n\ell}$ therefore yields mode-dependent effective potentials. 
The GR piece is related to the standard Regge--Wheeler/Zerilli
potentials by
\begin{equation}
V^{o/e,\,\mathrm{GR}}_{\ell}(r)
= \frac{V^{\mathrm{RW/Z}}_{\ell}(r)}{f(r)}\,,
\end{equation}
while the coefficient of the EFT-induced $\omega_{n \ell}^2$-term is
\begin{equation}
\Delta c(r)
=144\,(2 d_{9} + d_{10})\,
\frac{f(r)}{(r/r_g)^5}\,.
\end{equation}

\subsection{Axial--polar isospectrality and superpotential}

 For Schwarzschild in GR ($\zeta=0$), the odd (axial) and even (polar) sectors are known to be isospectral: for each $(\ell,n)$ the axial and polar QNM frequencies coincide.

The standard Chandrasekhar transformation relates the RW master
and the Zerilli master functions via a superpotential $W_{\rm GR}(r)$. In our
normalization \cite{chandrasekhar_1983,Glampedakis_2017},  
\begin{equation}
W_{\rm GR}(r) = \frac{3 r_g(r_g - r)}{r^2\bigl(3 r_g + (\lambda_\ell -2) r\bigr)} 
       - \frac{(\lambda_\ell -2)\lambda_\ell}{6 r_g},
\label{eq:W_GR_def}
\end{equation}
where $\lambda_\ell \equiv \ell(\ell+1)\,$.

In the presence of the null-test operators $\mathcal{O}_5$ and
$\mathcal{O}_8$, the axial and polar potentials receive EFT
corrections, but they remain isospectral to first order in $\zeta$.
More precisely, for $\mathcal{O}_5$ and $\mathcal{O}_8$ alone the
axial and polar sectors continue to be related by a superpotential
whose integration constant is unchanged and given by
$(2\lambda_\ell - \lambda_\ell^2)/(6 r_g)$. The superpotential
itself acquires a $\mathcal{O}(\zeta)$ correction:
\begin{equation}
\begin{aligned}
W(r)
  &= W_{\rm GR}(r) + \zeta\,d_{58}\,W_{58}(r)\,,\\
   W_{58}(r)&= \frac{9 r_g^7  (-6 r_g^2 - 3 r r_g (-3 + \lambda_{\ell}) + 
   2 r^2 (-2 + \lambda_{\ell}))}{(r^8 (3 r_g + r (-2 + \lambda_{\ell}))^2)}
  \end{aligned}
\end{equation}

However, while isospectrality between the axial and polar sectors in GR follows from the Chandrasekhar transformation (equivalently, a Darboux/SUSY partner structure of the RW and Zerilli potentials), the corresponding statements in the EFT-deformed problems are more subtle.

It is not straightforward to exhibit an explicit analytic Darboux map that connects the GR spectral
problem to the $\zeta$-deformed problems (or, more nontrivially, maps the $\mathcal{O}_5$ problem to the
$\mathcal{O}_8$ one) in a way that makes these spectral equivalences manifest within the standard
Darboux transformation framework \cite{chandrasekhar_1983,Glampedakis_2017}. For this reason, we test the expected spectral relations numerically.

The situation changes qualitatively once genuine curvature corrections are switched on through
$\mathcal{O}_9$ and/or $\mathcal{O}_{10}$: odd/even (parity) isospectrality is then expected to be
broken. Indeed, the EFT results of \cite{deRham_EFT_GW_2020} show that the odd- and even-parity
fractional frequency shifts induced by these operators differ, providing explicit numerical evidence
that parity isospectrality does not survive in the presence of $\mathcal{O}_9$ and $\mathcal{O}_{10}$.

\subsection{Two complementary QNM calculation strategies}

We employ two independent approaches to compute the QNM corrections due to the
EFT-deformed background: EVP method and Leaver method.

\subsubsection{EVP method}

We follow a two-parameter bookkeeping scheme common in modified
Teukolsky and EFT treatments of QNMs. As defined briefly in Sec. \ref{introduction}, the parameter $\epsilon$ tracks
the linear GW perturbation order, while $\zeta$ tracks the EFT
deformation. We retain terms up to $\mathcal{O}(\zeta^1\epsilon^1)$ and
neglect $\mathcal{O}(\zeta^2)$ corrections.

Expanding the background and fields to first order in $\zeta$ we
write
\begin{gather}
A(r) = f(r) + \zeta\,A_1(r),\quad
B(r) = f(r) + \zeta\,B_1(r),\\
\sqrt{A(r) B(r)} = f(r) + \zeta\,\bar f(r),\quad
\bar f(r) \equiv \frac{A_1(r)+B_1(r)}{2},\\
\frac{\mathrm{d}r}{\mathrm{d}\bar{r}_*} 
= f(r) + \zeta\,\bar f(r),\\
\Psi^{o/e}_{n\ell} 
= \Psi^{o/e\,(0,1)}_{n\ell} + \zeta\,\Psi^{o/e,(1,1)}_{n\ell},\\
\omega_{n \ell} = \omega_{0\, n \ell} + \zeta\,\omega_{1\, n \ell}.
\end{gather}
Here $\omega_{0\, n \ell}$ is the GR QNM frequency, and the superscripts
$(0,1)$ and $(1,1)$ denote orders
$\mathcal{O}(\zeta^0\epsilon^1)$ and $\mathcal{O}(\zeta^1\epsilon^1)$,
respectively.

Substituting these expansions into Eq.~\eqref{eq:master} and
collecting terms at $\mathcal{O}(\zeta^0\epsilon^1)$ yields the GR
homogeneous problem
\begin{equation}
\begin{gathered}
\mathcal{L}^{(0)}_{o/e,n\ell}
\equiv
\frac{\mathrm{d}^2}{\mathrm{d} r_*^2}
+ \omega_{0\,n \ell}^2 - f(r)\,V^{o/e,\,\mathrm{GR}}_{\ell}(r),\\[4pt]
\mathcal{L}^{(0)}_{o/e,n\ell}\,\Psi^{o/e\,(0,1)}_{n\ell} = 0\,.
\end{gathered}
\end{equation}
At $\mathcal{O}(\zeta^1\epsilon^1)$ we obtain an inhomogeneous EVP
equation of the form
\begin{equation}
\mathcal{L}^{(0)}_{o/e,n\ell}\,
\Psi^{o/e,(1,1)}_{n\ell}
\;=\;
\mathcal{S}^{(1,1)}_{o/e,n\ell}
\big[\Psi^{o/e\,(0,1)}_{n\ell};\,\omega_{1 \, n \ell}\big]\,,
\label{eq:EVP}
\end{equation}
where $\mathcal{S}^{(1,1)}_{o/e,n\ell}$ encodes all EFT-induced
source terms built from the GR background, the GR eigenfunction
$\Psi^{o/e\,(0,1)}_{n\ell}$, and the frequency correction $\omega_{1\, n \ell}$.

We extract $\omega_{1\, n \ell}$ by using the bilinear form that renders
$\mathcal{L}^{(0)}_{o/e,n\ell}$ formally self-adjoint under the
appropriate QNM inner product. Schematically,
\begin{equation}
\begin{aligned}
\omega_{1\, n \ell}
  &= -\frac{
  \big\langle \Psi^{o/e\,(0,1)}_{n\ell}\big|\,
  \delta\mathcal{L}^{(1)}_{o/e,n\ell}
  \big[\Psi^{o/e\,(0,1)}_{n\ell}\big]
  \big\rangle}
  {\big\langle \Psi^{o/e\,(0,1)}_{n\ell}\big|\,
  \partial_\omega\mathcal{L}^{(0)}_{o/e,n\ell}(\omega_{0\, n \ell})\,
  \Psi^{o/e\,(0,1)}_{n\ell}
  \big\rangle}\\&\equiv\frac{N_{o/e}}{D_{o/e}}\,
  \end{aligned}
\end{equation}
where $\delta\mathcal{L}^{(1)}_{o/e,n\ell}$ is the $\mathcal{O}(\zeta)$
EFT correction to the master operator, and the numerator and
denominator $N_{o/e},D_{o/e}$ are implemented as contour integrals
over the complex $r$--plane in our EVP approach. The detailed
contour representation of the bilinear form, and its numerical
discretization, are described in Sec.~\ref{subsec:evp_contour}.

\subsubsection{Scalar Leaver method on the EFT master equation}

For the EFT-corrected master operator in Eq.~\eqref{eq:master}, the
effective potentials $\bar{V}^{o/e}_{n\ell}(r;\omega)$ vanish near $r_H$. At spatial infinity they decay with the same asymptotic
behaviour as in GR. Consequently, the asymptotic solutions still
behave as $e^{\pm i\omega \bar{r}_*}$, and we impose standard black
hole boundary conditions: purely ingoing waves at the perturbed
horizon $r_H$ and purely outgoing waves as $r\rightarrow\infty$.

The perturbed horizon radius is defined as the common positive root of
$A(r)=B(r)=0$. To linear order in $\zeta$ we write
\begin{equation}
  r_H \;=\; r_g\,\beta, 
  \qquad 
  \beta \;=\; 1 - \zeta\left(3 d_{58} + 10 d_9 + \frac{1}{2} d_{10}\right),
  \label{eq:rHbeta}
\end{equation}
and introduce the EFT-corrected surface-gravity radius
\begin{equation}
  \frac{1}{2 \kappa} \equiv r_{\mathrm{SG}} 
  \;\equiv\; r_g\bigl(1 - 2(d_{10} + 2 d_9)\,\zeta\bigr),
  \label{eq:rSG_def}
\end{equation}
which controls the phase of the near-horizon behaviour.

Following Leaver~\cite{Leaver_1985_CFM}, we peel off the non-analytic
asymptotics at the horizon and at infinity with the ansatz
\begin{equation}
  \Psi^{o/e}_{n\ell}(r) 
  \;=\; \phi(r)\,\tilde{\Psi}^{o/e}_{n\ell}(r),
  \label{eq:psi_factorization}
\end{equation}
where
\begin{equation}
  \phi(r) \;=\; 
  (r-r_H)^{- i \omega r_{\mathrm{SG}}}\,
  e^{i\omega (r - r_H)}\,
  \left(\frac{r}{r_H}\right)^{i\,\omega \,(r_g + r_{\mathrm{SG}})}.
  \label{eq:phi_prefactor}
\end{equation}
Here and throughout, when it is unambiguous we suppress some mode labels (typically $n$ and/or $\ell$) for notational clarity. We reinstate the full index set whenever needed. By construction, $\phi(r)$ enforces the boundary conditions. The
remaining factor $\tilde{\Psi}^{o/e}_\ell(r)$ is analytic at
$r=r_H$ and can be expanded in a Frobenius series. Introducing the
dimensionless radial coordinate
\begin{equation}
  x \equiv 1 - \frac{r_H}{r}\,,
\end{equation}
we write
\begin{equation}
  \tilde{\Psi}^{o/e}_{n\ell}(r) 
  \;=\; \sum_{m=0}^{\infty} a^{o/e}_m\,x^m,
  \label{eq:frobenius_series}
\end{equation}
with $a^{o/e}_0$ fixed by normalization.

Substituting Eqs.~\eqref{eq:psi_factorization}--\eqref{eq:frobenius_series}
into the master equation~\eqref{eq:master} and expanding about $x=0$
gives, for each parity sector, a linear recurrence for the coefficients
$a^{o/e}_m$. For the even and odd sectors we obtain, respectively,
\begin{align}
  \sum_{k=-10}^{+1} 
    \Gamma^{(e)}_k(m)\,a^{(e)}_{m+k} \;=\; 0,
  \label{eq:even_12term}\\
  \sum_{k=-7}^{+1} 
    \Gamma^{(o)}_{k}(m)\,a^{(o)}_{m+k} \;=\; 0,
  \label{eq:odd_9term}
\end{align}
where $m$ is the Frobenius index. The coefficients $\Gamma^{(e/o)}_k(m)$
are at most quadratic polynomials in $m$, linear in $\zeta$ and in
the couplings $\{d_5,d_8,d_9,d_{10}\}$, and they depend on the
frequency $\omega_{n\ell}$ and on the angular index $\ell$. Explicit expressions for all $\Gamma^{(e/o)}_k(m)$ are compiled in a companion \textit{Mathematica} notebook and will be made publicly available upon publication.

These ``fat'' $m$-band recurrences (with $m=12$ for even and
$m=9$ for odd) could in principle be treated using Hill determinants
or a tensorial generalization of Leaver's continued-fraction method \cite{Matyjasek_2021,alves2025criticalmassesnumericalcomputation,Rosa_2012,Kaminski_2010}. Instead, we perform an exact
numerical band reduction from the $m$-term system to an equivalent
$3$-term recurrence of the form
\begin{equation}
  \tilde{\alpha}(m)\,a_{m-1}
  + \tilde{\beta}(m)\,a_m
  + \tilde{\gamma}(m)\,a_{m+1}
  \;=\; 0,
  \label{eq:3term_reduced}
\end{equation}
for both parity sectors. Here, $(\tilde{\alpha}(m),\tilde{\beta}(m),\tilde{\gamma}(m))$ labels the $3$-term band coefficients after reduction.

We then solve for the minimal solution using a scalar Leaver continued-fraction algorithm \cite{Leaver_1985_CFM}.
Implementation details and numerical checks are discussed in Sec.~\ref{subsec:leaver_pipeline} (see also Appendix~\ref{More_Details_On_Numerical_Methods}).

\begin{widetext}
        

\newcommand{\cellfour}[4]{%
  \ensuremath{%
    \begin{array}{@{}r|r@{}}
      #1 & #2\\\hline
      #3 & #4
    \end{array}%
  }%
}

\newcommand{\paritycell}{%
  \ensuremath{%
    \begin{array}{@{}c@{}}
      \text{\textit{ even}}\\\hline
      \text{\textit{odd}}
    \end{array}%
  }%
}

\renewcommand{\arraystretch}{1.3}


\begin{table}[htbp]
  \centering
  \caption{Fractional shifts using EVP method $\delta_{\mathrm{EVP}}$ for $(n,\ell)$ with $\ell=2$.
  In each data cell, the top row corresponds to the even sector and
  the bottom row to the odd sector. The left and right entries are 
  $\Re(\delta_{\mathrm{EVP}})$ and  $\Im(\delta_{\mathrm{EVP}})$ respectively. }
  \begin{tabular}{|c|c|c|c|c|c|}
    \hline
    & & $(0,2)$ & $(1,2)$ & $(2,2)$ & $(3,2)$ \\
    \hline
    $\mathcal{O}_5$ & \paritycell &
    \cellfour{3.42\times10^{-18}}{-1.84\times10^{-17}}{2.23\times10^{-18}}{-1.18\times10^{-17}} &
    \cellfour{-8.18\times10^{-14}}{-3.64\times10^{-14}}{-5.39\times10^{-14}}{-1.85\times10^{-14}} &
    \cellfour{4.88\times10^{-11}}{5.60\times10^{-11}}{4.03\times10^{-11}}{3.34\times10^{-11}} &
    \cellfour{1.04\times10^{-8}}{-1.88\times10^{-9}}{6.34\times10^{-9}}{-1.88\times10^{-9}} \\
    \hline
    $\mathcal{O}_8$ & \paritycell &
    \cellfour{8.55\times10^{-19}}{-4.60\times10^{-18}}{5.57\times10^{-19}}{-2.94\times10^{-18}} &
    \cellfour{-2.04\times10^{-14}}{-9.10\times10^{-15}}{-1.35\times10^{-14}}{-4.63\times10^{-15}} &
    \cellfour{1.22\times10^{-11}}{1.40\times10^{-11}}{1.01\times10^{-11}}{8.35\times10^{-12}} &
    \cellfour{2.60\times10^{-9}}{-4.71\times10^{-10}}{1.59\times10^{-9}}{-4.69\times10^{-10}} \\
    \hline
    $\mathcal{O}_9$ & \paritycell &
    \cellfour{12.34}{58.12}{-21.07}{-47.53} &
    \cellfour{-10.76}{55.82}{-16.36}{-70.57} &
    \cellfour{-75.80}{54.48}{51.79}{-115.18} &
    \cellfour{-229.39}{61.44}{392.92}{-183.67} \\
    \hline
    $\mathcal{O}_{10}$ & \paritycell &
    \cellfour{6.17}{29.06}{-10.54}{-23.76} &
    \cellfour{-5.38}{27.91}{-8.18}{-35.29} &
    \cellfour{-37.90}{27.24}{25.90}{-57.59} &
    \cellfour{-114.70}{30.72}{196.46}{-91.83} \\
    \hline
  \end{tabular}
  \label{l=2_Table_EVP}
\end{table}

\begin{table}[htbp]
  \centering
  \caption{Fractional shifts $\delta_{\mathrm{EVP}}$ for $(n,\ell)$ with $\ell=3$.
  Layout as in the $\ell=2$ table.}
  \begin{tabular}{|c|c|c|c|c|c|}
    \hline
    & & $(0,3)$ & $(1,3)$ & $(2,3)$ & $(3,3)$ \\
    \hline
    $\mathcal{O}_5$ & \paritycell &
    \cellfour{1.50\times10^{-23}}{5.02\times10^{-23}}{1.46\times10^{-23}}{3.45\times10^{-23}} &
    \cellfour{3.38\times10^{-20}}{1.95\times10^{-19}}{4.50\times10^{-20}}{1.70\times10^{-19}} &
    \cellfour{1.99\times10^{-17}}{1.64\times10^{-16}}{4.42\times10^{-17}}{1.50\times10^{-16}} &
    \cellfour{3.35\times10^{-14}}{3.95\times10^{-14}}{4.37\times10^{-14}}{3.27\times10^{-14}} \\
    \hline
    $\mathcal{O}_8$ & \paritycell &
    \cellfour{3.76\times10^{-24}}{1.26\times10^{-23}}{3.66\times10^{-24}}{8.62\times10^{-24}} &
    \cellfour{8.45\times10^{-21}}{4.88\times10^{-20}}{1.13\times10^{-20}}{4.24\times10^{-20}} &
    \cellfour{4.97\times10^{-18}}{4.11\times10^{-17}}{1.10\times10^{-17}}{3.74\times10^{-17}} &
    \cellfour{8.37\times10^{-15}}{9.89\times10^{-15}}{1.09\times10^{-14}}{8.17\times10^{-15}} \\
    \hline
    $\mathcal{O}_9$ & \paritycell &
    \cellfour{14.15}{51.82}{-18.29}{-46.27} &
    \cellfour{7.99}{53.53}{-15.65}{-51.39} &
    \cellfour{-8.03}{57.09}{-5.25}{-61.31} &
    \cellfour{-42.34}{62.89}{25.14}{-75.63} \\
    \hline
    $\mathcal{O}_{10}$ & \paritycell &
    \cellfour{7.07}{25.91}{-9.14}{-23.13} &
    \cellfour{4.00}{26.77}{-7.82}{-25.70} &
    \cellfour{-4.02}{28.55}{-2.63}{-30.65} &
    \cellfour{-21.17}{31.44}{12.57}{-37.82} \\
    \hline
  \end{tabular}
  \label{l=3_Table_EVP}
\end{table}

\begin{table}[htbp]
  \centering
  \caption{Fractional shifts $\delta_{\mathrm{EVP}}$ for $(n,\ell)$ with $\ell=4$.
  Layout as in the $\ell=2$ table.}
  \begin{tabular}{|c|c|c|c|c|c|}
    \hline
    & & $(0,4)$ & $(1,4)$ & $(2,4)$ & $(3,4)$ \\
    \hline
    $\mathcal{O}_5$ & \paritycell &
    \cellfour{-1.25\times10^{-27}}{-1.05\times10^{-26}}{-1.36\times10^{-27}}{-9.11\times10^{-27}} &
    \cellfour{5.77\times10^{-24}}{-4.42\times10^{-24}}{5.48\times10^{-24}}{-6.36\times10^{-24}} &
    \cellfour{6.38\times10^{-22}}{1.01\times10^{-20}}{1.50\times10^{-21}}{9.84\times10^{-21}} &
    \cellfour{-2.18\times10^{-18}}{9.44\times10^{-19}}{-2.06\times10^{-18}}{1.31\times10^{-18}} \\
    \hline
    $\mathcal{O}_8$ & \paritycell &
    \cellfour{-3.13\times10^{-28}}{-2.63\times10^{-27}}{-3.41\times10^{-28}}{-2.28\times10^{-27}} &
    \cellfour{1.44\times10^{-24}}{-1.10\times10^{-24}}{1.37\times10^{-24}}{-1.59\times10^{-24}} &
    \cellfour{1.60\times10^{-22}}{2.51\times10^{-21}}{3.75\times10^{-22}}{2.46\times10^{-21}} &
    \cellfour{-5.45\times10^{-19}}{2.36\times10^{-19}}{-5.14\times10^{-19}}{3.28\times10^{-19}} \\
    \hline
    $\mathcal{O}_9$ & \paritycell &
    \cellfour{14.46}{50.16}{-17.55}{-45.70} &
    \cellfour{11.45}{51.35}{-15.80}{-47.80} &
    \cellfour{4.28}{53.74}{-11.07}{-51.93} &
    \cellfour{-9.49}{57.39}{-0.56}{-58.02} \\
    \hline
    $\mathcal{O}_{10}$ & \paritycell &
    \cellfour{7.23}{25.08}{-8.77}{-22.85} &
    \cellfour{5.72}{25.68}{-7.90}{-23.90} &
    \cellfour{2.14}{26.87}{-5.53}{-25.96} &
    \cellfour{-4.75}{28.70}{-0.28}{-29.01} \\
    \hline
  \end{tabular}
  \label{l=4_Table_EVP}
\end{table}
\end{widetext}

\section{Numerical methods}
\label{subsec:numerics}
In Ref.~\cite{deRham_EFT_GW_2020}, QNM shifts were evaluated using a frequency-basis expansion in which the EFT-induced potential deformation is written asymptotically as
\begin{equation}
\delta V = r_H^{-2}\sum_{j\ge 0}z_j (r_H/r)^j,
\end{equation}
so that, at linear order, the QNM frequency can be expressed as
\begin{equation}
\omega=\omega_{0}+\sum_{j\ge 0} z_j e_j,
\end{equation}
where $\delta V$ represents the $\mathcal{O}(\zeta)$ deformation of the master potential (in our notation, the $\zeta$-dependent part of $\bar V$). Here the complex coefficients \(e_j\) are precomputed numerically for each \(\ell\) and (in practice) provided up to a finite cutoff \(j_{\max}\) (e.g. \(j_{\max}=50\) in the implementation used in Ref.~\cite{deRham_EFT_GW_2020}, following ~\cite{METHOD_RHAM}) and $z_j$ are the corresponding expansion coefficients. 

This strategy is efficient for estimating physical shifts, but two systematics become critical for \emph{null} cases where the numerical floor is the answer. 
First, for even-parity perturbations the far-zone series contains infinitely many terms, so truncating the sum at finite \(j_{\max}\) generically leaves a residual that can dominate the numerical floor (for instance, Ref.~\cite{deRham_EFT_GW_2020} estimates an even-parity truncation floor of order $\mathcal{O}(10^{-3})$ for $\ell=2$ when $j_{\max}=50$); by contrast, the odd-parity deformation is a finite polynomial in \(r_H/r\) in the cases of interest, so null residuals are not truncation-dominated. 
Second, because \(e_j\) are numerically tabulated at finite precision, their use imposes an intrinsic roundoff floor: null results are cancellation-dominated residuals, and any finite-precision error in \(\alpha_j\) and/or \(e_j\) can propagate into an apparent nonzero \(\delta\omega\).

The key practical difference in our approach is that we do not rely on a truncated far-zone power-law expansion calibrated by fixed tabulated coefficients. 
Instead, we solve the (perturbative) eigenvalue problem directly using arbitrary-precision arithmetic and independent numerical approaches (EVP and generalized Leaver-type continued fractions), and we quantify the numerical floor by systematic convergence under increased precision and tightened truncation controls. 
This is essential for establishing robust tests for null operators.

In this section we summarize the two numerical approaches used in this work.

\subsection{EVP method}
\label{subsec:evp_contour}

The EVP approach proceeds as follows:
\begin{enumerate}
\item compute the GR QNM frequency $\omega_{0\, n \ell}$ and series coefficients using a continued fraction with a Nollert-type asymptotic tail \cite{Leaver_1985_CFM,Nollert1993,Nollert1999,Zhidenko2006,KonoplyaZhidenko2011}, see App.~\ref{app:nollert_tail} for more details;
\item reconstruct the RW and Zerilli (as well as Teukolsky functions for cross-check) on a complex $r$--contour, using branch-tracked prefactors. See implementation details in App.~\ref{app:series_branch_residuals};
\item evaluate a contour bilinear form to obtain the first-order frequency shift $\omega_{1\, n \ell}$ due to the EFT source operators, see Sec.~\ref{sec:setup} for EVP expressions.
\end{enumerate}

\paragraph{Complex contour and Chandrasekhar map.}

The EVP bilinear form is evaluated along a complex $r$--contour
$\mathcal{C}$. It is convenient to parameterize this contour by a
real parameter $\lambda\in[-\lambda_{\max},\lambda_{\max}]$.
We introduce a dimensionless coordinate
\begin{equation}
\rho(\lambda) \equiv \frac{r(\lambda)}{r_g}
= 1 + \frac{\lambda}{1+\lambda^2} + i\bigl(\lambda^2 - 2\bigr).
\end{equation}
For large $|\lambda|$ the contour extends to $\mathrm{Im}\,r\to +\infty$. It crosses the real axis at $\lambda=\pm\sqrt{2}$ and reaches its minimum imaginary part at $\lambda=0$, where $\mathrm{Im}\,r=-2r_g$. The derivative
$\mathrm{d}r/\mathrm{d}\lambda$ is computed spectrally on the
$\lambda$-grid and is used both in the quadrature weights and in
propagating the branch-tracked prefactors.

The Zerilli function $\Psi^{e\,(0,1)}$ is obtained via the Chandrasekhar
transformation built from the GR superpotential $W_{\rm GR}(r)$ in
Eq.~\eqref{eq:W_GR_def}.

\paragraph{Bilinear form and normalization denominator.}

For each parity sector we define the EVP denominator as
\begin{equation}
D_{\rm o/e}
= \oint_{\mathcal{C}} 2 \omega_{0}\, \bigl[\psi^{o/e}_{\rm GR}(r)\bigr]^2\,\mathrm{d}r_\ast,
\end{equation}
where the QNM inner product is understood in the sense of complex
contour integration. Parameterizing the contour by $\lambda$, we
write
\begin{equation}
\frac{\mathrm{d}r_\ast}{\mathrm{d}\lambda}
= \frac{1}{f(r(\lambda))}\,\frac{\mathrm{d}r}{\mathrm{d}\lambda}
\equiv w_r(\lambda),
\end{equation}
and approximate the integral by a trapezoidal rule on a uniform
$\lambda$--grid:
\begin{equation}
D_{\rm o/e}
\simeq \sum_{i} 2\omega_{0}\, \bigl(\psi^{o/e}_{\rm GR}\bigr)^2_{i}\,
w_{r,i}\,\Delta\lambda,
\end{equation}
with $w_{r,i}=w_r(\lambda_i)$.

The numerator $N_{\rm o/e}$ is evaluated similarly. The integrand is
assembled pointwise on the contour using the RW/Zerilli/Teukolsky
fields and their radial derivatives.

\subsubsection{EVP Discretization choices and convergence}\label{EVP Discretization choices and convergence}

The EVP contour integrals are computed on a uniform grid in
$\lambda\in[-\lambda_{\max},\lambda_{\max}]$ with $\lambda_{\max}=5$, using $N_\lambda = 2000$
sampling points and a trapezoidal rule for all contour integrals.

For the Frobenius expansions of both the RW and Teukolsky
solutions we use $N_{\rm max} = 900$ terms. The continued fraction for
the GR problem is evaluated with a Nollert-type asymptotic tail, with
base index $N_{\rm tail} = 700$, step $\Delta N = 60$, and exponent
$p = \tfrac{1}{2}$. We have also checked that slightly deeper tails
(for instance $N_{\rm tail} = 800$, $\Delta N = 100$) do not change
the results within the quoted digits.

All EVP runs use high-precision arithmetic with $100$ digits, and the GR QNM roots are converged to a
tolerance $|\delta\omega| \lesssim 10^{-30}$. As a sanity check we
evaluate the GR residuals on a real radial grid
$r \in [r_g + 10^{-5}, 100\,r_g]$ with $\mathcal{O}(10^2)$ points
(for example $N_r \simeq 160$), and verify that the residual
$\infty$--norm remains small and stable under reasonable variations of
$N_{\rm max}$, $(-\lambda_{\max},\lambda_{\max},N_\lambda)$, and
$(N_{\rm tail},\Delta N,p)$.

\subsection{Leaver method + band reduction}\label{subsec:leaver_pipeline}

Our second approach provides an independent cross-check by solving the fully EFT-corrected recurrence directly,
without contour integration. We use a band-reduced Leaver solver\footnote{Our implementation follows the logic of the public \texttt{qnm} package \cite{Stein_2019mop}, but is reimplemented end-to-end in
\texttt{mpmath}, together with an additional reduction step that handles the $m$-term banded recurrences that arise in the modified equations.
This enables arbitrary-precision complex arithmetic and allows us to push the numerical floor tests well below
that of \texttt{qnm}. By contrast, \texttt{qnm} is optimized for speed using \texttt{numpy} and \texttt{numba} \cite{Stein_2019mop}, which operate at machine precision, whereas \texttt{mpmath} targets arbitrary precision. The tradeoff is increased runtime that grows with the requested precision and reduced benefit from low-level vectorization speedups; we overcome those cons by parallelizing over different cores and caching roots.}. The implementation proceeds in two main steps:
\begin{enumerate}
\item \emph{Reduce to a three--term recurrence.}
We perform an exact numerical band reduction, via banded Gaussian
elimination, from the $m$-term recurrences to an equivalent
$3$--term recurrence.

\item \emph{Evaluate the continued fraction and find the root.}
We apply a scalar Leaver continued-fraction solver to the reduced
recurrence, using a modified Lentz's method \cite{Lentz1976,Press2007NR} for the infinite tail
and a complex secant--Newton root finder for $\omega_{n\ell}$.
\end{enumerate}

For a chosen inversion index $m_{\mathrm{inv}} \ge 1$, we write the
spectral function as
\begin{equation}
F(\omega)
  = \tilde{\beta}_{m_{\mathrm{inv}}}
    - \tilde{\gamma}_{m_{\mathrm{inv}}} C_{m_{\mathrm{inv}}-1}(\omega)
    + \tilde{\gamma}_{m_{\mathrm{inv}}} T_{m_{\mathrm{inv}}}(\omega),
\label{eq:F_def_num}
\end{equation}
where $C_{m_{\mathrm{inv}}-1}$ is the contribution from the finite
prefix of the continued fraction down to $m=0$, and
$T_{m_{\mathrm{inv}}}$ is the infinite tail starting at
$m_{\mathrm{inv}}$.

The finite prefix is computed recursively as
\begin{equation}
C_{-1} = 0, \qquad
C_k = \frac{\tilde{\alpha}_k}{\tilde{\beta}_k - \tilde{\gamma}_k C_{k-1}},
\quad k = 0,1,\dots,m_{\mathrm{inv}}-1.
\end{equation}
The tail $T_{m_{\mathrm{inv}}}$ is evaluated using a modified Lentz
algorithm for the associated $J$--fraction. In terms of the
tri-diagonal coefficients, we define
\begin{align}
A_j &= -\frac{\tilde{\alpha}_{m_{\mathrm{inv}}+j}}{\tilde{\gamma}_{m_{\mathrm{inv}}+j}},
&
B_j &= \frac{\tilde{\beta}_{m_{\mathrm{inv}}+j+1}}{\tilde{\gamma}_{m_{\mathrm{inv}}+j+1}},
\end{align}
and iterate the standard Lentz update for $f_j$, $C_j$, $D_j$
and $\Delta_j$ until the fractional change falls within the chosen
tolerance. This yields an evaluation of $F(\omega)$ together with a
tail error estimate $\varepsilon_{\text{tail}}$.

To better align the inversion index with the minimal solution of the
recurrence, we perform a cheap ``neighbor-pick'' step: given a seed
$m_{\mathrm{inv}}$, we briefly evaluate $|F(\omega)|$ for a small
window of neighboring inversion indices and choose the one that
minimizes $|F(\omega)|$. This step is inexpensive and improves the
robustness of the solver.

The core QNM condition $F(\omega)=0$ is solved using a complex
secant method with an optional Newton polish. Starting from two
jittered seeds $\omega_{0\, n \ell}$ and $\omega_{1\, n \ell}$ around a reference
frequency $\omega_{\mathrm{seed}}$, we iterate
\begin{equation}
\omega_{k+1}
  = \omega_k
    - F(\omega_k)\,\frac{\omega_k - \omega_{k-1}}{F(\omega_k)-F(\omega_{k-1})},
\end{equation}
until $|F(\omega_k)|$ falls below a target root tolerance. If the
secant denominator becomes too small, we temporarily switch to a
single Newton step with a numerically estimated derivative,
\begin{equation}
F'(\omega) \approx \frac{F(\omega+h) - F(\omega-h)}{2h},
\end{equation}
where $h$ is chosen adaptively based on the target precision and
$|\omega|$. After the secant phase, we perform one additional Newton polish on the final iterate.

We estimate the uncertainty $\delta\omega$ in the extracted
frequency by combining the nonlinear root-finding residual and the
tail truncation error. Defining $
F_{\mathrm{residual}}(\omega) \equiv F(\omega),
$ evaluated at the final iterate, we use
\begin{equation}
\delta \omega
  \sim \frac{|F_{\mathrm{residual}}(\omega)| + \varepsilon_{\text{tail}}}{|F'(\omega)|}.
\end{equation}
This provides a conservative internal error estimate for the
Leaver+reduction solver, which we compare against the EVP results and
the expected null-test behaviour of $\mathcal{O}_5$ and
$\mathcal{O}_8$.

\subsubsection{Leaver Discretization choices and convergence}\label{Leaver Discretization choices and convergence}

All runs use $80$ digits high-precision arithmetic, and strict continued-fraction and root-finding tolerances set to $10^{-60}$. The reduced tridiagonal recurrence starts from an initial depth of $10^{4}$ and is allowed to grow up to a ceiling of $2.5\times 10^{5}$ with a multiplicative growth factor of $1.6$. The solver can also shrink the working depth again when a significant fraction of the array is unused, effectively reducing the depth once more than about $40\%$ of the current depth is not needed.

For each mode, the initial guess is taken once from the tabulated Schwarzschild spectrum provided by the \texttt{qnm} library and a second time from a simple eikonal estimate, in order to compare robustness against different seed choices. The root finder uses a complex secant method followed by up to $3$ Newton iterations. The effective convergence threshold on the residual is defined as the maximum of the nominal root tolerance, a floor tied to the continued-fraction truncation error, and the intrinsic precision floor, which is approximately $10^{-78}$.
\begin{widetext}

\begin{figure}[t]
    \centering
    \includegraphics[width=\textwidth]{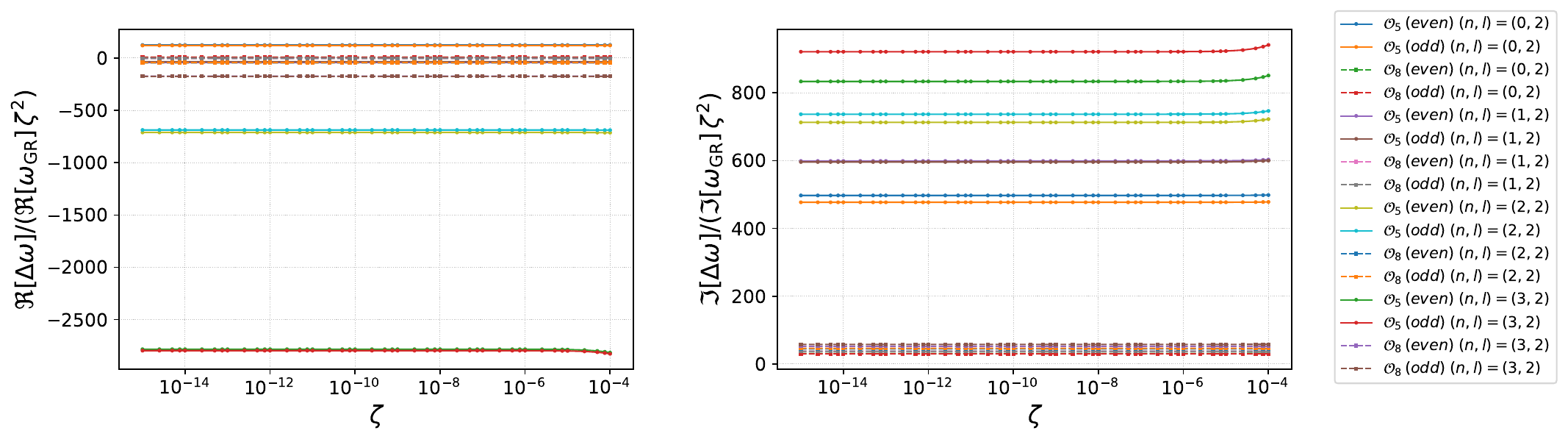}
    \caption{Scaling test of the EFT-induced frequency shift for $\mathcal{O}_{5}$ and $\mathcal{O}_{8}$ at $\ell=2$. The left (right) panel shows the real (imaginary) part of the reduced quantity $\delta_{\rm Leaver}/\zeta^{2}$ as a function of the coupling $\zeta$. The near-constancy of these curves over a wide range of $\zeta$ indicates that the leading correction scales quadratically, $\delta_{\rm Leaver}\propto \zeta^{2}$. Deviations from flatness at the largest couplings signal the onset of non-asymptotic (higher-order) contributions beyond the leading power.}
    \label{fig:null58_l2}
\end{figure}

\begin{figure}[t]
    \centering
    \includegraphics[width=\textwidth]{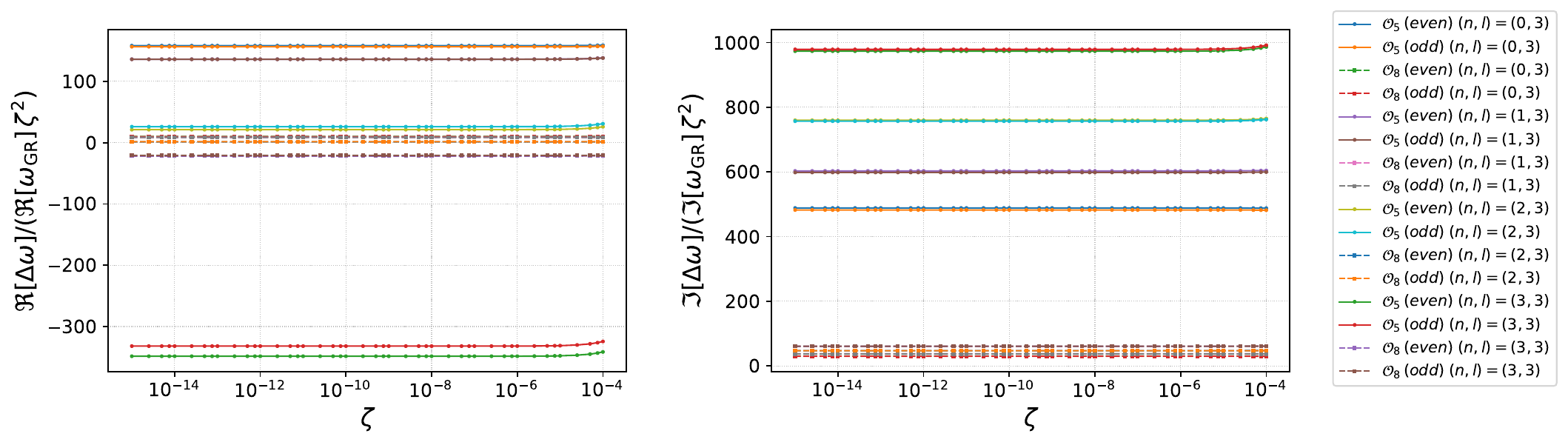}
    \caption{
        Same as Fig.~\ref{fig:null58_l2} but for $\ell = 3$.
    }
    \label{fig:null58_l3}
\end{figure}

\begin{figure}[t]
    \centering
    \includegraphics[width=\textwidth]{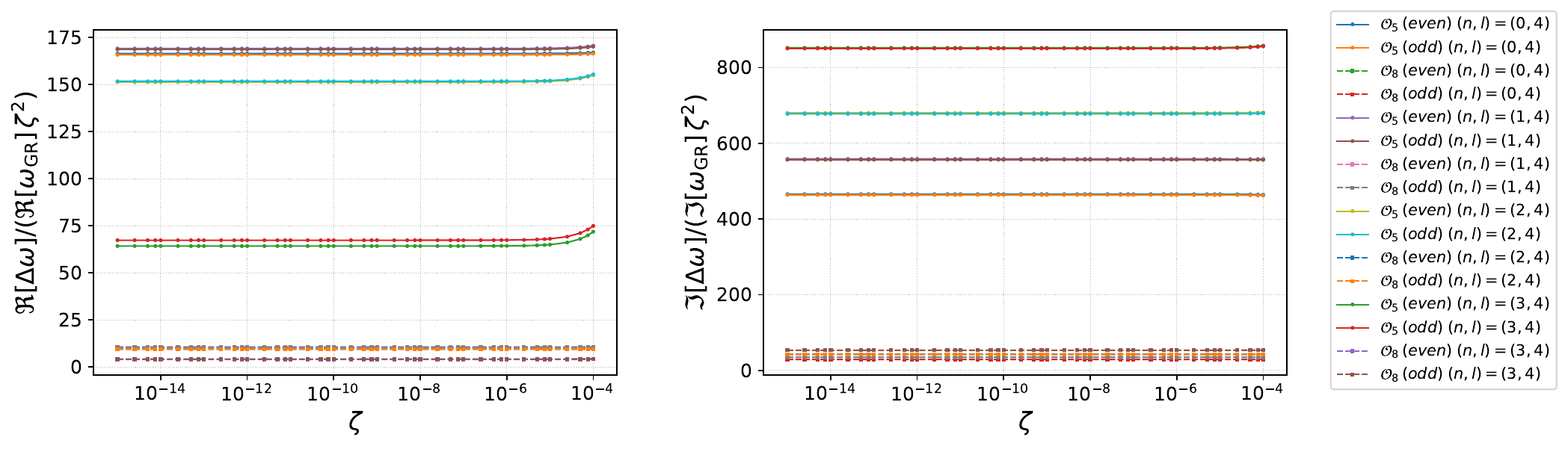}
    \caption{
        Same as Fig.~\ref{fig:null58_l2} but for $\ell = 4$.
    }
    \label{fig:null58_l4}
\end{figure}

\end{widetext}

\section{Numerical Results}\label{numerical_results}

In this section, we present QNM shifts for overtones $n=0,1,2,3$ and multipoles $\ell=2,3,4$ in both even and odd sectors. We work in units $r_g=1$ throughout, so that the $n=0$ modes can be compared directly with results of Ref.~\cite{deRham_EFT_GW_2020}. 

\subsection{EVP}
Using the EVP approach on the $\lambda$–-contour described in Sec.~\ref{subsec:evp_contour}, we compute the first–-order fractional shifts $\delta_{\mathrm{EVP}}$ for all modes in our grid. Following Eq.~(4.7) of Ref.~\cite{deRham_EFT_GW_2020}, we report
the dimensionless \emph{fractional shift}
\begin{equation}
\delta_{\mathrm{EVP}} \equiv \left( \frac{\Re(\omega_{1})}{\Re(\omega_{0})} , \frac{\Im(\omega_{1})}{\Im(\omega_{0})} \right).
\label{eq:delta_def}
\end{equation}
The numerical values are reported in Tables~\ref{l=2_Table_EVP}, \ref{l=3_Table_EVP} and~\ref{l=4_Table_EVP}.

\subsubsection{Null operators}

In theory, for the null operators $\mathcal{O}_5$ and $\mathcal{O}_8$, the linear response of the spectrum in vacuum should vanish. Analyzing $\delta_{{\mathrm{EVP}}}$ for    all $\ell$ and $n$, we find
\begin{equation}
\begin{aligned}
|\delta_{{\mathrm{EVP}},\min}| &\sim 10^{-28},\\
|\delta_{{\mathrm{EVP}},\mathrm{median}}| &\sim 10^{-18},\\ 
|\delta_{{\mathrm{EVP}},\max}| &\lesssim 10^{-8}.
\end{aligned}
\end{equation}
The even and odd sectors show the same qualitative behaviour, with the odd sector differing only by factors of order unity.

Two systematic trends are apparent. First, at fixed $\ell$, $|\delta_{\mathrm{EVP}}^{(5,8)}|$ increases as one moves to higher overtones $n$. Second, at fixed overtone $n$, $|\delta_{\mathrm{EVP}}^{(5,8)}|$ decreases as $\ell$ increases. In other words, the largest deviations from the ideal null result occur in the corner of the grid $(\ell=2,n=3)$, while higher-$\ell$ fundamentals are closest to zero.

The first pattern is consistent with the fact that higher overtones are intrinsically more challenging: they are more sensitive to root-finding errors, and small inaccuracies in $\omega_{0}$ propagate into the reconstruction of the scalar fields and the contour integrals, leading to cumulative systematic errors. We do not attempt to separate these contributions in detail here; instead we interpret the observed $|\delta_{ EVP}^{(5,8)}|$ as a practical measure of the numerical floor of the EVP method.

It is instructive to compare with the results of Ref.~\cite{deRham_EFT_GW_2020} that uses method developed in \cite{METHOD_RHAM} to compute the QNMs. They have focused on the fundamental modes ($n=0$) with $\ell=2,3,4$ and reported maximal null-operator shifts of order $10^{-4}$ in the even sector and $10^{-6}$ in the odd sector. Over the same set of modes, our EVP implementation reduces these residuals by several orders of magnitude, yielding $|\delta_{\mathrm{EVP}}|\lesssim 10^{-17}$ while simultaneously extending the null test to overtones up to $n=3$ without loss of control. We speculate that the improvement is primarily due to avoiding the series truncation error used In Ref.~\cite{deRham_EFT_GW_2020} and ~\cite{METHOD_RHAM}: in EVP we compute the shift directly from the perturbed eigenvalue problem at high precision, and verify convergence so the residual in null cases is driven down to a much lower numerical floor.

\subsubsection{Control Operators}
Up to an overall sign (that can be traced to a difference in
conventions), the EVP results for $\mathcal{O}_9$ and $\mathcal{O}_{10}$ operators gives a
fractional shifts $\delta_{\mathrm{EVP}}$ are $\mathcal{O}(10^0-10^2)$. They agree with those of
Ref.~\cite{deRham_EFT_GW_2020} in magnitude for both the real and
imaginary parts upon rounding to the two decimal places reported in their Table~I. 

Moreover, across all ratios between $\omega_{1}$ due to $\mathcal{O}_{9}$ to $\mathcal{O}_{10}$ are essentially $2$, again with maximum absolute deviation observed is $\max |\mathcal{R}_{9/10}-2|\sim 10^{-11}$. Generally, the deviation was increasing with higher overtone.

Finally, we stress that the EVP computation only probes the linear
response of the spectrum, encoded in the first--order coefficients $\omega_{1}$
and hence in $\delta_{\mathrm{EVP}}$.  By construction, this method cannot determine the range in which the expansion
\begin{equation}
\omega(\zeta) = \omega_{0} + \zeta\,\omega_{1} + \zeta^{2}\omega_{2} + \mathcal{O}(\zeta^{3})
\end{equation}
is well-approximated by its linear truncation.

\begin{widetext}
\begin{table*}[t]
\centering
\caption{Power-scaling summary for $\zeta_{\max}=10^{-3}$, odd parity, and null operators.
Each $p$ entry is $p_{\rm eff}\pm\sigma_p$ with the chosen discrete $p_\star$ shown as $[p_\star]$.
$C$ lists the fitted coefficient components $\Re(C)\,|\,\Im(C)$ and $\sigma/\mu$ is the flatness ratio.}
\label{tab:power_1Em3_odd_null}
\setlength{\tabcolsep}{4pt}
\renewcommand{\arraystretch}{1.15}
\begin{tabular}{cc||ccc|ccc}
\toprule
 &  & \multicolumn{3}{c}{$\mathcal{O}_{5}$} & \multicolumn{3}{c}{$\mathcal{O}_{8}$} \\
$\ell$ & $n$ & $p$ & $C$ & $\sigma/\mu$ & $p$ & $C$ & $\sigma/\mu$ \\
\midrule
2 & 0 & $2.0002\pm0.0007\,[2]$ & $\makebox[4em][r]{\ensuremath{121.79}}\,\vert\,\makebox[4em][l]{\ensuremath{479.64}}$ & $1.08\times10^{-3}$ & $2.0001\pm0.0002\,[2]$ & $\makebox[4em][r]{\ensuremath{7.52}}\,\vert\,\makebox[4em][l]{\ensuremath{29.86}}$ & $2.72\times10^{-4}$ \\
2 & 1 & $2.0006\pm0.0019\,[2]$ & $\makebox[4em][r]{\ensuremath{-31.52}}\,\vert\,\makebox[4em][l]{\ensuremath{605.84}}$ & $2.96\times10^{-3}$ & $2.0001\pm0.0005\,[2]$ & $\makebox[4em][r]{\ensuremath{-2.17}}\,\vert\,\makebox[4em][l]{\ensuremath{37.40}}$ & $7.33\times10^{-4}$ \\
2 & 2 & $2.0007\pm0.0023\,[2]$ & $\makebox[4em][r]{\ensuremath{-690.29}}\,\vert\,\makebox[4em][l]{\ensuremath{761.17}}$ & $3.99\times10^{-3}$ & $2.0002\pm0.0006\,[2]$ & $\makebox[4em][r]{\ensuremath{-43.01}}\,\vert\,\makebox[4em][l]{\ensuremath{46.43}}$ & $9.83\times10^{-4}$ \\
2 & 3 & $2.0010\pm0.0034\,[2]$ & $\makebox[4em][r]{\ensuremath{-2872.44}}\,\vert\,\makebox[4em][l]{\ensuremath{971.75}}$ & $5.11\times10^{-3}$ & $2.0003\pm0.0009\,[2]$ & $\makebox[4em][r]{\ensuremath{-175.94}}\,\vert\,\makebox[4em][l]{\ensuremath{58.34}}$ & $1.26\times10^{-3}$ \\
3 & 0 & $1.9999\pm0.0002\,[2]$ & $\makebox[4em][r]{\ensuremath{158.36}}\,\vert\,\makebox[4em][l]{\ensuremath{481.01}}$ & $7.61\times10^{-4}$ & $2.0000\pm0.0001\,[2]$ & $\makebox[4em][r]{\ensuremath{9.82}}\,\vert\,\makebox[4em][l]{\ensuremath{30.14}}$ & $1.86\times10^{-4}$ \\
3 & 1 & $2.0002\pm0.0008\,[2]$ & $\makebox[4em][r]{\ensuremath{141.23}}\,\vert\,\makebox[4em][l]{\ensuremath{601.60}}$ & $1.60\times10^{-3}$ & $2.0001\pm0.0002\,[2]$ & $\makebox[4em][r]{\ensuremath{8.59}}\,\vert\,\makebox[4em][l]{\ensuremath{37.46}}$ & $4.01\times10^{-4}$ \\
3 & 2 & $2.0006\pm0.0020\,[2]$ & $\makebox[4em][r]{\ensuremath{38.11}}\,\vert\,\makebox[4em][l]{\ensuremath{769.90}}$ & $3.79\times10^{-3}$ & $2.0002\pm0.0005\,[2]$ & $\makebox[4em][r]{\ensuremath{1.82}}\,\vert\,\makebox[4em][l]{\ensuremath{47.53}}$ & $9.38\times10^{-4}$ \\
3 & 3 & $2.0008\pm0.0027\,[2]$ & $\makebox[4em][r]{\ensuremath{-312.50}}\,\vert\,\makebox[4em][l]{\ensuremath{1011.00}}$ & $5.94\times10^{-3}$ & $2.0002\pm0.0007\,[2]$ & $\makebox[4em][r]{\ensuremath{-20.46}}\,\vert\,\makebox[4em][l]{\ensuremath{61.69}}$ & $1.45\times10^{-3}$ \\
4 & 0 & $1.9998\pm0.0007\,[2]$ & $\makebox[4em][r]{\ensuremath{167.22}}\,\vert\,\makebox[4em][l]{\ensuremath{459.85}}$ & $1.29\times10^{-3}$ & $1.9999\pm0.0002\,[2]$ & $\makebox[4em][r]{\ensuremath{10.39}}\,\vert\,\makebox[4em][l]{\ensuremath{28.91}}$ & $3.17\times10^{-4}$ \\
4 & 1 & $2.0000\pm0.0000\,[2]$ & $\makebox[4em][r]{\ensuremath{172.36}}\,\vert\,\makebox[4em][l]{\ensuremath{555.50}}$ & $1.09\times10^{-3}$ & $2.0000\pm0.0000\,[2]$ & $\makebox[4em][r]{\ensuremath{10.60}}\,\vert\,\makebox[4em][l]{\ensuremath{34.77}}$ & $2.69\times10^{-4}$ \\
4 & 2 & $2.0003\pm0.0009\,[2]$ & $\makebox[4em][r]{\ensuremath{160.86}}\,\vert\,\makebox[4em][l]{\ensuremath{681.24}}$ & $2.33\times10^{-3}$ & $2.0001\pm0.0002\,[2]$ & $\makebox[4em][r]{\ensuremath{9.63}}\,\vert\,\makebox[4em][l]{\ensuremath{42.41}}$ & $5.80\times10^{-4}$ \\
4 & 3 & $2.0006\pm0.0021\,[2]$ & $\makebox[4em][r]{\ensuremath{86.46}}\,\vert\,\makebox[4em][l]{\ensuremath{864.01}}$ & $4.53\times10^{-3}$ & $2.0002\pm0.0005\,[2]$ & $\makebox[4em][r]{\ensuremath{4.50}}\,\vert\,\makebox[4em][l]{\ensuremath{53.37}}$ & $1.12\times10^{-3}$ \\
\bottomrule
\end{tabular}
\end{table*}
\begin{table*}[t]
\centering
\caption{Power-scaling summary for $\zeta_{\max}=10^{-3}$, $\mathrm{even}$ parity, and null operators. Layout as in \ref{tab:power_1Em3_odd_null}.}
\label{tab:power_1Em3_even_null}
\setlength{\tabcolsep}{4pt}
\renewcommand{\arraystretch}{1.15}
\begin{tabular}{cc||ccc|ccc}
\toprule
 &  & \multicolumn{3}{c}{$\mathcal{O}_{5}$} & \multicolumn{3}{c}{$\mathcal{O}_{8}$} \\
$\ell$ & $n$ & $p$ & $C$ & $\sigma/\mu$ & $p$ & $C$ & $\sigma/\mu$ \\
\midrule
2 & 0 & $2.0002\pm0.0007\,[2]$ & $\makebox[4em][r]{\ensuremath{126.58}}\,\vert\,\makebox[4em][l]{\ensuremath{499.91}}$ & $1.09\times10^{-3}$ & $2.0001\pm0.0002\,[2]$ & $\makebox[4em][r]{\ensuremath{7.81}}\,\vert\,\makebox[4em][l]{\ensuremath{31.12}}$ & $2.74\times10^{-4}$ \\
2 & 1 & $2.0006\pm0.0019\,[2]$ & $\makebox[4em][r]{\ensuremath{-39.11}}\,\vert\,\makebox[4em][l]{\ensuremath{609.46}}$ & $3.08\times10^{-3}$ & $2.0001\pm0.0005\,[2]$ & $\makebox[4em][r]{\ensuremath{-2.64}}\,\vert\,\makebox[4em][l]{\ensuremath{37.60}}$ & $7.64\times10^{-4}$ \\
2 & 2 & $2.0007\pm0.0024\,[2]$ & $\makebox[4em][r]{\ensuremath{-716.55}}\,\vert\,\makebox[4em][l]{\ensuremath{736.74}}$ & $4.02\times10^{-3}$ & $2.0002\pm0.0006\,[2]$ & $\makebox[4em][r]{\ensuremath{-44.53}}\,\vert\,\makebox[4em][l]{\ensuremath{44.92}}$ & $9.89\times10^{-4}$ \\
2 & 3 & $2.0011\pm0.0036\,[2]$ & $\makebox[4em][r]{\ensuremath{-2863.94}}\,\vert\,\makebox[4em][l]{\ensuremath{877.71}}$ & $5.24\times10^{-3}$ & $2.0003\pm0.0009\,[2]$ & $\makebox[4em][r]{\ensuremath{-175.17}}\,\vert\,\makebox[4em][l]{\ensuremath{52.77}}$ & $1.29\times10^{-3}$ \\
3 & 0 & $1.9999\pm0.0002\,[2]$ & $\makebox[4em][r]{\ensuremath{160.04}}\,\vert\,\makebox[4em][l]{\ensuremath{486.96}}$ & $7.76\times10^{-4}$ & $2.0000\pm0.0001\,[2]$ & $\makebox[4em][r]{\ensuremath{9.92}}\,\vert\,\makebox[4em][l]{\ensuremath{30.52}}$ & $1.89\times10^{-4}$ \\
3 & 1 & $2.0002\pm0.0008\,[2]$ & $\makebox[4em][r]{\ensuremath{141.43}}\,\vert\,\makebox[4em][l]{\ensuremath{606.48}}$ & $1.63\times10^{-3}$ & $2.0001\pm0.0002\,[2]$ & $\makebox[4em][r]{\ensuremath{8.59}}\,\vert\,\makebox[4em][l]{\ensuremath{37.76}}$ & $4.08\times10^{-4}$ \\
3 & 2 & $2.0006\pm0.0020\,[2]$ & $\makebox[4em][r]{\ensuremath{33.12}}\,\vert\,\makebox[4em][l]{\ensuremath{772.96}}$ & $3.83\times10^{-3}$ & $2.0002\pm0.0005\,[2]$ & $\makebox[4em][r]{\ensuremath{1.51}}\,\vert\,\makebox[4em][l]{\ensuremath{47.70}}$ & $9.47\times10^{-4}$ \\
3 & 3 & $2.0008\pm0.0027\,[2]$ & $\makebox[4em][r]{\ensuremath{-330.29}}\,\vert\,\makebox[4em][l]{\ensuremath{1005.92}}$ & $5.87\times10^{-3}$ & $2.0002\pm0.0007\,[2]$ & $\makebox[4em][r]{\ensuremath{-21.51}}\,\vert\,\makebox[4em][l]{\ensuremath{61.36}}$ & $1.43\times10^{-3}$ \\
4 & 0 & $1.9998\pm0.0007\,[2]$ & $\makebox[4em][r]{\ensuremath{167.87}}\,\vert\,\makebox[4em][l]{\ensuremath{461.82}}$ & $1.31\times10^{-3}$ & $1.9999\pm0.0002\,[2]$ & $\makebox[4em][r]{\ensuremath{10.43}}\,\vert\,\makebox[4em][l]{\ensuremath{29.04}}$ & $3.21\times10^{-4}$ \\
4 & 1 & $2.0000\pm0.0000\,[2]$ & $\makebox[4em][r]{\ensuremath{172.87}}\,\vert\,\makebox[4em][l]{\ensuremath{557.51}}$ & $1.09\times10^{-3}$ & $2.0000\pm0.0000\,[2]$ & $\makebox[4em][r]{\ensuremath{10.63}}\,\vert\,\makebox[4em][l]{\ensuremath{34.90}}$ & $2.70\times10^{-4}$ \\
4 & 2 & $2.0003\pm0.0009\,[2]$ & $\makebox[4em][r]{\ensuremath{160.51}}\,\vert\,\makebox[4em][l]{\ensuremath{683.69}}$ & $2.34\times10^{-3}$ & $2.0001\pm0.0002\,[2]$ & $\makebox[4em][r]{\ensuremath{9.60}}\,\vert\,\makebox[4em][l]{\ensuremath{42.56}}$ & $5.84\times10^{-4}$ \\
4 & 3 & $2.0006\pm0.0021\,[2]$ & $\makebox[4em][r]{\ensuremath{83.30}}\,\vert\,\makebox[4em][l]{\ensuremath{865.99}}$ & $4.54\times10^{-3}$ & $2.0002\pm0.0005\,[2]$ & $\makebox[4em][r]{\ensuremath{4.30}}\,\vert\,\makebox[4em][l]{\ensuremath{53.48}}$ & $1.12\times10^{-3}$ \\
\bottomrule
\end{tabular}
\end{table*}
\begin{table*}[t]
\centering
\caption{Power-scaling summary for $\zeta_{\max}=10^{-3}$, odd parity, and control operators. Layout as in \ref{tab:power_1Em3_odd_null}.}
\label{tab:power_1Em3_odd_control}
\setlength{\tabcolsep}{4pt}
\renewcommand{\arraystretch}{1.15}
\begin{tabular}{cc||ccc|ccc}
\toprule
 &  & \multicolumn{3}{c}{$\mathcal{O}_{9}$} & \multicolumn{3}{c}{$\mathcal{O}_{10}$} \\
$\ell$ & $n$ & $p$ & $C$ & $\sigma/\mu$ & $p$ & $C$ & $\sigma/\mu$ \\
\midrule
2 & 0 & $0.9992\pm0.0027\,[1]$ & $\makebox[4em][r]{\ensuremath{21.25}}\,\vert\,\makebox[4em][l]{\ensuremath{46.31}}$ & $4.54\times10^{-3}$ & $0.9997\pm0.0011\,[1]$ & $\makebox[4em][r]{\ensuremath{10.53}}\,\vert\,\makebox[4em][l]{\ensuremath{23.54}}$ & $1.66\times10^{-3}$ \\
2 & 1 & $0.9999\pm0.0006\,[1]$ & $\makebox[4em][r]{\ensuremath{20.40}}\,\vert\,\makebox[4em][l]{\ensuremath{69.28}}$ & $1.12\times10^{-2}$ & $0.9997\pm0.0009\,[1]$ & $\makebox[4em][r]{\ensuremath{8.95}}\,\vert\,\makebox[4em][l]{\ensuremath{34.85}}$ & $4.66\times10^{-3}$ \\
2 & 2 & $1.0015\pm0.0049\,[1]$ & $\makebox[4em][r]{\ensuremath{-16.24}}\,\vert\,\makebox[4em][l]{\ensuremath{129.38}}$ & $5.74\times10^{-2}$ & $1.0004\pm0.0012\,[1]$ & $\makebox[4em][r]{\ensuremath{-18.80}}\,\vert\,\makebox[4em][l]{\ensuremath{60.84}}$ & $2.35\times10^{-2}$ \\
2 & 3 & $0.9896\pm0.0568\,[1]$ & $\makebox[4em][r]{\ensuremath{-58.05}}\,\vert\,\makebox[4em][l]{\ensuremath{284.37}}$ & $1.54\times10^{-1}$ & $0.9982\pm0.0126\,[1]$ & $\makebox[4em][r]{\ensuremath{-162.20}}\,\vert\,\makebox[4em][l]{\ensuremath{124.76}}$ & $4.11\times10^{-2}$ \\
3 & 0 & $0.9991\pm0.0031\,[1]$ & $\makebox[4em][r]{\ensuremath{18.35}}\,\vert\,\makebox[4em][l]{\ensuremath{45.04}}$ & $4.74\times10^{-3}$ & $0.9997\pm0.0010\,[1]$ & $\makebox[4em][r]{\ensuremath{9.13}}\,\vert\,\makebox[4em][l]{\ensuremath{22.94}}$ & $1.49\times10^{-3}$ \\
3 & 1 & $0.9989\pm0.0038\,[1]$ & $\makebox[4em][r]{\ensuremath{16.52}}\,\vert\,\makebox[4em][l]{\ensuremath{49.59}}$ & $7.16\times10^{-3}$ & $0.9994\pm0.0020\,[1]$ & $\makebox[4em][r]{\ensuremath{7.92}}\,\vert\,\makebox[4em][l]{\ensuremath{25.26}}$ & $3.18\times10^{-3}$ \\
3 & 2 & $0.9999\pm0.0010\,[1]$ & $\makebox[4em][r]{\ensuremath{10.54}}\,\vert\,\makebox[4em][l]{\ensuremath{60.32}}$ & $1.68\times10^{-2}$ & $0.9997\pm0.0012\,[1]$ & $\makebox[4em][r]{\ensuremath{3.65}}\,\vert\,\makebox[4em][l]{\ensuremath{30.29}}$ & $6.80\times10^{-3}$ \\
3 & 3 & $1.0011\pm0.0031\,[1]$ & $\makebox[4em][r]{\ensuremath{-5.21}}\,\vert\,\makebox[4em][l]{\ensuremath{81.24}}$ & $4.95\times10^{-2}$ & $1.0001\pm0.0003\,[1]$ & $\makebox[4em][r]{\ensuremath{-8.51}}\,\vert\,\makebox[4em][l]{\ensuremath{39.02}}$ & $2.03\times10^{-2}$ \\
4 & 0 & $0.9991\pm0.0031\,[1]$ & $\makebox[4em][r]{\ensuremath{17.57}}\,\vert\,\makebox[4em][l]{\ensuremath{44.51}}$ & $4.68\times10^{-3}$ & $0.9997\pm0.0009\,[1]$ & $\makebox[4em][r]{\ensuremath{8.76}}\,\vert\,\makebox[4em][l]{\ensuremath{22.68}}$ & $1.34\times10^{-3}$ \\
4 & 1 & $0.9988\pm0.0041\,[1]$ & $\makebox[4em][r]{\ensuremath{16.13}}\,\vert\,\makebox[4em][l]{\ensuremath{46.10}}$ & $6.62\times10^{-3}$ & $0.9995\pm0.0018\,[1]$ & $\makebox[4em][r]{\ensuremath{7.91}}\,\vert\,\makebox[4em][l]{\ensuremath{23.54}}$ & $2.71\times10^{-3}$ \\
4 & 2 & $0.9988\pm0.0043\,[1]$ & $\makebox[4em][r]{\ensuremath{12.90}}\,\vert\,\makebox[4em][l]{\ensuremath{49.84}}$ & $1.00\times10^{-2}$ & $0.9993\pm0.0024\,[1]$ & $\makebox[4em][r]{\ensuremath{5.82}}\,\vert\,\makebox[4em][l]{\ensuremath{25.42}}$ & $4.43\times10^{-3}$ \\
4 & 3 & $0.9996\pm0.0020\,[1]$ & $\makebox[4em][r]{\ensuremath{7.06}}\,\vert\,\makebox[4em][l]{\ensuremath{56.95}}$ & $2.18\times10^{-2}$ & $0.9996\pm0.0016\,[1]$ & $\makebox[4em][r]{\ensuremath{1.57}}\,\vert\,\makebox[4em][l]{\ensuremath{28.64}}$ & $8.89\times10^{-3}$ \\
\bottomrule
\end{tabular}
\end{table*}
\begin{table*}[t]
\centering
\caption{Power-scaling summary for $\zeta_{\max}=10^{-3}$, $\mathrm{even}$ parity, and control operators. Layout as in \ref{tab:power_1Em3_odd_null}.}
\label{tab:power_1Em3_even_control}
\setlength{\tabcolsep}{4pt}
\renewcommand{\arraystretch}{1.15}
\begin{tabular}{cc||ccc|ccc}
\toprule
 &  & \multicolumn{3}{c}{$\mathcal{O}_{9}$} & \multicolumn{3}{c}{$\mathcal{O}_{10}$} \\
$\ell$ & $n$ & $p$ & $C$ & $\sigma/\mu$ & $p$ & $C$ & $\sigma/\mu$ \\
\midrule
2 & 0 & $1.0000\pm0.0001\,[1]$ & $\makebox[4em][r]{\ensuremath{-12.38}}\,\vert\,\makebox[4em][l]{\ensuremath{-58.15}}$ & $1.73\times10^{-4}$ & $1.0000\pm0.0001\,[1]$ & $\makebox[4em][r]{\ensuremath{-6.15}}\,\vert\,\makebox[4em][l]{\ensuremath{-29.08}}$ & $1.97\times10^{-4}$ \\
2 & 1 & $1.0003\pm0.0009\,[1]$ & $\makebox[4em][r]{\ensuremath{11.36}}\,\vert\,\makebox[4em][l]{\ensuremath{-56.10}}$ & $2.27\times10^{-3}$ & $1.0000\pm0.0001\,[1]$ & $\makebox[4em][r]{\ensuremath{5.63}}\,\vert\,\makebox[4em][l]{\ensuremath{-27.83}}$ & $1.79\times10^{-3}$ \\
2 & 2 & $1.0001\pm0.0003\,[1]$ & $\makebox[4em][r]{\ensuremath{78.47}}\,\vert\,\makebox[4em][l]{\ensuremath{-50.92}}$ & $9.23\times10^{-3}$ & $0.9998\pm0.0006\,[1]$ & $\makebox[4em][r]{\ensuremath{38.56}}\,\vert\,\makebox[4em][l]{\ensuremath{-25.92}}$ & $6.10\times10^{-3}$ \\
2 & 3 & $0.9963\pm0.0122\,[1]$ & $\makebox[4em][r]{\ensuremath{211.66}}\,\vert\,\makebox[4em][l]{\ensuremath{-48.07}}$ & $1.82\times10^{-2}$ & $0.9976\pm0.0080\,[1]$ & $\makebox[4em][r]{\ensuremath{108.75}}\,\vert\,\makebox[4em][l]{\ensuremath{-26.40}}$ & $1.20\times10^{-2}$ \\
3 & 0 & $0.9998\pm0.0006\,[1]$ & $\makebox[4em][r]{\ensuremath{-14.19}}\,\vert\,\makebox[4em][l]{\ensuremath{-51.56}}$ & $9.37\times10^{-4}$ & $1.0000\pm0.0001\,[1]$ & $\makebox[4em][r]{\ensuremath{-7.07}}\,\vert\,\makebox[4em][l]{\ensuremath{-25.93}}$ & $1.91\times10^{-4}$ \\
3 & 1 & $1.0005\pm0.0017\,[1]$ & $\makebox[4em][r]{\ensuremath{-8.10}}\,\vert\,\makebox[4em][l]{\ensuremath{-54.22}}$ & $2.46\times10^{-3}$ & $1.0002\pm0.0008\,[1]$ & $\makebox[4em][r]{\ensuremath{-3.94}}\,\vert\,\makebox[4em][l]{\ensuremath{-26.93}}$ & $1.25\times10^{-3}$ \\
3 & 2 & $1.0007\pm0.0023\,[1]$ & $\makebox[4em][r]{\ensuremath{9.42}}\,\vert\,\makebox[4em][l]{\ensuremath{-57.91}}$ & $5.36\times10^{-3}$ & $1.0002\pm0.0005\,[1]$ & $\makebox[4em][r]{\ensuremath{4.54}}\,\vert\,\makebox[4em][l]{\ensuremath{-28.58}}$ & $3.49\times10^{-3}$ \\
3 & 3 & $1.0006\pm0.0019\,[1]$ & $\makebox[4em][r]{\ensuremath{47.62}}\,\vert\,\makebox[4em][l]{\ensuremath{-60.43}}$ & $1.48\times10^{-2}$ & $1.0001\pm0.0002\,[1]$ & $\makebox[4em][r]{\ensuremath{22.74}}\,\vert\,\makebox[4em][l]{\ensuremath{-30.42}}$ & $9.53\times10^{-3}$ \\
4 & 0 & $0.9997\pm0.0012\,[1]$ & $\makebox[4em][r]{\ensuremath{-14.48}}\,\vert\,\makebox[4em][l]{\ensuremath{-49.70}}$ & $1.70\times10^{-3}$ & $1.0000\pm0.0000\,[1]$ & $\makebox[4em][r]{\ensuremath{-7.23}}\,\vert\,\makebox[4em][l]{\ensuremath{-25.08}}$ & $1.78\times10^{-5}$ \\
4 & 1 & $1.0003\pm0.0009\,[1]$ & $\makebox[4em][r]{\ensuremath{-11.66}}\,\vert\,\makebox[4em][l]{\ensuremath{-51.66}}$ & $1.36\times10^{-3}$ & $1.0002\pm0.0007\,[1]$ & $\makebox[4em][r]{\ensuremath{-5.72}}\,\vert\,\makebox[4em][l]{\ensuremath{-25.81}}$ & $1.01\times10^{-3}$ \\
4 & 2 & $1.0009\pm0.0029\,[1]$ & $\makebox[4em][r]{\ensuremath{-4.16}}\,\vert\,\makebox[4em][l]{\ensuremath{-54.94}}$ & $4.29\times10^{-3}$ & $1.0004\pm0.0012\,[1]$ & $\makebox[4em][r]{\ensuremath{-1.98}}\,\vert\,\makebox[4em][l]{\ensuremath{-27.13}}$ & $2.15\times10^{-3}$ \\
4 & 3 & $1.0009\pm0.0030\,[1]$ & $\makebox[4em][r]{\ensuremath{11.53}}\,\vert\,\makebox[4em][l]{\ensuremath{-58.38}}$ & $7.46\times10^{-3}$ & $1.0003\pm0.0008\,[1]$ & $\makebox[4em][r]{\ensuremath{5.47}}\,\vert\,\makebox[4em][l]{\ensuremath{-28.76}}$ & $4.80\times10^{-3}$ \\
\bottomrule
\end{tabular}
\end{table*}

\begin{figure}[t]
    \centering
    \includegraphics[width=\textwidth]{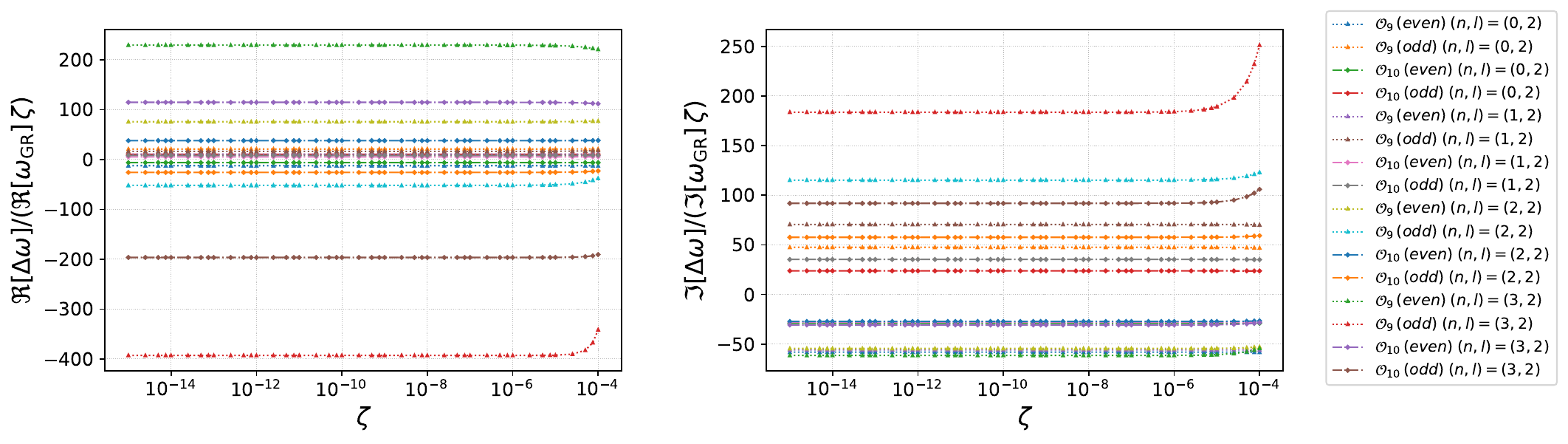}
    \caption{
Scaling test of the EFT frequency shift for $\mathcal{O}_{9}$ and $\mathcal{O}_{10}$ at $\ell=2$.
The left (right) panel shows $\Re\!\left(\delta_{\rm Leaver}/\zeta\right)$ ($\Im\!\left(\delta_{\rm Leaver}/\zeta\right)$) versus the coupling $\zeta$.
The near-flat behavior supports a leading linear dependence on $\zeta$, while departures at the largest $\zeta$ indicate the onset of higher-order contamination.
}
    \label{fig:control910_l2}
\end{figure}

\begin{figure}[t]
    \centering
    \includegraphics[width=\textwidth]{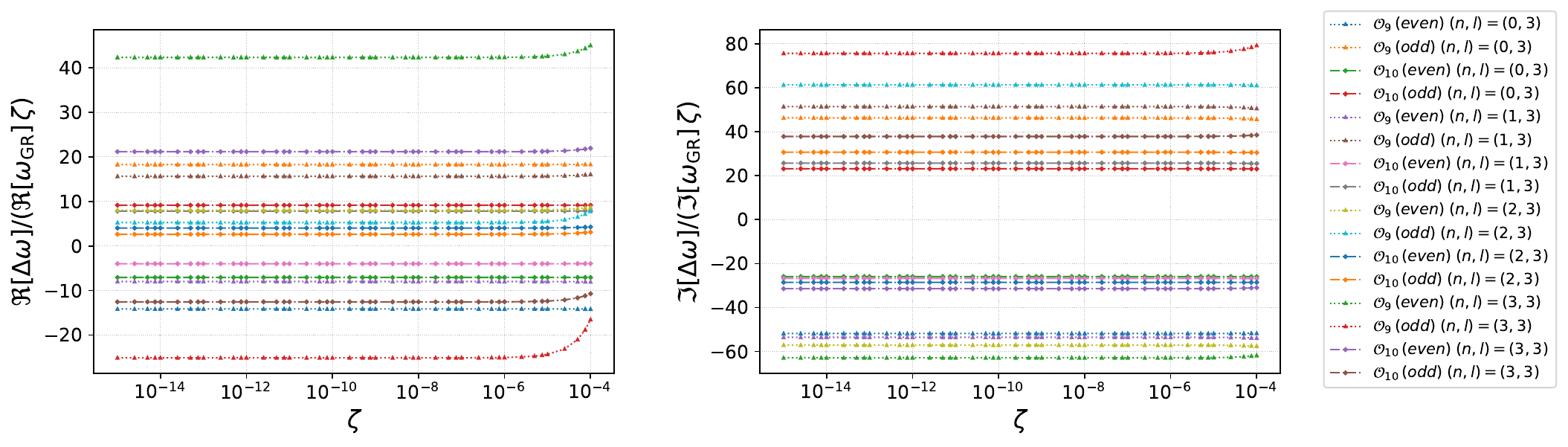}
    \caption{Same as Fig.~\ref{fig:control910_l2}, but for $\ell=3$.}
    \label{fig:control910_l3}
\end{figure}

\begin{figure}[t]
    \centering
    \includegraphics[width=\textwidth]{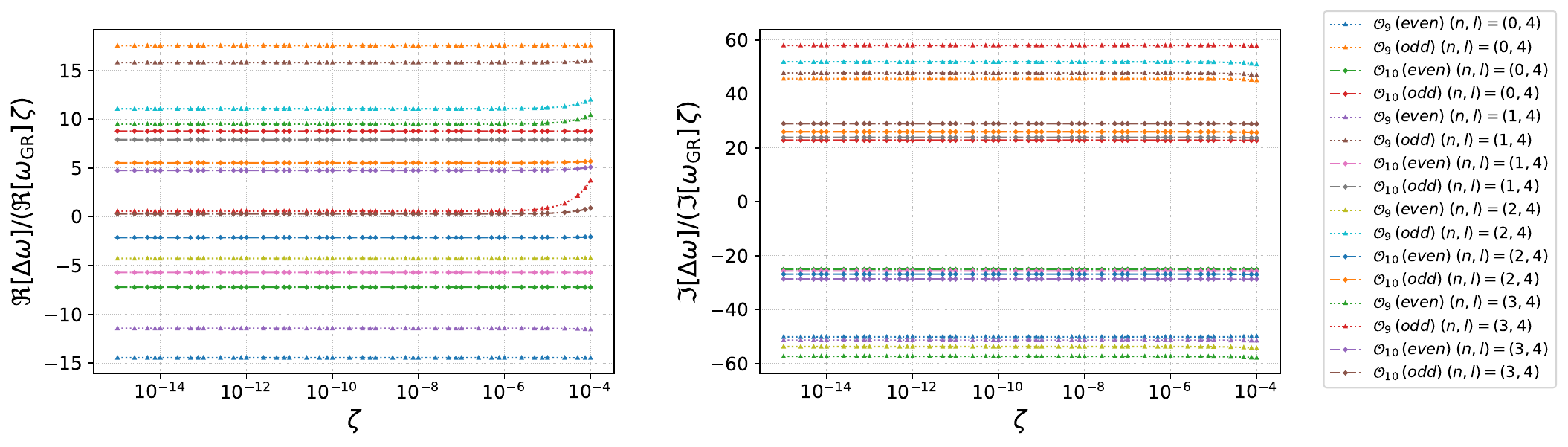}
    \caption{
       Same as Fig.~\ref{fig:control910_l2}, but for $\ell=4$.}
    \label{fig:control910_l4}
\end{figure}

\end{widetext}

\subsection{Leaver method + band reduction}

Now we compute the EFT-corrected frequency $\omega(\zeta)$ for some values $\zeta=\alpha\,10^{-p}$ with
$p\in\{3,4,\ldots,15\}$ and $\alpha\in\left\{1,\frac34,\frac12,\frac14\right\}$, keeping only
$\zeta\in[10^{-15},10^{-3}]$.

\paragraph{EFT validity and choice of \texorpdfstring{$\zeta$}{zeta} range.}
The EFT linear expansion in $\zeta$ is only meaningful as long as the background is
not strongly deformed.  An out-of-envelope diagnostic is the correction to the horizon
radius $r_H$.  For example, we
find that at $d_5\zeta=10^{-2}$ the horizon is shifted by about $12\%$ from $r_g$,
while for $d_9\zeta=10^{-2}$ the shift is about $10\%$.\footnote{ Pushing to
$d_i\zeta=10^{-1}$ is not even physically meaningful, since $r_H$ would formally become negative.}  To remain safely within the linear regime, in what follows we therefore take
$\zeta_{\max}=10^{-4}$ as our default upper limit, and treat
$\zeta=10^{-3}$ only as a reference point illustrating the onset of
linear perturbation failure\footnote{In what follows, Wilson coefficients $d_i\in\{0,1\}$ is a switch that selects the $i$th EFT operator, while $\zeta$ is the explicit perturbation parameter.}.

\paragraph{Higher-order contamination in the numerical extraction.}
Although the underlying field equations are linearized in $\zeta$, the numerical procedure
used to extract $\omega(\zeta)$ via Leaver's method can effectively introduce
higher-order dependence on $\zeta$.
Two immediate sources are: (i) the Gaussian-elimination band-reduction step, which combines $\zeta$-dependent terms and therefore generates products that promote higher powers of $\zeta$ \footnote{A strictly linear-in-$\zeta$ reduction can be enforced by truncating to $\mathcal{O}(\zeta)$ after each elimination step, nevertheless, we do not impose this truncation in our implementation.}, and
(ii) the nonlinear Newton solver employed to locate the roots of the continued fraction,
which can also imprint nonlinear $\zeta$-dependence through its iterative updates.
As a result, the numerically extracted frequency is more accurately viewed as an expansion
\[
\Delta\omega(\zeta)\equiv\omega(\zeta)-\omega(0)=a_1\,\zeta+a^{(1)}_{2}\,\zeta^2+a^{(1)}_{3}\,\zeta^3+\cdots,
\]
\footnote{Here in this subsection, we re-use $\omega(\zeta)$ to denote the Leaver extracted frequency.}
 where $a_1$ is the linear coefficient, while $a^{(1)}_{2},a^{(1)}_{3},\ldots$
denote effective higher-order contributions induced by the Leaver extraction procedure,
even when the input perturbation is truncated at $\mathcal{O}(\zeta)$.
Consequently, estimating $a_1$---especially in cases where the linear contribution is
suppressed or vanishes for a given operator---is a subtle task.

\paragraph{Diagnostics.} Hence, to characterize the $\zeta$-dependence of $\omega(\zeta)$ for both the null and control operators, we employ the following diagnostics:
\begin{itemize}
\item \textbf{Leading-power diagnostic:} determine the dominant scaling $\Delta\omega(\zeta)\propto \zeta^{\,p}$ in the different-$\zeta$ regime. 
\item \textbf{Polynomial fitting:} fit the shift $\Delta\omega(\zeta)$ with a polynomial in $\zeta$ to extract the linear coefficient and quantify higher-order contamination.
\end{itemize}

\subsubsection{What is the leading power?}

We model the deviation as a power law in the coupling range set by a chosen cutoff $\zeta_{\max}$,
\begin{align}
\delta_{\mathrm{Leaver}}(\zeta)
&\equiv \left(
\frac{\Re\!\big(\Delta\omega(\zeta)\big)}{\Re\!\big(\omega(0)\big)}
,\,\frac{\Im\!\big(\Delta\omega(\zeta)\big)}{\Im\!\big(\omega(0)\big)} \right)
\approx C\,\zeta^{\,p},
\end{align}

where the coefficient $C\in\mathbb{C}$ are generally complex.

On a chosen coupling window $[\zeta_{\min},\zeta_{\max}]$, we then infer the leading scaling via two complementary steps:
(i) estimate a continuous effective exponent $p_{\rm eff}$ from log--log slopes of the amplitude $\|\delta_{\mathrm{Leaver}}\|$ across adjacent $\zeta$ values, and we then average these local slope estimates across the $\zeta$ points within the chosen fitting window $[\zeta_{\min},\zeta_{\max}]$ to obtain the reported $p_{\rm eff}$ (with scatter across the window used to define $\sigma_p$), and
(ii) test candidate integers $p\in\{1,2,3\}$ by rescaling the data by $\zeta^{p}$ and selecting the $p_\star$ that makes the rescaled complex series as flat as possible. Finally, with $p_\star$ fixed, we obtain the corresponding complex coefficient $C$ from a least-squares fit using the flatness ratio $\sigma/\mu$ reported in the tables where $N$, $\mu$ and $\sigma$ is the sample population, mean and standard deviation respectively inside the fit window.

\paragraph{Null operators.}
Figures~\ref{fig:null58_l2}--\ref{fig:null58_l4} show scaled $\delta_{\mathrm{Leaver}}$ across our $\zeta$ sampling and visually indicate a leading quadratic behaviour for the null combinations. To quantify the leading power, we apply the windowed diagnostics described above, using data restricted to $\zeta\le \zeta_{\max}$.

At the widest diagnostic window $\zeta_{\max}=10^{-3}$ (Tables~\ref{tab:power_1Em3_odd_null} and~\ref{tab:power_1Em3_even_null}), the null-sector data select $p_\star=2$ for both $\mathcal{O}_5$ and $\mathcal{O}_8$ across all listed modes and both parities. The effective exponents are tightly clustered around $2$: for $\mathcal{O}_5$ and $\mathcal{O}_8$, $|p_{\rm eff}-2|\lesssim10^{-3}$) and $\lesssim10^{-4}$ respectively. The complex flatness ratios are small and mode-dependent: $\sigma/\mu\simeq 10^{-4}\text{--}10^{-3}$ for $\mathcal{O}_5$ and $\mathcal{O}_8$. Together, these diagnostics support a leading quadratic scaling for both null operators in both parities.

Upon tightening the window to $\zeta_{\max}=10^{-4}$ and $10^{-5}$ (as shown in Tables \ref{tab:power_1Em4_odd_58}, \ref{tab:power_1Em4_even_58}, \ref{tab:power_1Em5_odd_58}, and \ref{tab:power_1Em5_even_58}), the same $p_\star=2$ conclusion is reinforced: $p_{\rm eff}$ moves closer to $2$ and $\sigma/\mu$ decreases, consistent with convergence toward the asymptotic quadratic regime. We therefore identify $p=2$ as the leading power for both null operators in both parities

\paragraph{Control operators.}
Figures~\ref{fig:control910_l2}--\ref{fig:control910_l4} display scaled $\delta_{\mathrm{Leaver}}$ over the full sampled $\zeta$ range and visually support a leading linear scaling for the control operators. As above, we quantify the leading power using the windowed diagnostics on data restricted to $\zeta\le \zeta_{\max}$.

At the widest diagnostic window $\zeta_{\max}=10^{-3}$ (Tables~\ref{tab:power_1Em3_odd_control} and~\ref{tab:power_1Em3_even_control}), both $\mathcal{O}_9$ and $\mathcal{O}_{10}$ select $p_\star=1$ for every listed mode and both parities. For $\mathcal{O}_9$ and $\mathcal{O}_{10}$, the effective exponents remain close to unity, with $|p_{\rm eff}-1| \lesssim10^{-3}$ and $\lesssim10^{-3}$ respectively. Yet, there are some higher-overtone outliers (down to $p_{\rm eff}=0.9963$ in even parity and $p_{\rm eff}=0.9896$ in odd parity). This non-asymptotic contamination is mirrored in the complex flatness: in odd parity, $\sigma/\mu\simeq 10^{-3}\text{--}10^{-1}$ for $\mathcal{O}_9$ and $\sigma/\mu\simeq 10^{-3}\text{--}10^{-2}$ for $\mathcal{O}_{10}$, with the largest scatter occurring at the highest listed overtone. It's also evident that even parity is systematically cleaner.

Tightening the window to $\zeta_{\max}=10^{-4}$ and $10^{-5}$ (as shown in Tables \ref{tab:power_1Em4_odd_910}, \ref{tab:power_1Em4_even_910}, \ref{tab:power_1Em5_odd_910}, and \ref{tab:power_1Em5_even_910}) preserves the same $p_\star=1$ selection while reducing uncertainties and significantly decreasing $\sigma/\mu$; the fitted complex coefficient $C$ (reported via $\Re(C)\,|\,\Im(C)$) varies only mildly across windows, consistent with a well-resolved linear leading term.

\subsubsection{Polynomial fits}
Our primary goal is to test whether the null operators produce a genuinely linear-in-$\zeta$ correction to the vacuum QNM spectrum. In practice, as discussed above, the Leaver-type numerical implementation introduces a mild nonlinear dependence in $\zeta$. Having already determined the leading power of $\zeta$ for all operators considered in this work, we now investigate the subleading structure by fitting the frequency shift $\Delta\omega(\zeta) $ with a polynomial in $\zeta$,
$$
\delta_{\mathrm{Leaver}}(\zeta) \approx \sum_{k=1}^{K} c^{(b)}_k \,\zeta^k,
$$
where the complex coefficients $c_k^{(b)}$ are obtained from least-squares fits over suitably chosen windows.

First, since higher-order terms become influential when the fit extends to relatively large values of $\zeta$, we restrict attention to a sweet-spot region that is well below our largest deformation, $\zeta = 10^{-3}$, but still contains enough data points to make all diagnostics meaningful. We choose to work with ten nested fitting windows whose upper endpoints are
\begin{equation}
\begin{aligned}
\zeta_{\max} \in \bigl\{
&2.5\times10^{-11},\,
  5\times10^{-11},\,
  7.5\times10^{-11},\,
  10^{-10},\\
&2.5\times10^{-10},\,
  5\times10^{-10},\,
  7.5\times10^{-10},\,
  10^{-9},\\
&2.5\times10^{-9},\,
  5\times10^{-9}
\bigr\}.
\end{aligned}
\end{equation}

Second, we deliberately restrict attention to three nested models $M_b$ with $b\in\{1,12,123\}$:
\begin{align*}
M_1 &: \quad \Delta\omega(\zeta) \approx c^{(1)}_1\,\zeta,\\
M_{12} &: \quad \Delta\omega(\zeta) \approx c^{(12)}_1\,\zeta + c^{(12)}_2\,\zeta^2,\\
M_{123} &: \quad \Delta\omega(\zeta) \approx c^{(123)}_1\,\zeta + c^{(123)}_2\,\zeta^2 + c^{(123)}_3\,\zeta^3,
\end{align*} 

where we use $M_{123}$ only for the null-operator cases.

Moreover, although our calculated QNMs have $\sim 60$ digits of precision and we could in principle include work with $M_{1234}$ as well, we won't to avoid overfitting \footnote{Nevertheless, when stated explicitly, we also fit $M_{1234}$ as a reference model; it is not used to set our reported bounds or conclusions.}. This is the same reason why $M_{123}$ is not going to be used for control cases.

As we will show below for null operators, the fitted values of $c^{(12),(123)}_1$ is expected to be much smaller than the actual values. In other words, stopping at $\zeta^3$ term is conservative and the linear coefficient $c^{(123)}_1$ can be interpreted as a lower bound on any true linear-in-$\zeta$ contribution.

Third, to decide which model to adopt and which coefficients to trust, we use two simple but strict criteria. For each model $M_b$ and coefficient index $k$, we denote by $c^{(b)}_{k,i}$ the value obtained in the $i$-th $\zeta_{\max}$ window and by $\bar c^{b}_k$ their mean. We then define the stability measure
$$
\Gamma_k^{(b)} \equiv
\frac{\max\limits_i \bigl|c^{(b)}_{k,i} - \bar c^{(b)}_k\bigr|}{\bigl|\bar c^{(b)}_k\bigr|}
< 10^{-4}.
$$
In practice, for null operators, $c^{(b)}_2$ and $c^{(b)}_3$ satisfy this criterion by many orders of magnitude, while $c^{(b)}_1$ typically does not, as expected. However, for control cases it will.

By progressively building up the polynomial models $M_1$, $M_{12}$, and $M_{123}$, we monitor both the coefficient stability and the root-mean-square (RMS) residual of the fit as the polynomial degree increases. We select the lowest-order model whose coefficients pass the stability criterion and for which the RMS error has already saturated. 

\paragraph{Null operators.}
For the null-operator cases the polynomial fits exhibit a clear hierarchy. As we enlarge the model from $M_1$ to $M_{12}$ the RMS residual decrease by roughly $10^{-10}$, and it decrease by a further $\sim 10^{-10}$ upon moving to $M_{123}$. Extending once more to $M_{1234}$ causes yet another $\sim 10^{-1}- 10^{-2}$ decrease, indicate a saturation in the fitting. At the same time, the additional parameters beyond cubic order fail our coefficient-stability requirement, indicating that these higher-degree models $c_1$ and $c_{>3}$ will start canceling each other.

A useful way to interpret this behavior is to inspect the relative contributions of the fitted terms to $\Delta\omega(\zeta)$ over the fitting window. In the null cases, the quadratic term dominates overwhelmingly: in the degree-$3$ model, $c^{(123)}_2 \zeta^2$ accounts for $\simeq 1-\mathcal{O}(10^{-7})$ of the reconstructed shift, whereas the linear and cubic pieces contribute only at the level of $\sim 10^{-14}\%$ and $\sim 10^{-5}\%$, respectively. This extreme suppression implies a large budget for still-higher powers. For example, in $M_{1234}$, a quartic contribution of order $10^{-11}\%$ will knock median$(|c_1^{1234}|)$ by $10^{-3}$ yet it stable only among fewer number of windows and it doesn't . Consequently, the fitted $c^{(123)}_1$ should be viewed as a conservative \emph{upper bound} rather than a measurement: we expect the true linear coefficient to satisfy $|c_1|\ll |c^{(123)}_1|$.

In practice, $M_{123}$ emerges as the best compromise: it is the highest-order model whose coefficients remain stable across the nested $\zeta_{\max}$ windows, while simultaneously yielding the lowest RMS residual among the stable models. Quantitatively, across all modes in Tables~\ref{tab:coeff_stab_deg3_even_odd_null_l2}, ~\ref{tab:coeff_stab_deg3_even_odd_null_l3}, and ~\ref{tab:coeff_stab_deg3_even_odd_null_l4} we find
\begin{equation}
\begin{aligned}
\min\!\left(|c^{(123)}_1|\right) &= 3.09\times10^{-24},\\
\mathrm{median}\!\left(|c^{(123)}_1|\right) &= 5.15\times10^{-22},\\
\max\!\left(|c^{(123)}_1|\right) &= 4.60\times10^{-20},
\end{aligned}
\end{equation}

\paragraph{Control operators.}
In the control data, the change $\Delta\omega(\zeta)$ is dominated by the linear term. In the $M_{12}$ fits, this is borne out quantitatively by the contribution audit: $|c^{(12)}_1\zeta|$ accounts for $1-\mathcal{O}(10^{-7})$ of the reconstructed shift over the fitting window, while the quadratic correction remains suppressed at the level $|c^{(12)}_2\zeta^2|\sim 10^{-8}\text{--}10^{-5}$. Consistently, $c^{(12)}_1$ is extremely stable across nested $\zeta_{\max}$ windows at the $\sim 11$--$14$ (~$10^{-11} \%$) digit level, whereas $c^{(12)}_2$ is stable at only $\sim 5$--$6$ digits (~$10^{-4}\%$). See Tables ~\ref{tab:coeff_stab_deg3_even_odd_control_l2}, ~\ref{tab:coeff_stab_deg3_even_odd_control_l3} and ~\ref{tab:coeff_stab_deg3_even_odd_control_l4} for further details. 

Furthermore, across all ratios between QNMs shift due to $\mathcal{O}_{9}$ to $\mathcal{O}_{10}$ are extremely close to $2$ with the largest absolute deviation is $\max |\mathcal{R}_{9/10}-2|\sim 10^{-11}$. As was the case for EVP case, the deviation increases with higher overtone.

We verified that the QNM frequencies for the physical operators extracted with the Leaver method are consistent with Ref.~\cite{deRham_EFT_GW_2020}: when rounded to the two decimal places reported in their Table~I, our values coincide with theirs.

\begin{widetext}
\begin{table*}[t]
\centering
\caption{Coefficient stability summary for degree $3$ ($\Delta\omega(\zeta)=\sum_{k=1}^3 c_k \zeta^k$), with real and imaginary parts for each mode.}
\label{tab:coeff_stab_deg3_even_odd_null_l2}
\setlength{\tabcolsep}{4pt}
\renewcommand{\arraystretch}{1.15}
\begin{tabular}{c|c|c|c|c|c|c|c|c|c|}
\toprule
\multirow{2}{*}{} & \multirow{2}{*}{$c_k$} & \multicolumn{2}{c}{$( 0,2 )$} & \multicolumn{2}{c}{$( 1,2 )$} & \multicolumn{2}{c}{$( 2,2 )$} & \multicolumn{2}{c}{$( 3,2 )$} \\
 &  & $\Re$ & $\Im$ & $\Re$ & $\Im$ & $\Re$ & $\Im$ & $\Re$ & $\Im$ \\
\midrule
\multirow{6}{*}{$\mathcal{O}_{5}$} & ${}_{\mathrm{o}}c_{1}$ & $3.7\times 10^{-22}$ & $-7.6\times 10^{-22}$ & $2.7\times 10^{-21}$ & $1.7\times 10^{-21}$ & $6.6\times 10^{-21}$ & $1.1\times 10^{-20}$ & $-2.4\times 10^{-20}$ & $3.9\times 10^{-20}$ \\
 & ${}_{\mathrm{o}}c_{2}$ & $119.9$ & $477.1$ & $-35.73$ & $596.0$ & $-687.3$ & $736.9$ & $-2796.0$ & $921.4$ \\
 & ${}_{\mathrm{o}}c_{3}$ & $7831.0$ & $1.065\times 10^{4}$ & $1.656\times 10^{4}$ & $3.98\times 10^{4}$ & $-1.409\times 10^{4}$ & $9.63\times 10^{4}$ & $-3.062\times 10^{5}$ & $1.96\times 10^{5}$ \\
\cline{2-10}
 & ${}_{\mathrm{e}}c_{1}$ & $4.1\times 10^{-22}$ & $-8.0\times 10^{-22}$ & $2.7\times 10^{-21}$ & $1.9\times 10^{-21}$ & $5.0\times 10^{-21}$ & $1.1\times 10^{-20}$ & $-3.1\times 10^{-20}$ & $3.4\times 10^{-20}$ \\
 & ${}_{\mathrm{e}}c_{2}$ & $124.5$ & $497.3$ & $-43.28$ & $599.0$ & $-711.0$ & $712.8$ & $-2783.0$ & $833.7$ \\
 & ${}_{\mathrm{e}}c_{3}$ & $8379.0$ & $1.107\times 10^{4}$ & $1.636\times 10^{4}$ & $4.212\times 10^{4}$ & $-2.386\times 10^{4}$ & $9.494\times 10^{4}$ & $-3.246\times 10^{5}$ & $1.712\times 10^{5}$ \\
\midrule
\multirow{6}{*}{$\mathcal{O}_{8}$} & ${}_{\mathrm{o}}c_{1}$ & $1.5\times 10^{-24}$ & $-2.7\times 10^{-24}$ & $1.1\times 10^{-23}$ & $6.8\times 10^{-24}$ & $2.5\times 10^{-23}$ & $4.4\times 10^{-23}$ & $-9.4\times 10^{-23}$ & $1.5\times 10^{-22}$ \\
 & ${}_{\mathrm{o}}c_{2}$ & $7.491$ & $29.82$ & $-2.233$ & $37.25$ & $-42.95$ & $46.06$ & $-174.8$ & $57.58$ \\
 & ${}_{\mathrm{o}}c_{3}$ & $122.4$ & $166.5$ & $258.8$ & $621.9$ & $-220.1$ & $1505.0$ & $-4784.0$ & $3062.0$ \\
\cline{2-10}
 & ${}_{\mathrm{e}}c_{1}$ & $1.7\times 10^{-24}$ & $-2.9\times 10^{-24}$ & $1.1\times 10^{-23}$ & $7.9\times 10^{-24}$ & $1.9\times 10^{-23}$ & $4.4\times 10^{-23}$ & $-1.2\times 10^{-22}$ & $1.4\times 10^{-22}$ \\
 & ${}_{\mathrm{e}}c_{2}$ & $7.782$ & $31.08$ & $-2.705$ & $37.44$ & $-44.44$ & $44.55$ & $-173.9$ & $52.11$ \\
 & ${}_{\mathrm{e}}c_{3}$ & $130.9$ & $173.0$ & $255.6$ & $658.1$ & $-372.8$ & $1483.0$ & $-5072.0$ & $2676.0$ \\
\bottomrule
\end{tabular}
\end{table*}
\begin{table*}[t]
\centering
\caption{Coefficient stability summary for degree $3$ ($\Delta\omega(\zeta)=\sum_{k=1}^3 c_k \zeta^k$), with real and imaginary parts for each mode.}
\label{tab:coeff_stab_deg3_even_odd_null_l3}
\setlength{\tabcolsep}{4pt}
\renewcommand{\arraystretch}{1.15}
\begin{tabular}{c|c|c|c|c|c|c|c|c|c|}
\toprule
\multirow{2}{*}{} & \multirow{2}{*}{$c_k$} & \multicolumn{2}{c}{$( 0,3 )$} & \multicolumn{2}{c}{$( 1,3 )$} & \multicolumn{2}{c}{$( 2,3 )$} & \multicolumn{2}{c}{$( 3,3 )$} \\
 &  & $\Re$ & $\Im$ & $\Re$ & $\Im$ & $\Re$ & $\Im$ & $\Re$ & $\Im$ \\
\midrule
\multirow{6}{*}{$\mathcal{O}_{5}$} & ${}_{\mathrm{o}}c_{1}$ & $8.5\times 10^{-23}$ & $-2.0\times 10^{-21}$ & $1.4\times 10^{-21}$ & $-2.3\times 10^{-21}$ & $7.5\times 10^{-21}$ & $-2.4\times 10^{-22}$ & $2.5\times 10^{-20}$ & $1.1\times 10^{-20}$ \\
 & ${}_{\mathrm{o}}c_{2}$ & $156.6$ & $482.6$ & $136.1$ & $598.6$ & $26.16$ & $757.2$ & $-332.0$ & $979.1$ \\
 & ${}_{\mathrm{o}}c_{3}$ & $7007.0$ & $-6011.0$ & $2.056\times 10^{4}$ & $1.296\times 10^{4}$ & $4.698\times 10^{4}$ & $5.185\times 10^{4}$ & $7.319\times 10^{4}$ & $1.276\times 10^{5}$ \\
\cline{2-10}
 & ${}_{\mathrm{e}}c_{1}$ & $9.0\times 10^{-23}$ & $-2.1\times 10^{-21}$ & $1.5\times 10^{-21}$ & $-2.2\times 10^{-21}$ & $7.6\times 10^{-21}$ & $-5.1\times 10^{-24}$ & $2.4\times 10^{-20}$ & $1.1\times 10^{-20}$ \\
 & ${}_{\mathrm{e}}c_{2}$ & $158.3$ & $488.6$ & $136.2$ & $603.3$ & $21.27$ & $759.9$ & $-348.5$ & $973.8$ \\
 & ${}_{\mathrm{e}}c_{3}$ & $7134.0$ & $-6324.0$ & $2.094\times 10^{4}$ & $1.356\times 10^{4}$ & $4.654\times 10^{4}$ & $5.346\times 10^{4}$ & $6.819\times 10^{4}$ & $1.285\times 10^{5}$ \\
\midrule
\multirow{6}{*}{$\mathcal{O}_{8}$} & ${}_{\mathrm{o}}c_{1}$ & $4.1\times 10^{-25}$ & $-7.7\times 10^{-24}$ & $5.6\times 10^{-24}$ & $-8.5\times 10^{-24}$ & $2.9\times 10^{-23}$ & $-5.8\times 10^{-25}$ & $9.8\times 10^{-23}$ & $4.3\times 10^{-23}$ \\
 & ${}_{\mathrm{o}}c_{2}$ & $9.79$ & $30.16$ & $8.507$ & $37.41$ & $1.635$ & $47.33$ & $-20.75$ & $61.2$ \\
 & ${}_{\mathrm{o}}c_{3}$ & $109.5$ & $-93.92$ & $321.3$ & $202.4$ & $734.1$ & $810.1$ & $1144.0$ & $1994.0$ \\
\cline{2-10}
 & ${}_{\mathrm{e}}c_{1}$ & $4.3\times 10^{-25}$ & $-7.9\times 10^{-24}$ & $5.8\times 10^{-24}$ & $-8.5\times 10^{-24}$ & $3.0\times 10^{-23}$ & $3.5\times 10^{-25}$ & $9.5\times 10^{-23}$ & $4.5\times 10^{-23}$ \\
 & ${}_{\mathrm{e}}c_{2}$ & $9.893$ & $30.54$ & $8.514$ & $37.71$ & $1.329$ & $47.49$ & $-21.78$ & $60.86$ \\
 & ${}_{\mathrm{e}}c_{3}$ & $111.5$ & $-98.81$ & $327.2$ & $211.9$ & $727.3$ & $835.4$ & $1066.0$ & $2008.0$ \\
\bottomrule
\end{tabular}
\end{table*}
\begin{table*}[t]
\centering
\caption{Coefficient stability summary for degree $3$ ($\Delta\omega(\zeta)=\sum_{k=1}^3 c_k \zeta^k$), with real and imaginary parts for each mode.}
\label{tab:coeff_stab_deg3_even_odd_null_l4}
\setlength{\tabcolsep}{4pt}
\renewcommand{\arraystretch}{1.15}
\begin{tabular}{c|c|c|c|c|c|c|c|c|c|}
\toprule
\multirow{2}{*}{} & \multirow{2}{*}{$c_k$} & \multicolumn{2}{c}{$( 0,4 )$} & \multicolumn{2}{c}{$( 1,4 )$} & \multicolumn{2}{c}{$( 2,4 )$} & \multicolumn{2}{c}{$( 3,4 )$} \\
 &  & $\Re$ & $\Im$ & $\Re$ & $\Im$ & $\Re$ & $\Im$ & $\Re$ & $\Im$ \\
\midrule
\multirow{6}{*}{$\mathcal{O}_{5}$} & ${}_{\mathrm{o}}c_{1}$ & $-4.9\times 10^{-23}$ & $-2.0\times 10^{-21}$ & $3.5\times 10^{-22}$ & $-3.1\times 10^{-21}$ & $2.9\times 10^{-21}$ & $-4.5\times 10^{-21}$ & $1.2\times 10^{-20}$ & $-3.8\times 10^{-21}$ \\
 & ${}_{\mathrm{o}}c_{2}$ & $165.8$ & $463.5$ & $168.7$ & $556.6$ & $151.8$ & $677.7$ & $67.34$ & $850.5$ \\
 & ${}_{\mathrm{o}}c_{3}$ & $5699.0$ & $-1.424\times 10^{4}$ & $1.492\times 10^{4}$ & $-3752.0$ & $3.647\times 10^{4}$ & $1.584\times 10^{4}$ & $7.512\times 10^{4}$ & $5.643\times 10^{4}$ \\
\cline{2-10}
 & ${}_{\mathrm{e}}c_{1}$ & $-5.0\times 10^{-23}$ & $-2.1\times 10^{-21}$ & $3.6\times 10^{-22}$ & $-3.1\times 10^{-21}$ & $3.0\times 10^{-21}$ & $-4.5\times 10^{-21}$ & $1.3\times 10^{-20}$ & $-3.6\times 10^{-21}$ \\
 & ${}_{\mathrm{e}}c_{2}$ & $166.5$ & $465.5$ & $169.2$ & $558.6$ & $151.4$ & $680.0$ & $64.25$ & $852.2$ \\
 & ${}_{\mathrm{e}}c_{3}$ & $5731.0$ & $-1.449\times 10^{4}$ & $1.507\times 10^{4}$ & $-3719.0$ & $3.668\times 10^{4}$ & $1.637\times 10^{4}$ & $7.48\times 10^{4}$ & $5.746\times 10^{4}$ \\
\midrule
\multirow{6}{*}{$\mathcal{O}_{8}$} & ${}_{\mathrm{o}}c_{1}$ & $-1.1\times 10^{-25}$ & $-7.7\times 10^{-24}$ & $1.4\times 10^{-24}$ & $-1.2\times 10^{-23}$ & $1.2\times 10^{-23}$ & $-1.7\times 10^{-23}$ & $4.9\times 10^{-23}$ & $-1.4\times 10^{-23}$ \\
 & ${}_{\mathrm{o}}c_{2}$ & $10.36$ & $28.97$ & $10.54$ & $34.79$ & $9.485$ & $42.35$ & $4.209$ & $53.15$ \\
 & ${}_{\mathrm{o}}c_{3}$ & $89.05$ & $-222.5$ & $233.1$ & $-58.63$ & $569.9$ & $247.5$ & $1174.0$ & $881.7$ \\
\cline{2-10}
 & ${}_{\mathrm{e}}c_{1}$ & $-1.1\times 10^{-25}$ & $-7.8\times 10^{-24}$ & $1.5\times 10^{-24}$ & $-1.2\times 10^{-23}$ & $1.2\times 10^{-23}$ & $-1.7\times 10^{-23}$ & $4.9\times 10^{-23}$ & $-1.4\times 10^{-23}$ \\
 & ${}_{\mathrm{e}}c_{2}$ & $10.4$ & $29.09$ & $10.57$ & $34.91$ & $9.46$ & $42.5$ & $4.015$ & $53.26$ \\
 & ${}_{\mathrm{e}}c_{3}$ & $89.55$ & $-226.5$ & $235.5$ & $-58.1$ & $573.1$ & $255.7$ & $1169.0$ & $897.8$ \\
\bottomrule
\end{tabular}
\end{table*}

\begin{table*}[t]
\centering
\caption{Coefficient stability summary for degree $2$ ($\Delta\omega(\zeta)=\sum_{k=1}^3 c_k \zeta^k$), with real and imaginary parts for each mode.}
\label{tab:coeff_stab_deg3_even_odd_control_l2}
\setlength{\tabcolsep}{4pt}
\renewcommand{\arraystretch}{1.15}
\begin{tabular}{c|c|c|c|c|c|c|c|c|c|}
\toprule
\multirow{2}{*}{} & \multirow{2}{*}{$c_k$} & \multicolumn{2}{c}{$( 0,2 )$} & \multicolumn{2}{c}{$( 1,2 )$} & \multicolumn{2}{c}{$( 2,2 )$} & \multicolumn{2}{c}{$( 3,2 )$} \\
 &  & $\Re$ & $\Im$ & $\Re$ & $\Im$ & $\Re$ & $\Im$ & $\Re$ & $\Im$ \\
\midrule
\multirow{4}{*}{$\mathcal{O}_{9}$} & ${}_{\mathrm{o}}c_{1}$ & $21.072$ & $47.529$ & $16.362$ & $70.574$ & $-51.795$ & $115.18$ & $-392.92$ & $183.67$ \\
 & ${}_{\mathrm{o}}c_{2}$ & $878.35$ & $-5724.0$ & $19342.0$ & $-3665.4$ & $1.2052\times 10^{5}$ & $82766.0$ & $25356.0$ & $5.4812\times 10^{5}$ \\
\cline{2-10}
 & ${}_{\mathrm{e}}c_{1}$ & $-12.339$ & $-58.117$ & $10.758$ & $-55.815$ & $75.805$ & $-54.477$ & $229.39$ & $-61.441$ \\
 & ${}_{\mathrm{e}}c_{2}$ & $-192.46$ & $-154.32$ & $2869.2$ & $-1587.4$ & $15666.0$ & $17652.0$ & $-83385.0$ & $79190.0$ \\
\midrule
\multirow{4}{*}{$\mathcal{O}_{10}$} & ${}_{\mathrm{o}}c_{1}$ & $10.536$ & $23.764$ & $8.1809$ & $35.287$ & $-25.897$ & $57.589$ & $-196.46$ & $91.834$ \\
 & ${}_{\mathrm{o}}c_{2}$ & $-17.661$ & $-1072.4$ & $3693.3$ & $-1852.9$ & $28896.0$ & $16718.0$ & $16490.0$ & $1.2713\times 10^{5}$ \\
\cline{2-10}
 & ${}_{\mathrm{e}}c_{1}$ & $-6.1697$ & $-29.059$ & $5.3792$ & $-27.908$ & $37.902$ & $-27.238$ & $114.7$ & $-30.72$ \\
 & ${}_{\mathrm{e}}c_{2}$ & $89.514$ & $-114.32$ & $1227.5$ & $315.55$ & $3680.5$ & $6463.4$ & $-28099.0$ & $23393.0$ \\
\bottomrule
\end{tabular}
\end{table*}
\begin{table*}[t]
\centering
\caption{Coefficient stability summary for degree $2$ ($\Delta\omega(\zeta)=\sum_{k=1}^3 c_k \zeta^k$), with real and imaginary parts for each mode.}
\label{tab:coeff_stab_deg3_even_odd_control_l3}
\setlength{\tabcolsep}{4pt}
\renewcommand{\arraystretch}{1.15}
\begin{tabular}{c|c|c|c|c|c|c|c|c|c|}
\toprule
\multirow{2}{*}{} & \multirow{2}{*}{$c_k$} & \multicolumn{2}{c}{$( 0,3 )$} & \multicolumn{2}{c}{$( 1,3 )$} & \multicolumn{2}{c}{$( 2,3 )$} & \multicolumn{2}{c}{$( 3,3 )$} \\
 &  & $\Re$ & $\Im$ & $\Re$ & $\Im$ & $\Re$ & $\Im$ & $\Re$ & $\Im$ \\
\midrule
\multirow{4}{*}{$\mathcal{O}_{9}$} & ${}_{\mathrm{o}}c_{1}$ & $18.289$ & $46.265$ & $15.647$ & $51.392$ & $5.2511$ & $61.309$ & $-25.138$ & $75.63$ \\
 & ${}_{\mathrm{o}}c_{2}$ & $329.92$ & $-5880.2$ & $4520.4$ & $-8128.7$ & $25401.0$ & $-679.63$ & $78221.0$ & $40799.0$ \\
\cline{2-10}
 & ${}_{\mathrm{e}}c_{1}$ & $-14.149$ & $-51.817$ & $-7.9918$ & $-53.533$ & $8.0346$ & $-57.094$ & $42.345$ & $-62.886$ \\
 & ${}_{\mathrm{e}}c_{2}$ & $-196.85$ & $1219.1$ & $-588.48$ & $-3160.8$ & $6344.8$ & $-4489.1$ & $28293.0$ & $9939.7$ \\
\midrule
\multirow{4}{*}{$\mathcal{O}_{10}$} & ${}_{\mathrm{o}}c_{1}$ & $9.1445$ & $23.133$ & $7.8237$ & $25.696$ & $2.6255$ & $30.655$ & $-12.569$ & $37.815$ \\
 & ${}_{\mathrm{o}}c_{2}$ & $-63.314$ & $-923.07$ & $515.92$ & $-2050.3$ & $4965.9$ & $-1378.1$ & $17745.0$ & $7043.9$ \\
\cline{2-10}
 & ${}_{\mathrm{e}}c_{1}$ & $-7.0743$ & $-25.908$ & $-3.9959$ & $-26.767$ & $4.0173$ & $-28.547$ & $21.172$ & $-31.443$ \\
 & ${}_{\mathrm{e}}c_{2}$ & $16.534$ & $-122.15$ & $249.11$ & $-804.15$ & $2487.2$ & $-291.16$ & $8141.7$ & $4624.4$ \\
\bottomrule
\end{tabular}
\end{table*}
\begin{table*}[t]
\centering
\caption{Coefficient stability summary for degree $2$ ($\Delta\omega(\zeta)=\sum_{k=1}^3 c_k \zeta^k$), with real and imaginary parts for each mode.}
\label{tab:coeff_stab_deg3_even_odd_control_l4}
\setlength{\tabcolsep}{4pt}
\renewcommand{\arraystretch}{1.15}
\begin{tabular}{c|c|c|c|c|c|c|c|c|c|}
\toprule
\multirow{2}{*}{} & \multirow{2}{*}{$c_k$} & \multicolumn{2}{c}{$( 0,4 )$} & \multicolumn{2}{c}{$( 1,4 )$} & \multicolumn{2}{c}{$( 2,4 )$} & \multicolumn{2}{c}{$( 3,4 )$} \\
 &  & $\Re$ & $\Im$ & $\Re$ & $\Im$ & $\Re$ & $\Im$ & $\Re$ & $\Im$ \\
\midrule
\multirow{4}{*}{$\mathcal{O}_{9}$} & ${}_{\mathrm{o}}c_{1}$ & $17.545$ & $45.703$ & $15.804$ & $47.798$ & $11.067$ & $51.928$ & $0.5627$ & $58.02$ \\
 & ${}_{\mathrm{o}}c_{2}$ & $124.05$ & $-5784.2$ & $1733.0$ & $-8123.4$ & $9596.9$ & $-8663.7$ & $31503.0$ & $693.51$ \\
\cline{2-10}
 & ${}_{\mathrm{e}}c_{1}$ & $-14.46$ & $-50.165$ & $-11.445$ & $-51.351$ & $-4.2793$ & $-53.74$ & $9.492$ & $-57.393$ \\
 & ${}_{\mathrm{e}}c_{2}$ & $-84.342$ & $2147.0$ & $-1001.1$ & $-1338.9$ & $320.76$ & $-5643.1$ & $9165.3$ & $-5740.8$ \\
\midrule
\multirow{4}{*}{$\mathcal{O}_{10}$} & ${}_{\mathrm{o}}c_{1}$ & $8.7725$ & $22.852$ & $7.9022$ & $23.899$ & $5.5334$ & $25.964$ & $0.28135$ & $29.01$ \\
 & ${}_{\mathrm{o}}c_{2}$ & $-77.687$ & $-814.44$ & $42.925$ & $-1710.9$ & $1472.9$ & $-2497.3$ & $6268.1$ & $-1214.2$ \\
\cline{2-10}
 & ${}_{\mathrm{e}}c_{1}$ & $-7.23$ & $-25.082$ & $-5.7226$ & $-25.676$ & $-2.1397$ & $-26.87$ & $4.746$ & $-28.697$ \\
 & ${}_{\mathrm{e}}c_{2}$ & $11.424$ & $1.0423$ & $-9.5121$ & $-646.49$ & $716.05$ & $-1247.6$ & $3427.0$ & $-504.68$ \\
\bottomrule
\end{tabular}
\end{table*}

\begin{table*}[h!]
\centering
\setlength{\tabcolsep}{3pt}
\label{ComarsionEVPLEAVERl2}
\renewcommand{\arraystretch}{1.2}
\begin{tabular}{|c|c|c|c|c|c|}
\hline
 &  & $(0,2)$ & $(1,2)$ & $(2,2)$ & $(3,2)$ \\
\hline
\multirow{2}{*}{$\mathcal{O}_9$} & even & $3.12\times10^{-16},\,1.11\times10^{-15}$ & $1.2\times10^{-13},\,4.08\times10^{-14}$ & $2.91\times10^{-10},\,2.43\times10^{-11}$ & $1.13\times10^{-6},\,1.59\times10^{-7}$ \\
\cline{2-6}
 & odd & $1.25\times10^{-14},\,5.72\times10^{-15}$ & $1.06\times10^{-13},\,1.19\times10^{-13}$ & $4.29\times10^{-10},\,2.5\times10^{-11}$ & $7.28\times10^{-7},\,5.31\times10^{-8}$ \\
\hline
\multirow{2}{*}{$\mathcal{O}_{10}$} & even & $9.59\times10^{-16},\,3.54\times10^{-16}$ & $1.38\times10^{-13},\,5.16\times10^{-14}$ & $2.91\times10^{-10},\,2.48\times10^{-11}$ & $1.13\times10^{-6},\,1.59\times10^{-7}$ \\
\cline{2-6}
 & odd & $1.65\times10^{-15},\,1.35\times10^{-15}$ & $7.35\times10^{-14},\,1.58\times10^{-14}$ & $4.32\times10^{-10},\,2.5\times10^{-11}$ & $7.28\times10^{-7},\,5.31\times10^{-8}$ \\
\hline
\end{tabular}
\caption{Symmetric relative errors between EVP and Leaver degree-2 fit using $c_1$ values for $\ell=2$, computed component-wise $\epsilon_X$. Each cell lists $\epsilon_{\Re},\,\epsilon_{\Im}$.}
\label{tab:evp_leaver_relerr_c1_deg2_l2}
\end{table*}
\begin{table*}[h!]
\centering
\label{ComarsionEVPLEAVERl3}
\setlength{\tabcolsep}{3pt}
\renewcommand{\arraystretch}{1.2}
\begin{tabular}{|c|c|c|c|c|c|}
\hline
 &  & $(0,3)$ & $(1,3)$ & $(2,3)$ & $(3,3)$ \\
\hline
\multirow{2}{*}{$\mathcal{O}_9$} & even & $2.08\times10^{-15},\,8.52\times10^{-16}$ & $3.26\times10^{-14},\,8.16\times10^{-15}$ & $1.6\times10^{-13},\,4.93\times10^{-14}$ & $3.51\times10^{-13},\,1.64\times10^{-13}$ \\
\cline{2-6}
 & odd & $8.28\times10^{-15},\,6.9\times10^{-15}$ & $1.08\times10^{-13},\,4.57\times10^{-14}$ & $1.98\times10^{-13},\,3.07\times10^{-13}$ & $3.69\times10^{-12},\,5.87\times10^{-13}$ \\
\hline
\multirow{2}{*}{$\mathcal{O}_{10}$} & even & $1.14\times10^{-16},\,6.84\times10^{-16}$ & $1.5\times10^{-14},\,1.13\times10^{-15}$ & $9.08\times10^{-17},\,2.2\times10^{-14}$ & $1.59\times10^{-13},\,4.24\times10^{-14}$ \\
\cline{2-6}
 & odd & $4.18\times10^{-16},\,2.52\times10^{-15}$ & $2.25\times10^{-14},\,1.01\times10^{-15}$ & $1.31\times10^{-13},\,5.5\times10^{-14}$ & $6.24\times10^{-13},\,1.41\times10^{-13}$ \\
\hline
\end{tabular}
\caption{Same Layout as in \ref{tab:coeff_stab_deg3_even_odd_control_l2} but for $\ell=3$.}
\label{tab:evp_leaver_relerr_c1_deg2_l3}
\end{table*}
\begin{table*}[h!]
\centering
\label{ComarsionEVPLEAVERl4}
\setlength{\tabcolsep}{3pt}
\renewcommand{\arraystretch}{1.2}
\begin{tabular}{|c|c|c|c|c|c|}
\hline
 &  & $(0,4)$ & $(1,4)$ & $(2,4)$ & $(3,4)$ \\
\hline
\multirow{2}{*}{$\mathcal{O}_9$} & even & $1.44\times10^{-15},\,4.94\times10^{-15}$ & $3.81\times10^{-15},\,1.12\times10^{-14}$ & $2.41\times10^{-13},\,7.42\times10^{-15}$ & $2.76\times10^{-13},\,7.52\times10^{-14}$ \\
\cline{2-6}
 & odd & $4.82\times10^{-15},\,1.46\times10^{-14}$ & $6.49\times10^{-14},\,2.06\times10^{-15}$ & $3.47\times10^{-13},\,1.3\times10^{-13}$ & $3.23\times10^{-12},\,4.75\times10^{-13}$ \\
\hline
\multirow{2}{*}{$\mathcal{O}_{10}$} & even & $1.81\times10^{-16},\,5.06\times10^{-16}$ & $4.38\times10^{-15},\,2.29\times10^{-15}$ & $8.08\times10^{-14},\,4.8\times10^{-15}$ & $1.76\times10^{-14},\,3.13\times10^{-14}$ \\
\cline{2-6}
 & odd & $3.54\times10^{-17},\,2.39\times10^{-15}$ & $8.98\times10^{-15},\,6.25\times10^{-15}$ & $8.31\times10^{-14},\,1.26\times10^{-14}$ & $1.51\times10^{-12},\,8.2\times10^{-14}$ \\
\hline
\end{tabular}
\caption{Same Layout as in \ref{tab:coeff_stab_deg3_even_odd_control_l2} but for $\ell=4$.}
\label{tab:evp_leaver_relerr_c1_deg2_l4}
\end{table*}

\end{widetext}
\subsection{Cross Check: EVP Vs Leaver}

We quantify agreement between the EVP results and the Leaver Model $M_{12}$ fit coefficient $c^{(12)}_1$, by using the symmetric relative difference
\begin{equation}
\epsilon_{X} \equiv 
\frac{2\left||\delta_{\rm EVP\, X}|-|\delta_{\rm Leaver\, X}|\right|}{|\delta_{\rm EVP\, X}|+|\delta_{\rm Leaver\, X}|},
\qquad X\in\{\Re,\Im\},
\end{equation}

computed mode-by-mode for $\ell=2,3,4$, $n=0,1,2,3$, both operators $\mathcal{O}_9,\mathcal{O}_{10}$,
and both parities. Aggregating all comparisons, we find
\begin{equation}
\begin{aligned}
\min(\epsilon_{\Re},\epsilon_{\Im}) &= (3.54\times10^{-17},\,3.54\times10^{-16}),\\
\mathrm{median}(\epsilon_{\Re},\epsilon_{\Im}) &= (9.48\times10^{-14},\,2.66\times10^{-14}),\\
\max(\epsilon_{\Re},\epsilon_{\Im}) &= (1.13\times10^{-6},\,1.59\times10^{-7}).
\end{aligned}
\end{equation}
Overall most of those discrepancies (but not all) are closely aligned with the numerical floor we estimated for the null operators QNMs using the EVP. This might suggest that the Leaver $M_{12}$ coefficient $c^{(12)}_1$
reproduces the EVP results to near the numerical floor set by the EVP extraction.

Two systematic trends mirror those seen in the EVP null-operator analysis: at fixed $\ell$ the disagreement
typically increases with overtone $n$, while at fixed $n$ it decreases with increasing $\ell$.
The largest mismatch occurs at $(\ell,n)=(2,3)$, whereas higher-$\ell$ fundamentals show the tightest agreement.
See Tables~\ref{ComarsionEVPLEAVERl2}--\ref{ComarsionEVPLEAVERl4}.

\section{Discussion and Conclusion}\label{conclusion}

Next-generation gravitational-wave observatories will enable high-SNR ringdown measurements and make black-hole QNMs increasingly sensitive probes of small, high-scale deviations from GR in the dynamical strong-field regime \cite{ET_Punturo_2010,CE_Reitze_2019,LISA_AmaroSeoane_2017,Berti_2006}. A controlled way to parameterize such deviations is the gravitational EFT, in which new physics is encoded in higher-curvature operators suppressed by a high scale \cite{deRham_EFT_GW_2020}. Before interpreting tiny EFT-induced QNM shifts as physical predictions, however, the framework and its numerical extraction must be stress-tested: (i) the problem is intrinsically two-parameter, involving both the GW perturbation order $\epsilon$ and the EFT deformation $\zeta$; and (ii) the expected signals can be comparable to the numerical floor, especially for overtones, unless precision and convergence are tightly controlled.

We therefore designed two null tests. First, we compute the QNM response to operators $\mathcal{O}_5$ and $\mathcal{O}_8$, which must not generate any first-order ($\mathcal{O}(\zeta)$) shift at vacuum; any sufficiently significant nonzero result directly diagnoses a breakdown of the scheme. Second, exploiting the Ricci-flat identity that ties $\mathcal{O}_{9}$ to $\mathcal{O}_{10}$ (up to linear and higher Ricci terms), we check numerically that—when the Wilson coefficients are set equal—the ratio between the extracted QNM shifts  should be $2$.

 We use two independent and complementary QNM strategies. First, we implement the EVP method to extract the linear frequency correction $\omega_{1}$ from a bilinear form evaluated along a complex contour, with the GR RW/Zerilli sector reconstructed at high precision. Second, we built a generalized Leaver approach that solves the EFT master equation directly: we derive the banded Frobenius recurrences, perform an exact band reduction to a three-term system, and apply a scalar continued-fraction solver. 

It is worth emphasizing that the two approaches are complementary. EVP has two practical advantages: (i) it targets the linear response by construction, extracting $\omega_{1}$ directly with no fitting needed to separate $\mathcal{O}(\zeta)$ from higher-order terms; and (ii) once the analytic structure is fixed (branch cuts and contour choice), it is relatively straightforward to implement numerically. By contrast, the generalized Leaver method is intentionally blind to the perturbative bookkeeping—it simply solves the EFT master equation at finite $\zeta$—and this is precisely its strength: it enables direct scans in $\zeta$ and provides a clean diagnostic for when higher-order contamination sets in, i.e., when $\omega(\zeta)$ departs from $\omega_{0}+\zeta\omega_{1} +\zeta^2 \omega_2+\mathcal{O}(\zeta^3)$.

\paragraph{Null Tests: (i) Null Operators}.
On a Ricci-flat background, the operators $\mathcal{O}_{5}$ and $\mathcal{O}_{8}$ are removable by local field redefinitions and therefore induce no first-order ($\mathcal{O}(\zeta)$) vacuum QNM shifts in theory \cite{deRham_EFT_GW_2020,Cano_2024_HD_Kerr_QNM}. This provides a sharp, operator-level null prediction against which both solvers can be stress-tested.

Using EVP across multipoles $\ell=2,3,4$ and overtones $n=0,1,2,3$, we find that the extracted first-order fractional shifts in the null sector are consistent with zero down to a mode-dependent numerical floor. Aggregating over the full $(n,\ell)$ grid yields
$|\delta_{\mathrm{EVP},\min}|\sim10^{-28}$,
$|\delta_{\mathrm{EVP},\mathrm{median}}|\sim10^{-18}$,
and
$|\delta_{\mathrm{EVP},\max}|\lesssim10^{-8}$,
with the largest residuals occurring at low $\ell$ and high $n$, consistent with the increased numerical sensitivity of overtone extraction. 

Independently, in the Leaver approach we confirm that, in the null sector, the dominant dependence of $\Delta\omega(\zeta)$ is quadratic (i.e., $\propto \zeta^{2}$). Moreover, fitting $\Delta\omega(\zeta)$ with polynomial models under harsh requirement shows that the cubic model $M_{123}$ is the most stable (lowest RMS and flattest fit coefficients $\{c_2^{(123)},c_3^{(123)}\}$ across $11$ different fitting windows) for the null-operator diagnostics. Scanning in $\zeta$ confirms that any apparent linear correction $c_1^{(123)}$ at $\mathcal{O}(\zeta)$ is strongly suppressed and consistent with numerical noise, with representative floors
$|\delta_{\mathrm{Leaver},\min}|\lesssim10^{-24}$,
$|\delta_{\mathrm{Leaver},\mathrm{median}}|\lesssim10^{-22}$,
and
$|\delta_{\mathrm{Leaver},\max}|\lesssim10^{-20}$. This makes Leaver estimates for the numerical floor is much more tighter, at same level of EVP diagnostics.

Focusing on the fundamental modes ($n=0$) and taking the magnitude of the complex shifts over $\ell=2,3,4$ using EVP (Tables~\ref{l=2_Table_EVP}--\ref{l=4_Table_EVP}), we obtain $|\delta_{\mathrm{EVP}}^{n=0}|_{\min,\mathrm{med},\max}\lesssim(10^{-26},\,10^{-23},\,10^{-17})$. While the corresponding linear coefficient in the $M_{123}$ fit, the $n=0$ results in Tables~\ref{tab:coeff_stab_deg3_even_odd_null_l2}--\ref{tab:coeff_stab_deg3_even_odd_null_l4} imply that the extracted ${}_{e/o}c_{1}$ values, we find
for $\mathcal{O}_{5}$:
$|{}_{\mathrm{o/e}}c_{1}^{n=0}|_{\min,\mathrm{med},\max}\sim(10^{-21},\,10^{-21},\,10^{-21})$,
while for $\mathcal{O}_{8}$:
$|{}_{\mathrm{o/e}}c_{1}^{n=0}|_{\min,\mathrm{med},\max}\sim(10^{-24},\,10^{-24},\,10^{-23})$. Thus, for the null operators our $n=0$ residuals are suppressed well below the levels reported in \cite{deRham_EFT_GW_2020} (quoted there at $\sim10^{-4}$ in the even sector and $\sim10^{-6}$ in the odd sector), improving the null-test precision with at least $\sim 15-17$ orders of magnitude depending on the solver and parity.

Working with Leaver method, further analysis fitting $\Delta\omega(\zeta)$ with higher-degree polynomials indicates a sizable budget for still-higher powers of $\zeta$ due to higher order contamination; accordingly, we interpret the quoted Leaver-based $c_{1}$ values as a conservative \emph{upper bound} on any residual $\mathcal{O}(\zeta)$ contamination since unmodeled higher-order terms will leak into lower-order coefficients in finite-window fits.

\paragraph{Null Tests: (ii) Ratio between $\mathcal{O}_9$ and $\mathcal{O}_{10}$ QNM shifts.}
We also extract QNM shifts sourced by the control operators $\mathcal{O}_9$ and $\mathcal{O}_{10}$, which are physical cubic invariants built from the Riemann tensor. On Ricci-flat backgrounds ($R_{\mu\nu}=0=R$), we use the identity $\mathcal{O}_{9}=2\mathcal{O}_{10}+\frac{3}{4}\mathcal{O}_{5}\,$,
so that—up to the (null) $\mathcal{O}_{5}$ contamination—setting the Wilson coefficients equal predicts a mode-by-mode ratio $\mathcal{R}_{9/10}=2$ for the first-order shifts. Computing $\mathcal{R}_{9/10}$ from both EVP and Leaver outputs, we find $\mathcal{R}_{9/10}$ is essentially $2$ across the full set of modes, with $\min|\mathcal{R}_{9/10}-2|\sim10^{-12}$ and $\max|\mathcal{R}_{9/10}-2|\lesssim10^{-9}$ across the full mode set.

Finally, comparing the odd and even sectors for both operators, we confirm that parity isospectrality is broken, with no significant systematic trend.

\paragraph{More on Control Operators.}
We use both EVP and Leaver method to compute fractional frequency shifts $\delta_{\mathrm{EVP}}$ and $\delta_{\mathrm{Leaver}}$ for control operators: $\mathcal{O}_9$ and $\mathcal{O}_{10}$. Our $\delta_{\mathrm{EVP}}$ and $\delta_{\mathrm{Leaver}}$ agrees in magnitude with Ref.~\cite{deRham_EFT_GW_2020} when rounded to the two decimal places reported in their Table~I. 

Moreover, comparing EVP shifts to the linear coefficient $c^{(12)}_1$ extracted from Leaver fits, we find tight mode-by-mode agreement quantified by the symmetric relative difference $\epsilon_X$ (computed separately for real $\epsilon_\Re$ and imaginary $\epsilon_\Im$ parts). Aggregating all modes and both control operators, we obtain
$\min(\epsilon_{\Re},\epsilon_{\Im})=(3.54\times10^{-17},\,3.54\times10^{-16})$,
$\mathrm{median}(\epsilon_{\Re},\epsilon_{\Im})=(9.48\times10^{-14},\,2.66\times10^{-14})$,
and $\max(\epsilon_{\Re},\epsilon_{\Im})=(1.13\times10^{-6},\,1.59\times10^{-7})$.
Almost, the same systematic trends appear here as in the null analysis: disagreement grows with overtone index $n$ and decreases with increasing $\ell$. Importantly, the worst-case mismatch occurs precisely where numerical sensitivity is greatest (low $\ell$, high $n$), and is broadly consistent with the EVP numerical floor inferred from null operators.

\paragraph{The main lesson from this work}. Viable EFT corrections can lead to \emph{tiny} effects, so modeling them demands end-to-end numerical approaches whose systematic errors are controlled well below the target signal. In this work we provided such a stress test for the MTE/MTF program in HDG. Taken together all the results listed above, we believe MTE has pass stringent null tests in this article, providing a robust foundation for interpreting future beyond-GR ringdown signatures as physical rather than numerical artifacts.

Looking ahead, the same program can be extended in several directions. First, one can move beyond Schwarzschild by developing slow-rotation and ultimately generic Kerr implementations, enabling high-precision cross-checks in HDG parity-violating sectors, and in the presence of additional matter fields (e.g., dCS) \cite{Cano_2024_HD_Kerr_QNM,Wagle_2024_dCS_SlowRot_Eqs,Li_2025_dCS_SlowRot_QNMs}. Second, finite-$\zeta$ scans can be used to chart the breakdown of linear response and to quantify the onset of higher-order contributions in a controlled manner, complementing EVP’s inherently linear sensitivity. In combination with next-generation instruments \cite{ET_Punturo_2010,CE_Reitze_2019,LISA_AmaroSeoane_2017}, these extensions will help turn EFT-motivated deformations into robust, model-agnostic strong-field tests using black-hole ringdown.

\begin{acknowledgments}
We acknowledge the partial support of Dr. D.S. through the US National Science Foundation under the Grant No.  PHY-2310363.  
This work was supported by the Natural Sciences and Engineering Research Council (NSERC) of Canada. LL
also thanks financial support via the Carlo Fidani Rainer Weiss Chair at Perimeter Institute and CIFAR. Research at Perimeter Institute is supported in part by the Government of Canada through
the Department of Innovation, Science and Economic De- velopment and by the Province of Ontario through the
Ministry of Colleges and Universities. D.~L. acknowledges support from the Simons Foundation (via Award No. 896696), the Simons Foundation International (via Grant No. SFI-MPS-BH-00012593-01), the NSF (via Grants No. PHY-2207650 and PHY-2512423), and the National Aeronautics and Space Administration through award 80NSSC22K0806. P.W. acknowledges funding from the Deutsche Forschungsgemeinschaft (DFG) - project number: 386119226.
\end{acknowledgments}

\bibliography{main}   

\appendix
\begin{widetext}
\section{Effective potentials}\label{Effective potentials}
It is convenient to introduce the dimensionless radius
\begin{equation}
y \;\equiv\; \frac{r}{r_g}\,,
\end{equation}
and to factor out an overall $1/r_g^2$ from all potentials. We define the coefficients $\alpha_\ell,\beta_\ell,\gamma_\ell,\delta_\ell$ as in the Zerilli sector.
\begin{equation}
    \begin{gathered}
        \alpha_{\ell}=(\frac{\lambda_{\ell}}{2} -1)^2 \frac{\lambda_{\ell}}{2} \qquad 
        \beta{\ell}=3/2(\frac{\lambda_{\ell}}{2} -1)^2 \qquad 
        \gamma{\ell}=\frac{9}{4}(\frac{\lambda_{\ell}}{2} -1)\qquad  
        \delta=\frac{9}{8}
    \end{gathered}
\end{equation}

For the odd/even  sectors we have
\begin{equation}
V^{\mathrm{GR}}_{\ell,\mathrm{odd}}(r)
=
\frac{\lambda_\ell}{r^2}
-\frac{3\,r_g}{r^3}
=
\frac{1}{r_g^2}\left(
\frac{\lambda_\ell}{y^2}
-\frac{3}{y^3}
\right), \qquad V^{\mathrm{GR}}_{\ell,\mathrm{even}}(r)
=
\frac{8}{r_g^2}
\,
\frac{
\displaystyle
\frac{\delta_\ell}{y^3}
+\frac{\gamma_\ell}{y^2}
+\frac{\beta_\ell}{y}
+\alpha_\ell}
{\bigl((\lambda_\ell -2)\,y+3\bigr)^2}\,.
\label{eq:VGR_odd}
\end{equation}

\paragraph{EFT corrections.} We keep $d_{58}$ as the shorthand used in the main text and treat all Wilson coefficients as symbolic.

For the odd-parity sector, the EFT correction (cf.\ \texttt{Vodd} in the code) is
\begin{equation}
\begin{aligned}
V^{\mathrm{EFT}}_{\ell,\mathrm{odd}}(r)
&=
\frac{3}{r_g^2\,y^9}
\Bigg[
d_{58}\,\Big(24 - 7\,\lambda_\ell\,y\Big)
+ 6 d_9\Big(80(\lambda_\ell-6)\,y^2
+ 5(230-19\lambda_\ell)\,y - 662\Big)
\\ &\hspace{5em}
+ \frac{3 d_{10}}{2}\Big(
160(\lambda_\ell-6)\,y^2
+ (2300-183\lambda_\ell)\,y - 1348
\Big)
\Bigg].
\end{aligned}
\label{eq:Vodd_EFT}
\end{equation}

For the even-parity sector, the EFT correction (cf.\ \texttt{Veven} in the code) takes the form
\begin{equation}
V^{\mathrm{EFT}}_{\ell,\mathrm{even}}(r)
=
-\frac{3}{2\,r_g^2\,y^9\bigl[(\lambda_\ell-2)\,y+3\bigr]^3}
\Big[
d_{58}\,\mathcal{P}^{(58)}_\ell(y)
+ d_9\,\mathcal{P}^{(9)}_\ell(y)
+ d_{10}\,\mathcal{P}^{(10)}_\ell(y)
\Big],
\label{eq:Veven_EFT}
\end{equation}
with the polynomials in $y$ given by
\begin{align}
\mathcal{P}^{(58)}_\ell(y) &= 
14(\lambda_\ell-2)^3\lambda_\ell\,y^4
+ 6(\lambda_\ell-2)^2(13\lambda_\ell-16)\,y^3
+ 270(\lambda_\ell-2)^2\,y^2
+ 558(\lambda_\ell-2)\,y
+ 432,
\label{eq:P58}\\[0.5em]
\mathcal{P}^{(9)}_\ell(y) &= 
480(\lambda_\ell-6)(\lambda_\ell-2)^3\,y^5
- 36(\lambda_\ell-2)^2(15\lambda_\ell^2-336\lambda_\ell+836)\,y^4
- 60(\lambda_\ell-2)(147\lambda_\ell^2-1304\lambda_\ell+2164)\,y^3
\nonumber\\ &\quad
- 36(\lambda_\ell-2)(1073\lambda_\ell-3988)\,y^2
- 12(5407\lambda_\ell-13460)\,y
- 36648,
\label{eq:P9}\\[0.5em]
\mathcal{P}^{(10)}_\ell(y) &= 
240(\lambda_\ell-6)(\lambda_\ell-2)^3\,y^5
- 3(\lambda_\ell-2)^2(97\lambda_\ell^2-2030\lambda_\ell+5016)\,y^4
- 3(\lambda_\ell-2)(1509\lambda_\ell^2-13166\lambda_\ell+21736)\,y^3
\nonumber\\ &\quad
- 9(\lambda_\ell-2)(2191\lambda_\ell-8066)\,y^2
- 3(11093\lambda_\ell-27478)\,y
- 18972.
\label{eq:P10}
\end{align}

\end{widetext}

\section{More Details On Numerical Methods }\label{More_Details_On_Numerical_Methods}      
\subsection{Continued fraction with Nollert-type tail.}\label{app:nollert_tail}

We denote by $\alpha^{GR}_n$, $\beta^{GR}_n$,
$\gamma^{GR}_n$ the three-term recurrence coefficients of the GR
Schwarzschild problem (either RW or Teukolsky). Introducing the
Riccati variable $R_n \equiv a_{n+1}/a_n$, the characteristic
equation can be written as~\cite{NollertReview}
\begin{equation}
\begin{aligned}
F(\omega) &\equiv \frac{\beta^{GR}_0}{\alpha^{GR}_0} - R_0(\omega) = 0,\\
R_n(\omega) &= \frac{\gamma^{GR}_n}
{\beta^{GR}_n - \alpha^{GR}_n\,R_{n+1}(\omega)}.
\end{aligned}
\end{equation}
At large $n$ we assume that the $R_n$ admits an
asymptotic expansion of the form
\begin{equation}
R_n(\omega)
  = R_\ast(\omega)
  + b(\omega)\,n^{-p}
  + \mathcal{O}\,\!\bigl(n^{-p-1}\bigr),
  \quad n\to\infty,
\end{equation}
where $R_\ast(\omega)$ is the large--$n$ limit of $R_n(\omega)$
and $b(\omega)$ is the complex coefficient of the leading
$n^{-p}$ correction. In practice we
fix a positive exponent $p$ (empirically $p=\tfrac12$ is used), and use this model to construct a Nollert-type tail.

For each trial $\omega_{n\ell}$ we choose a base index $N_{\rm tail}$
(tail start) and two deeper indices
$N_1 = N_{\rm tail} + \Delta N$ and
$N_2 = N_{\rm tail} + 2\Delta N$. Starting from an initial guess
for $R_{N_1+1}$ and $R_{N_2+1}$ (we use $-1$ for both), we
propagate backwards using to obtain two approximations $R^{(1)}_{N_{\rm tail}+1}$ and $R^{(2)}_{N_{\rm tail}+1}$. Assuming
\begin{equation}
R_{N_k}(\omega) \simeq R_\ast(\omega) + a(\omega)\,N_k^{-p},
\quad k=1,2,
\end{equation}
we solve algebraically for $R_\ast(\omega)$ and use it as the seed for the backward sweep:
\begin{equation}
\begin{gathered}
R_{N_{\rm tail}+1}(\omega) \equiv R_\ast(\omega).
\end{gathered}
\end{equation}
This gives a numerically stable approximation to $R_0(\omega)$, and
we then solve the QNM condition $F(\omega)=0$ using a two-point
complex secant method. The precise numerical tolerances and parameter
choices for $N_{\rm tail}$, $\Delta N$, and $p$ are summarized
in the \ref{Leaver Discretization choices and convergence}.

\subsection{Series solution, branch tracking, and residuals.}\label{app:series_branch_residuals}

Once $\omega_{0}$ is fixed, we construct truncated Frobenius series
for both the RW and Teukolsky solutions. For RW we use the
horizon-based coordinate
\begin{equation}
x_{\rm RW} = 1 - \frac{r_g}{r},
\end{equation}
and write the master function as
\begin{equation}
\Psi_{\rm RW}(r)
  = \mathcal{P}_{\rm RW}(r,\omega_{0})
    \sum_{n=0}^{N_{\rm max}} a_n x_{\rm RW}^n,
\end{equation}
where the prefactor $\mathcal{P}_{\rm RW}$ enforces the ingoing
horizon and outgoing infinity behaviour,
\begin{equation}
\mathcal{P}_{\rm RW}(r,\omega_{0})
= e^{-i\omega_{0}\,\ln(r-r_g) + 2 i \omega_{0}\,\ln r + i\omega_{0} (r-r_g)}.
\end{equation}
In the code this prefactor is evaluated using a branch-tracking system
that keeps track of the phases of $\ln r$ and $\ln(r-r_g)$ along a
chosen path in the complex $r$--plane. As $r$ is varied, the
arguments are adjusted by multiples of $2\pi$ so that individual
phase jumps are kept below $\pi$. A normalization factor
$\mathcal{N}$ is fixed at an anchor point $r=r_{\rm anc}=r_g$ so
that the prefactor is continuous and has a fixed phase at that
location:
\begin{equation}
\mathcal{P}_{\rm RW}(r,\omega_{0}) = \frac{
e^{-i\omega_{0}\,\log_{(r-r_g)}(r-r_g) + 2 i\omega_{0}\,\log_{(r)}(r)
+ i\omega_{0} (r-r_g)}
}{\mathcal{N}},
\end{equation}
with $\mathcal{N}$ chosen such that the tracked logarithms coincide
with the principal logarithms at $r=r_{\rm anc}$. This guarantees a
consistent asymptotic phase for $\Psi_{\rm RW}$ both along the real
axis and along the complex contour. The Teukolsky master function is
treated analogously.

Radial derivatives of the series are obtained analytically, and along
the contour we also use spectral differentiation in the contour
parameter (see below) to compute $\mathrm{d}r/\mathrm{d}\lambda$ and
$\mathrm{d}r_*/\mathrm{d}\lambda$.

As a sanity check on the series representation, we evaluate the RW
and Teukolsky differential operators on a real radial grid
$r\in[r_g+\varepsilon, R_{\rm max}]$, with a small buffer
$\varepsilon$ above the horizon and a finite outer radius
$R_{\rm max}$. This yields residuals $\mathcal{R}(r)$ for each
equation, and we monitor the $\infty$--norm
\begin{equation}
\|\mathcal{R}\|_\infty = \max_r|\mathcal{R}(r)|,
\end{equation}
for both RW and Teukolsky. In practice, this norm is many orders of
magnitude smaller than unity and is stable under variations of
$N_{\rm max}$ and the Nollert tail parameters, confirming that both
the tail and the branch-tracked prefactors are correctly capturing
the desired boundary conditions.

\subsection{Consistency check in GR.}
Before turning on the EFT couplings we perform a stringent self–consistency
test of the Leaver+reduction approach.  We build a ``fat''
$12$–term Frobenius recurrence for both the RW and
Zerilli cases, and then reduce it to a $3$--term band by the
same high–precision reducer used in the EFT runs.  We then solve for the GR
frequencies $\omega^{\mathrm{GR}}_{\mathrm{RW}}$ and
$\omega^{\mathrm{GR}}_{\mathrm{Z}}$ using the scalar Leaver solver and
compare
\begin{equation}
  \Delta\omega^{\mathrm{GR}}
  \equiv \omega^{\mathrm{GR}}_{\mathrm{RW}}
       - \omega^{\mathrm{GR}}_{\mathrm{Z}} \,.
\end{equation}
Across the tested modes we find
$|\Delta\omega^{\mathrm{GR}}| \lesssim 10^{-61}$,
comfortably below the continued–fraction truncation error and intrinsic
precision floor of the solver.  This confirms that the Leaver method and the
band–reduction algebra are internally consistent, and that over the entire
$\zeta$ grid our EFT frequency shifts are measured against GR with an error
budget of order $10^{-61}$.

\begin{table*}[t]
\centering
\caption{Power-scaling summary for $\zeta_{\max}=10^{-4}$ and odd parity.
Each $p$ entry is $p_{\rm eff}\pm\sigma_p$ with the chosen discrete $p_\star$ shown as $[p_\star]$.
$C$ lists the fitted coefficient components $\Re(C)\,|\,\Im(C)$ and $\sigma/\mu$ is the flatness ratio.}
\label{tab:power_1Em4_odd_910}
\setlength{\tabcolsep}{4pt}
\renewcommand{\arraystretch}{1.15}
\begin{tabular}{cc||ccc|ccc}
\toprule
 &  & \multicolumn{3}{c}{$\mathcal{O}_{9}$} & \multicolumn{3}{c}{$\mathcal{O}_{10}$} \\
$\ell$ & $n$ & $p$ & $C$ & $\sigma/\mu$ & $p$ & $C$ & $\sigma/\mu$ \\
\midrule
2 & 0 & $0.9995\pm0.0016\,[1]$ & $\makebox[4em][r]{\ensuremath{21.14}}\,\vert\,\makebox[4em][l]{\ensuremath{47.05}}$ & $2.19\times10^{-3}$ & $0.9998\pm0.0006\,[1]$ & $\makebox[4em][r]{\ensuremath{10.53}}\,\vert\,\makebox[4em][l]{\ensuremath{23.68}}$ & $8.05\times10^{-4}$ \\
2 & 1 & $1.0000\pm0.0001\,[1]$ & $\makebox[4em][r]{\ensuremath{17.96}}\,\vert\,\makebox[4em][l]{\ensuremath{70.20}}$ & $5.36\times10^{-3}$ & $0.9999\pm0.0005\,[1]$ & $\makebox[4em][r]{\ensuremath{8.48}}\,\vert\,\makebox[4em][l]{\ensuremath{35.13}}$ & $2.24\times10^{-3}$ \\
2 & 2 & $1.0010\pm0.0031\,[1]$ & $\makebox[4em][r]{\ensuremath{-40.36}}\,\vert\,\makebox[4em][l]{\ensuremath{121.76}}$ & $2.46\times10^{-2}$ & $1.0002\pm0.0008\,[1]$ & $\makebox[4em][r]{\ensuremath{-23.37}}\,\vert\,\makebox[4em][l]{\ensuremath{58.94}}$ & $1.07\times10^{-2}$ \\
2 & 3 & $0.9979\pm0.0133\,[1]$ & $\makebox[4em][r]{\ensuremath{-356.62}}\,\vert\,\makebox[4em][l]{\ensuremath{238.34}}$ & $3.55\times10^{-2}$ & $1.0001\pm0.0010\,[1]$ & $\makebox[4em][r]{\ensuremath{-192.09}}\,\vert\,\makebox[4em][l]{\ensuremath{103.37}}$ & $1.34\times10^{-2}$ \\
3 & 0 & $0.9995\pm0.0018\,[1]$ & $\makebox[4em][r]{\ensuremath{18.32}}\,\vert\,\makebox[4em][l]{\ensuremath{45.78}}$ & $2.31\times10^{-3}$ & $0.9998\pm0.0006\,[1]$ & $\makebox[4em][r]{\ensuremath{9.14}}\,\vert\,\makebox[4em][l]{\ensuremath{23.06}}$ & $7.24\times10^{-4}$ \\
3 & 1 & $0.9994\pm0.0021\,[1]$ & $\makebox[4em][r]{\ensuremath{16.01}}\,\vert\,\makebox[4em][l]{\ensuremath{50.71}}$ & $3.42\times10^{-3}$ & $0.9997\pm0.0011\,[1]$ & $\makebox[4em][r]{\ensuremath{7.87}}\,\vert\,\makebox[4em][l]{\ensuremath{25.53}}$ & $1.54\times10^{-3}$ \\
3 & 2 & $1.0000\pm0.0001\,[1]$ & $\makebox[4em][r]{\ensuremath{7.35}}\,\vert\,\makebox[4em][l]{\ensuremath{61.13}}$ & $8.11\times10^{-3}$ & $0.9998\pm0.0006\,[1]$ & $\makebox[4em][r]{\ensuremath{3.03}}\,\vert\,\makebox[4em][l]{\ensuremath{30.53}}$ & $3.28\times10^{-3}$ \\
3 & 3 & $1.0008\pm0.0023\,[1]$ & $\makebox[4em][r]{\ensuremath{-18.09}}\,\vert\,\makebox[4em][l]{\ensuremath{78.65}}$ & $2.27\times10^{-2}$ & $1.0001\pm0.0003\,[1]$ & $\makebox[4em][r]{\ensuremath{-11.05}}\,\vert\,\makebox[4em][l]{\ensuremath{38.36}}$ & $9.56\times10^{-3}$ \\
4 & 0 & $0.9994\pm0.0018\,[1]$ & $\makebox[4em][r]{\ensuremath{17.55}}\,\vert\,\makebox[4em][l]{\ensuremath{45.23}}$ & $2.29\times10^{-3}$ & $0.9998\pm0.0005\,[1]$ & $\makebox[4em][r]{\ensuremath{8.77}}\,\vert\,\makebox[4em][l]{\ensuremath{22.78}}$ & $6.51\times10^{-4}$ \\
4 & 1 & $0.9993\pm0.0024\,[1]$ & $\makebox[4em][r]{\ensuremath{15.94}}\,\vert\,\makebox[4em][l]{\ensuremath{47.13}}$ & $3.22\times10^{-3}$ & $0.9997\pm0.0010\,[1]$ & $\makebox[4em][r]{\ensuremath{7.91}}\,\vert\,\makebox[4em][l]{\ensuremath{23.76}}$ & $1.32\times10^{-3}$ \\
4 & 2 & $0.9993\pm0.0023\,[1]$ & $\makebox[4em][r]{\ensuremath{11.83}}\,\vert\,\makebox[4em][l]{\ensuremath{51.17}}$ & $4.81\times10^{-3}$ & $0.9996\pm0.0014\,[1]$ & $\makebox[4em][r]{\ensuremath{5.65}}\,\vert\,\makebox[4em][l]{\ensuremath{25.76}}$ & $2.14\times10^{-3}$ \\
4 & 3 & $0.9999\pm0.0005\,[1]$ & $\makebox[4em][r]{\ensuremath{3.16}}\,\vert\,\makebox[4em][l]{\ensuremath{57.89}}$ & $1.06\times10^{-2}$ & $0.9998\pm0.0008\,[1]$ & $\makebox[4em][r]{\ensuremath{0.80}}\,\vert\,\makebox[4em][l]{\ensuremath{28.89}}$ & $4.31\times10^{-3}$ \\
\bottomrule
\end{tabular}
\end{table*}
\begin{table*}[t]
\centering
\caption{Power-scaling summary for $\zeta_{\max}=10^{-4}$ and $\mathrm{even}$ parity.
Each $p$ entry is $p_{\rm eff}\pm\sigma_p$ with the chosen discrete $p_\star$ shown as $[p_\star]$.
$C$ lists the fitted coefficient components $\Re(C)\,|\,\Im(C)$ and $\sigma/\mu$ is the flatness ratio.}
\label{tab:power_1Em4_even_910}
\setlength{\tabcolsep}{4pt}
\renewcommand{\arraystretch}{1.15}
\begin{tabular}{cc||ccc|ccc}
\toprule
 &  & \multicolumn{3}{c}{$\mathcal{O}_{9}$} & \multicolumn{3}{c}{$\mathcal{O}_{10}$} \\
$\ell$ & $n$ & $p$ & $C$ & $\sigma/\mu$ & $p$ & $C$ & $\sigma/\mu$ \\
\midrule
2 & 0 & $1.0000\pm0.0001\,[1]$ & $\makebox[4em][r]{\ensuremath{-12.36}}\,\vert\,\makebox[4em][l]{\ensuremath{-58.13}}$ & $8.22\times10^{-5}$ & $1.0000\pm0.0001\,[1]$ & $\makebox[4em][r]{\ensuremath{-6.16}}\,\vert\,\makebox[4em][l]{\ensuremath{-29.07}}$ & $9.54\times10^{-5}$ \\
2 & 1 & $1.0002\pm0.0006\,[1]$ & $\makebox[4em][r]{\ensuremath{11.00}}\,\vert\,\makebox[4em][l]{\ensuremath{-55.94}}$ & $1.12\times10^{-3}$ & $1.0000\pm0.0001\,[1]$ & $\makebox[4em][r]{\ensuremath{5.48}}\,\vert\,\makebox[4em][l]{\ensuremath{-27.88}}$ & $8.70\times10^{-4}$ \\
2 & 2 & $1.0001\pm0.0003\,[1]$ & $\makebox[4em][r]{\ensuremath{77.00}}\,\vert\,\makebox[4em][l]{\ensuremath{-53.04}}$ & $4.74\times10^{-3}$ & $0.9999\pm0.0003\,[1]$ & $\makebox[4em][r]{\ensuremath{38.19}}\,\vert\,\makebox[4em][l]{\ensuremath{-26.71}}$ & $3.05\times10^{-3}$ \\
2 & 3 & $0.9978\pm0.0071\,[1]$ & $\makebox[4em][r]{\ensuremath{222.39}}\,\vert\,\makebox[4em][l]{\ensuremath{-55.47}}$ & $9.22\times10^{-3}$ & $0.9986\pm0.0046\,[1]$ & $\makebox[4em][r]{\ensuremath{112.36}}\,\vert\,\makebox[4em][l]{\ensuremath{-28.89}}$ & $5.94\times10^{-3}$ \\
3 & 0 & $0.9999\pm0.0003\,[1]$ & $\makebox[4em][r]{\ensuremath{-14.17}}\,\vert\,\makebox[4em][l]{\ensuremath{-51.72}}$ & $4.51\times10^{-4}$ & $1.0000\pm0.0001\,[1]$ & $\makebox[4em][r]{\ensuremath{-7.07}}\,\vert\,\makebox[4em][l]{\ensuremath{-25.92}}$ & $9.08\times10^{-5}$ \\
3 & 1 & $1.0003\pm0.0010\,[1]$ & $\makebox[4em][r]{\ensuremath{-8.04}}\,\vert\,\makebox[4em][l]{\ensuremath{-53.80}}$ & $1.17\times10^{-3}$ & $1.0001\pm0.0005\,[1]$ & $\makebox[4em][r]{\ensuremath{-3.97}}\,\vert\,\makebox[4em][l]{\ensuremath{-26.83}}$ & $6.08\times10^{-4}$ \\
3 & 2 & $1.0004\pm0.0014\,[1]$ & $\makebox[4em][r]{\ensuremath{8.57}}\,\vert\,\makebox[4em][l]{\ensuremath{-57.45}}$ & $2.62\times10^{-3}$ & $1.0001\pm0.0003\,[1]$ & $\makebox[4em][r]{\ensuremath{4.22}}\,\vert\,\makebox[4em][l]{\ensuremath{-28.57}}$ & $1.70\times10^{-3}$ \\
3 & 3 & $1.0004\pm0.0013\,[1]$ & $\makebox[4em][r]{\ensuremath{44.58}}\,\vert\,\makebox[4em][l]{\ensuremath{-62.00}}$ & $7.52\times10^{-3}$ & $1.0001\pm0.0002\,[1]$ & $\makebox[4em][r]{\ensuremath{21.82}}\,\vert\,\makebox[4em][l]{\ensuremath{-31.05}}$ & $4.74\times10^{-3}$ \\
4 & 0 & $0.9998\pm0.0007\,[1]$ & $\makebox[4em][r]{\ensuremath{-14.47}}\,\vert\,\makebox[4em][l]{\ensuremath{-49.99}}$ & $8.12\times10^{-4}$ & $1.0000\pm0.0000\,[1]$ & $\makebox[4em][r]{\ensuremath{-7.23}}\,\vert\,\makebox[4em][l]{\ensuremath{-25.08}}$ & $8.51\times10^{-6}$ \\
4 & 1 & $1.0002\pm0.0005\,[1]$ & $\makebox[4em][r]{\ensuremath{-11.53}}\,\vert\,\makebox[4em][l]{\ensuremath{-51.47}}$ & $6.35\times10^{-4}$ & $1.0001\pm0.0004\,[1]$ & $\makebox[4em][r]{\ensuremath{-5.72}}\,\vert\,\makebox[4em][l]{\ensuremath{-25.73}}$ & $4.84\times10^{-4}$ \\
4 & 2 & $1.0005\pm0.0017\,[1]$ & $\makebox[4em][r]{\ensuremath{-4.25}}\,\vert\,\makebox[4em][l]{\ensuremath{-54.21}}$ & $2.06\times10^{-3}$ & $1.0002\pm0.0007\,[1]$ & $\makebox[4em][r]{\ensuremath{-2.08}}\,\vert\,\makebox[4em][l]{\ensuremath{-26.97}}$ & $1.04\times10^{-3}$ \\
4 & 3 & $1.0006\pm0.0019\,[1]$ & $\makebox[4em][r]{\ensuremath{10.27}}\,\vert\,\makebox[4em][l]{\ensuremath{-57.84}}$ & $3.63\times10^{-3}$ & $1.0002\pm0.0005\,[1]$ & $\makebox[4em][r]{\ensuremath{5.03}}\,\vert\,\makebox[4em][l]{\ensuremath{-28.73}}$ & $2.33\times10^{-3}$ \\
\bottomrule
\end{tabular}
\end{table*}
\begin{table*}[t]
\centering
\caption{Power-scaling summary for $\zeta_{\max}=10^{-5}$ and odd parity.
Each $p$ entry is $p_{\rm eff}\pm\sigma_p$ with the chosen discrete $p_\star$ shown as $[p_\star]$.
$C$ lists the fitted coefficient components $\Re(C)\,|\,\Im(C)$ and $\sigma/\mu$ is the flatness ratio.}
\label{tab:power_1Em5_odd_910}
\setlength{\tabcolsep}{4pt}
\renewcommand{\arraystretch}{1.15}
\begin{tabular}{cc||ccc|ccc}
\toprule
 &  & \multicolumn{3}{c}{$\mathcal{O}_{9}$} & \multicolumn{3}{c}{$\mathcal{O}_{10}$} \\
$\ell$ & $n$ & $p$ & $C$ & $\sigma/\mu$ & $p$ & $C$ & $\sigma/\mu$ \\
\midrule
2 & 0 & $0.9999\pm0.0002\,[1]$ & $\makebox[4em][r]{\ensuremath{21.08}}\,\vert\,\makebox[4em][l]{\ensuremath{47.48}}$ & $2.27\times10^{-4}$ & $1.0000\pm0.0001\,[1]$ & $\makebox[4em][r]{\ensuremath{10.54}}\,\vert\,\makebox[4em][l]{\ensuremath{23.76}}$ & $8.41\times10^{-5}$ \\
2 & 1 & $1.0000\pm0.0000\,[1]$ & $\makebox[4em][r]{\ensuremath{16.52}}\,\vert\,\makebox[4em][l]{\ensuremath{70.54}}$ & $5.55\times10^{-4}$ & $1.0000\pm0.0000\,[1]$ & $\makebox[4em][r]{\ensuremath{8.21}}\,\vert\,\makebox[4em][l]{\ensuremath{35.27}}$ & $2.33\times10^{-4}$ \\
2 & 2 & $1.0001\pm0.0004\,[1]$ & $\makebox[4em][r]{\ensuremath{-50.79}}\,\vert\,\makebox[4em][l]{\ensuremath{115.86}}$ & $2.38\times10^{-3}$ & $1.0000\pm0.0001\,[1]$ & $\makebox[4em][r]{\ensuremath{-25.66}}\,\vert\,\makebox[4em][l]{\ensuremath{57.73}}$ & $1.08\times10^{-3}$ \\
2 & 3 & $1.0002\pm0.0007\,[1]$ & $\makebox[4em][r]{\ensuremath{-392.50}}\,\vert\,\makebox[4em][l]{\ensuremath{188.29}}$ & $2.64\times10^{-3}$ & $1.0001\pm0.0003\,[1]$ & $\makebox[4em][r]{\ensuremath{-196.30}}\,\vert\,\makebox[4em][l]{\ensuremath{92.89}}$ & $1.22\times10^{-3}$ \\
3 & 0 & $0.9999\pm0.0002\,[1]$ & $\makebox[4em][r]{\ensuremath{18.29}}\,\vert\,\makebox[4em][l]{\ensuremath{46.22}}$ & $2.41\times10^{-4}$ & $1.0000\pm0.0001\,[1]$ & $\makebox[4em][r]{\ensuremath{9.14}}\,\vert\,\makebox[4em][l]{\ensuremath{23.13}}$ & $7.58\times10^{-5}$ \\
3 & 1 & $0.9999\pm0.0002\,[1]$ & $\makebox[4em][r]{\ensuremath{15.68}}\,\vert\,\makebox[4em][l]{\ensuremath{51.32}}$ & $3.54\times10^{-4}$ & $1.0000\pm0.0001\,[1]$ & $\makebox[4em][r]{\ensuremath{7.83}}\,\vert\,\makebox[4em][l]{\ensuremath{25.68}}$ & $1.61\times10^{-4}$ \\
3 & 2 & $1.0000\pm0.0000\,[1]$ & $\makebox[4em][r]{\ensuremath{5.46}}\,\vert\,\makebox[4em][l]{\ensuremath{61.30}}$ & $8.43\times10^{-4}$ & $1.0000\pm0.0001\,[1]$ & $\makebox[4em][r]{\ensuremath{2.67}}\,\vert\,\makebox[4em][l]{\ensuremath{30.64}}$ & $3.42\times10^{-4}$ \\
3 & 3 & $1.0001\pm0.0003\,[1]$ & $\makebox[4em][r]{\ensuremath{-24.49}}\,\vert\,\makebox[4em][l]{\ensuremath{75.96}}$ & $2.27\times10^{-3}$ & $1.0000\pm0.0000\,[1]$ & $\makebox[4em][r]{\ensuremath{-12.42}}\,\vert\,\makebox[4em][l]{\ensuremath{37.87}}$ & $9.79\times10^{-4}$ \\
4 & 0 & $0.9999\pm0.0002\,[1]$ & $\makebox[4em][r]{\ensuremath{17.55}}\,\vert\,\makebox[4em][l]{\ensuremath{45.66}}$ & $2.41\times10^{-4}$ & $1.0000\pm0.0001\,[1]$ & $\makebox[4em][r]{\ensuremath{8.77}}\,\vert\,\makebox[4em][l]{\ensuremath{22.85}}$ & $6.81\times10^{-5}$ \\
4 & 1 & $0.9999\pm0.0002\,[1]$ & $\makebox[4em][r]{\ensuremath{15.82}}\,\vert\,\makebox[4em][l]{\ensuremath{47.73}}$ & $3.36\times10^{-4}$ & $1.0000\pm0.0001\,[1]$ & $\makebox[4em][r]{\ensuremath{7.90}}\,\vert\,\makebox[4em][l]{\ensuremath{23.88}}$ & $1.39\times10^{-4}$ \\
4 & 2 & $0.9999\pm0.0002\,[1]$ & $\makebox[4em][r]{\ensuremath{11.15}}\,\vert\,\makebox[4em][l]{\ensuremath{51.86}}$ & $4.97\times10^{-4}$ & $1.0000\pm0.0001\,[1]$ & $\makebox[4em][r]{\ensuremath{5.55}}\,\vert\,\makebox[4em][l]{\ensuremath{25.94}}$ & $2.23\times10^{-4}$ \\
4 & 3 & $1.0000\pm0.0000\,[1]$ & $\makebox[4em][r]{\ensuremath{0.82}}\,\vert\,\makebox[4em][l]{\ensuremath{58.02}}$ & $1.11\times10^{-3}$ & $1.0000\pm0.0001\,[1]$ & $\makebox[4em][r]{\ensuremath{0.33}}\,\vert\,\makebox[4em][l]{\ensuremath{29.00}}$ & $4.49\times10^{-4}$ \\
\bottomrule
\end{tabular}
\end{table*}
\begin{table*}[t]
\centering
\caption{Power-scaling summary for $\zeta_{\max}=10^{-5}$ and $\mathrm{even}$ parity.
Each $p$ entry is $p_{\rm eff}\pm\sigma_p$ with the chosen discrete $p_\star$ shown as $[p_\star]$.
$C$ lists the fitted coefficient components $\Re(C)\,|\,\Im(C)$ and $\sigma/\mu$ is the flatness ratio.}
\label{tab:power_1Em5_even_910}
\setlength{\tabcolsep}{4pt}
\renewcommand{\arraystretch}{1.15}
\begin{tabular}{cc||ccc|ccc}
\toprule
 &  & \multicolumn{3}{c}{$\mathcal{O}_{9}$} & \multicolumn{3}{c}{$\mathcal{O}_{10}$} \\
$\ell$ & $n$ & $p$ & $C$ & $\sigma/\mu$ & $p$ & $C$ & $\sigma/\mu$ \\
\midrule
2 & 0 & $1.0000\pm0.0000\,[1]$ & $\makebox[4em][r]{\ensuremath{-12.34}}\,\vert\,\makebox[4em][l]{\ensuremath{-58.12}}$ & $8.48\times10^{-6}$ & $1.0000\pm0.0000\,[1]$ & $\makebox[4em][r]{\ensuremath{-6.17}}\,\vert\,\makebox[4em][l]{\ensuremath{-29.06}}$ & $9.97\times10^{-6}$ \\
2 & 1 & $1.0000\pm0.0001\,[1]$ & $\makebox[4em][r]{\ensuremath{10.78}}\,\vert\,\makebox[4em][l]{\ensuremath{-55.83}}$ & $1.18\times10^{-4}$ & $1.0000\pm0.0000\,[1]$ & $\makebox[4em][r]{\ensuremath{5.39}}\,\vert\,\makebox[4em][l]{\ensuremath{-27.90}}$ & $9.09\times10^{-5}$ \\
2 & 2 & $1.0000\pm0.0000\,[1]$ & $\makebox[4em][r]{\ensuremath{75.93}}\,\vert\,\makebox[4em][l]{\ensuremath{-54.33}}$ & $5.14\times10^{-4}$ & $1.0000\pm0.0000\,[1]$ & $\makebox[4em][r]{\ensuremath{37.93}}\,\vert\,\makebox[4em][l]{\ensuremath{-27.19}}$ & $3.24\times10^{-4}$ \\
2 & 3 & $0.9998\pm0.0007\,[1]$ & $\makebox[4em][r]{\ensuremath{228.70}}\,\vert\,\makebox[4em][l]{\ensuremath{-60.79}}$ & $9.85\times10^{-4}$ & $0.9998\pm0.0005\,[1]$ & $\makebox[4em][r]{\ensuremath{114.46}}\,\vert\,\makebox[4em][l]{\ensuremath{-30.53}}$ & $6.27\times10^{-4}$ \\
3 & 0 & $1.0000\pm0.0000\,[1]$ & $\makebox[4em][r]{\ensuremath{-14.15}}\,\vert\,\makebox[4em][l]{\ensuremath{-51.81}}$ & $4.69\times10^{-5}$ & $1.0000\pm0.0000\,[1]$ & $\makebox[4em][r]{\ensuremath{-7.07}}\,\vert\,\makebox[4em][l]{\ensuremath{-25.91}}$ & $9.37\times10^{-6}$ \\
3 & 1 & $1.0000\pm0.0001\,[1]$ & $\makebox[4em][r]{\ensuremath{-8.00}}\,\vert\,\makebox[4em][l]{\ensuremath{-53.56}}$ & $1.21\times10^{-4}$ & $1.0000\pm0.0000\,[1]$ & $\makebox[4em][r]{\ensuremath{-3.99}}\,\vert\,\makebox[4em][l]{\ensuremath{-26.77}}$ & $6.34\times10^{-5}$ \\
3 & 2 & $1.0001\pm0.0002\,[1]$ & $\makebox[4em][r]{\ensuremath{8.09}}\,\vert\,\makebox[4em][l]{\ensuremath{-57.13}}$ & $2.75\times10^{-4}$ & $1.0000\pm0.0000\,[1]$ & $\makebox[4em][r]{\ensuremath{4.04}}\,\vert\,\makebox[4em][l]{\ensuremath{-28.55}}$ & $1.77\times10^{-4}$ \\
3 & 3 & $1.0001\pm0.0002\,[1]$ & $\makebox[4em][r]{\ensuremath{42.58}}\,\vert\,\makebox[4em][l]{\ensuremath{-62.80}}$ & $8.05\times10^{-4}$ & $1.0000\pm0.0000\,[1]$ & $\makebox[4em][r]{\ensuremath{21.24}}\,\vert\,\makebox[4em][l]{\ensuremath{-31.40}}$ & $5.03\times10^{-4}$ \\
4 & 0 & $1.0000\pm0.0001\,[1]$ & $\makebox[4em][r]{\ensuremath{-14.46}}\,\vert\,\makebox[4em][l]{\ensuremath{-50.15}}$ & $8.40\times10^{-5}$ & $1.0000\pm0.0000\,[1]$ & $\makebox[4em][r]{\ensuremath{-7.23}}\,\vert\,\makebox[4em][l]{\ensuremath{-25.08}}$ & $8.95\times10^{-7}$ \\
4 & 1 & $1.0000\pm0.0001\,[1]$ & $\makebox[4em][r]{\ensuremath{-11.45}}\,\vert\,\makebox[4em][l]{\ensuremath{-51.36}}$ & $6.50\times10^{-5}$ & $1.0000\pm0.0000\,[1]$ & $\makebox[4em][r]{\ensuremath{-5.72}}\,\vert\,\makebox[4em][l]{\ensuremath{-25.68}}$ & $5.02\times10^{-5}$ \\
4 & 2 & $1.0001\pm0.0002\,[1]$ & $\makebox[4em][r]{\ensuremath{-4.28}}\,\vert\,\makebox[4em][l]{\ensuremath{-53.79}}$ & $2.14\times10^{-4}$ & $1.0000\pm0.0001\,[1]$ & $\makebox[4em][r]{\ensuremath{-2.13}}\,\vert\,\makebox[4em][l]{\ensuremath{-26.88}}$ & $1.09\times10^{-4}$ \\
4 & 3 & $1.0001\pm0.0002\,[1]$ & $\makebox[4em][r]{\ensuremath{9.57}}\,\vert\,\makebox[4em][l]{\ensuremath{-57.44}}$ & $3.79\times10^{-4}$ & $1.0000\pm0.0001\,[1]$ & $\makebox[4em][r]{\ensuremath{4.77}}\,\vert\,\makebox[4em][l]{\ensuremath{-28.70}}$ & $2.43\times10^{-4}$ \\
\bottomrule
\end{tabular}
\end{table*}

\begin{table*}[t]
\centering
\caption{Power-scaling summary for $\zeta_{\max}=10^{-4}$ and odd parity.
Each $p$ entry is $p_{\rm eff}\pm\sigma_p$ with the chosen discrete $p_\star$ shown as $[p_\star]$.
$C$ lists the fitted coefficient components $\Re(C)\,|\,\Im(C)$ and $\sigma/\mu$ is the flatness ratio.}
\label{tab:power_1Em4_odd_58}
\setlength{\tabcolsep}{4pt}
\renewcommand{\arraystretch}{1.15}
\begin{tabular}{cc||ccc|ccc}
\toprule
 &  & \multicolumn{3}{c}{$\mathcal{O}_{5}$} & \multicolumn{3}{c}{$\mathcal{O}_{8}$} \\
$\ell$ & $n$ & $p$ & $C$ & $\sigma/\mu$ & $p$ & $C$ & $\sigma/\mu$ \\
\midrule
2 & 0 & $2.0001\pm0.0004\,[2]$ & $\makebox[4em][r]{\ensuremath{120.58}}\,\vert\,\makebox[4em][l]{\ensuremath{478.06}}$ & $5.24\times10^{-4}$ & $2.0000\pm0.0001\,[2]$ & $\makebox[4em][r]{\ensuremath{7.50}}\,\vert\,\makebox[4em][l]{\ensuremath{29.83}}$ & $1.31\times10^{-4}$ \\
2 & 1 & $2.0003\pm0.0011\,[2]$ & $\makebox[4em][r]{\ensuremath{-34.19}}\,\vert\,\makebox[4em][l]{\ensuremath{599.67}}$ & $1.42\times10^{-3}$ & $2.0001\pm0.0003\,[2]$ & $\makebox[4em][r]{\ensuremath{-2.21}}\,\vert\,\makebox[4em][l]{\ensuremath{37.31}}$ & $3.54\times10^{-4}$ \\
2 & 2 & $2.0004\pm0.0013\,[2]$ & $\makebox[4em][r]{\ensuremath{-688.49}}\,\vert\,\makebox[4em][l]{\ensuremath{745.87}}$ & $1.91\times10^{-3}$ & $2.0001\pm0.0003\,[2]$ & $\makebox[4em][r]{\ensuremath{-42.97}}\,\vert\,\makebox[4em][l]{\ensuremath{46.20}}$ & $4.73\times10^{-4}$ \\
2 & 3 & $2.0006\pm0.0020\,[2]$ & $\makebox[4em][r]{\ensuremath{-2824.48}}\,\vert\,\makebox[4em][l]{\ensuremath{939.68}}$ & $2.44\times10^{-3}$ & $2.0002\pm0.0005\,[2]$ & $\makebox[4em][r]{\ensuremath{-175.20}}\,\vert\,\makebox[4em][l]{\ensuremath{57.87}}$ & $6.05\times10^{-4}$ \\
3 & 0 & $2.0000\pm0.0001\,[2]$ & $\makebox[4em][r]{\ensuremath{157.29}}\,\vert\,\makebox[4em][l]{\ensuremath{482.03}}$ & $3.60\times10^{-4}$ & $2.0000\pm0.0000\,[2]$ & $\makebox[4em][r]{\ensuremath{9.80}}\,\vert\,\makebox[4em][l]{\ensuremath{30.15}}$ & $8.93\times10^{-5}$ \\
3 & 1 & $2.0001\pm0.0005\,[2]$ & $\makebox[4em][r]{\ensuremath{138.01}}\,\vert\,\makebox[4em][l]{\ensuremath{599.74}}$ & $7.75\times10^{-4}$ & $2.0000\pm0.0001\,[2]$ & $\makebox[4em][r]{\ensuremath{8.54}}\,\vert\,\makebox[4em][l]{\ensuremath{37.43}}$ & $1.94\times10^{-4}$ \\
3 & 2 & $2.0004\pm0.0012\,[2]$ & $\makebox[4em][r]{\ensuremath{30.54}}\,\vert\,\makebox[4em][l]{\ensuremath{762.00}}$ & $1.82\times10^{-3}$ & $2.0001\pm0.0003\,[2]$ & $\makebox[4em][r]{\ensuremath{1.70}}\,\vert\,\makebox[4em][l]{\ensuremath{47.40}}$ & $4.52\times10^{-4}$ \\
3 & 3 & $2.0005\pm0.0016\,[2]$ & $\makebox[4em][r]{\ensuremath{-325.04}}\,\vert\,\makebox[4em][l]{\ensuremath{990.95}}$ & $2.82\times10^{-3}$ & $2.0001\pm0.0004\,[2]$ & $\makebox[4em][r]{\ensuremath{-20.64}}\,\vert\,\makebox[4em][l]{\ensuremath{61.38}}$ & $6.98\times10^{-4}$ \\
4 & 0 & $1.9999\pm0.0004\,[2]$ & $\makebox[4em][r]{\ensuremath{166.35}}\,\vert\,\makebox[4em][l]{\ensuremath{462.13}}$ & $6.16\times10^{-4}$ & $2.0000\pm0.0001\,[2]$ & $\makebox[4em][r]{\ensuremath{10.37}}\,\vert\,\makebox[4em][l]{\ensuremath{28.95}}$ & $1.53\times10^{-4}$ \\
4 & 1 & $2.0000\pm0.0000\,[2]$ & $\makebox[4em][r]{\ensuremath{170.06}}\,\vert\,\makebox[4em][l]{\ensuremath{556.24}}$ & $5.21\times10^{-4}$ & $2.0000\pm0.0000\,[2]$ & $\makebox[4em][r]{\ensuremath{10.56}}\,\vert\,\makebox[4em][l]{\ensuremath{34.78}}$ & $1.30\times10^{-4}$ \\
4 & 2 & $2.0002\pm0.0005\,[2]$ & $\makebox[4em][r]{\ensuremath{155.13}}\,\vert\,\makebox[4em][l]{\ensuremath{679.07}}$ & $1.12\times10^{-3}$ & $2.0000\pm0.0001\,[2]$ & $\makebox[4em][r]{\ensuremath{9.54}}\,\vert\,\makebox[4em][l]{\ensuremath{42.38}}$ & $2.80\times10^{-4}$ \\
4 & 3 & $2.0004\pm0.0012\,[2]$ & $\makebox[4em][r]{\ensuremath{74.34}}\,\vert\,\makebox[4em][l]{\ensuremath{855.61}}$ & $2.17\times10^{-3}$ & $2.0001\pm0.0003\,[2]$ & $\makebox[4em][r]{\ensuremath{4.32}}\,\vert\,\makebox[4em][l]{\ensuremath{53.23}}$ & $5.39\times10^{-4}$ \\
\bottomrule
\end{tabular}
\end{table*}
\begin{table*}[t]
\centering
\caption{Power-scaling summary for $\zeta_{\max}=10^{-4}$ and $\mathrm{even}$ parity.
Each $p$ entry is $p_{\rm eff}\pm\sigma_p$ with the chosen discrete $p_\star$ shown as $[p_\star]$.
$C$ lists the fitted coefficient components $\Re(C)\,|\,\Im(C)$ and $\sigma/\mu$ is the flatness ratio.}
\label{tab:power_1Em4_even_58}
\setlength{\tabcolsep}{4pt}
\renewcommand{\arraystretch}{1.15}
\begin{tabular}{cc||ccc|ccc}
\toprule
 &  & \multicolumn{3}{c}{$\mathcal{O}_{5}$} & \multicolumn{3}{c}{$\mathcal{O}_{8}$} \\
$\ell$ & $n$ & $p$ & $C$ & $\sigma/\mu$ & $p$ & $C$ & $\sigma/\mu$ \\
\midrule
2 & 0 & $2.0001\pm0.0004\,[2]$ & $\makebox[4em][r]{\ensuremath{125.28}}\,\vert\,\makebox[4em][l]{\ensuremath{498.26}}$ & $5.29\times10^{-4}$ & $2.0000\pm0.0001\,[2]$ & $\makebox[4em][r]{\ensuremath{7.79}}\,\vert\,\makebox[4em][l]{\ensuremath{31.09}}$ & $1.32\times10^{-4}$ \\
2 & 1 & $2.0003\pm0.0011\,[2]$ & $\makebox[4em][r]{\ensuremath{-41.76}}\,\vert\,\makebox[4em][l]{\ensuremath{602.93}}$ & $1.48\times10^{-3}$ & $2.0001\pm0.0003\,[2]$ & $\makebox[4em][r]{\ensuremath{-2.68}}\,\vert\,\makebox[4em][l]{\ensuremath{37.50}}$ & $3.68\times10^{-4}$ \\
2 & 2 & $2.0004\pm0.0014\,[2]$ & $\makebox[4em][r]{\ensuremath{-713.18}}\,\vert\,\makebox[4em][l]{\ensuremath{721.64}}$ & $1.92\times10^{-3}$ & $2.0001\pm0.0003\,[2]$ & $\makebox[4em][r]{\ensuremath{-44.47}}\,\vert\,\makebox[4em][l]{\ensuremath{44.69}}$ & $4.76\times10^{-4}$ \\
2 & 3 & $2.0006\pm0.0021\,[2]$ & $\makebox[4em][r]{\ensuremath{-2812.74}}\,\vert\,\makebox[4em][l]{\ensuremath{849.70}}$ & $2.49\times10^{-3}$ & $2.0002\pm0.0005\,[2]$ & $\makebox[4em][r]{\ensuremath{-174.38}}\,\vert\,\makebox[4em][l]{\ensuremath{52.35}}$ & $6.19\times10^{-4}$ \\
3 & 0 & $2.0000\pm0.0001\,[2]$ & $\makebox[4em][r]{\ensuremath{158.94}}\,\vert\,\makebox[4em][l]{\ensuremath{488.04}}$ & $3.68\times10^{-4}$ & $2.0000\pm0.0000\,[2]$ & $\makebox[4em][r]{\ensuremath{9.90}}\,\vert\,\makebox[4em][l]{\ensuremath{30.53}}$ & $9.11\times10^{-5}$ \\
3 & 1 & $2.0001\pm0.0005\,[2]$ & $\makebox[4em][r]{\ensuremath{138.16}}\,\vert\,\makebox[4em][l]{\ensuremath{604.53}}$ & $7.89\times10^{-4}$ & $2.0000\pm0.0001\,[2]$ & $\makebox[4em][r]{\ensuremath{8.54}}\,\vert\,\makebox[4em][l]{\ensuremath{37.73}}$ & $1.97\times10^{-4}$ \\
3 & 2 & $2.0004\pm0.0012\,[2]$ & $\makebox[4em][r]{\ensuremath{25.61}}\,\vert\,\makebox[4em][l]{\ensuremath{764.80}}$ & $1.83\times10^{-3}$ & $2.0001\pm0.0003\,[2]$ & $\makebox[4em][r]{\ensuremath{1.40}}\,\vert\,\makebox[4em][l]{\ensuremath{47.57}}$ & $4.57\times10^{-4}$ \\
3 & 3 & $2.0005\pm0.0016\,[2]$ & $\makebox[4em][r]{\ensuremath{-342.02}}\,\vert\,\makebox[4em][l]{\ensuremath{985.70}}$ & $2.78\times10^{-3}$ & $2.0001\pm0.0004\,[2]$ & $\makebox[4em][r]{\ensuremath{-21.68}}\,\vert\,\makebox[4em][l]{\ensuremath{61.05}}$ & $6.90\times10^{-4}$ \\
4 & 0 & $1.9999\pm0.0004\,[2]$ & $\makebox[4em][r]{\ensuremath{167.00}}\,\vert\,\makebox[4em][l]{\ensuremath{464.14}}$ & $6.23\times10^{-4}$ & $2.0000\pm0.0001\,[2]$ & $\makebox[4em][r]{\ensuremath{10.41}}\,\vert\,\makebox[4em][l]{\ensuremath{29.07}}$ & $1.55\times10^{-4}$ \\
4 & 1 & $2.0000\pm0.0000\,[2]$ & $\makebox[4em][r]{\ensuremath{170.55}}\,\vert\,\makebox[4em][l]{\ensuremath{558.24}}$ & $5.23\times10^{-4}$ & $2.0000\pm0.0000\,[2]$ & $\makebox[4em][r]{\ensuremath{10.59}}\,\vert\,\makebox[4em][l]{\ensuremath{34.91}}$ & $1.30\times10^{-4}$ \\
4 & 2 & $2.0002\pm0.0006\,[2]$ & $\makebox[4em][r]{\ensuremath{154.75}}\,\vert\,\makebox[4em][l]{\ensuremath{681.43}}$ & $1.13\times10^{-3}$ & $2.0000\pm0.0001\,[2]$ & $\makebox[4em][r]{\ensuremath{9.51}}\,\vert\,\makebox[4em][l]{\ensuremath{42.52}}$ & $2.82\times10^{-4}$ \\
4 & 3 & $2.0004\pm0.0012\,[2]$ & $\makebox[4em][r]{\ensuremath{71.22}}\,\vert\,\makebox[4em][l]{\ensuremath{857.42}}$ & $2.17\times10^{-3}$ & $2.0001\pm0.0003\,[2]$ & $\makebox[4em][r]{\ensuremath{4.12}}\,\vert\,\makebox[4em][l]{\ensuremath{53.34}}$ & $5.41\times10^{-4}$ \\
\bottomrule
\end{tabular}
\end{table*}
\begin{table*}[t]
\centering
\caption{Power-scaling summary for $\zeta_{\max}=10^{-5}$ and odd parity.
Each $p$ entry is $p_{\rm eff}\pm\sigma_p$ with the chosen discrete $p_\star$ shown as $[p_\star]$.
$C$ lists the fitted coefficient components $\Re(C)\,|\,\Im(C)$ and $\sigma/\mu$ is the flatness ratio.}
\label{tab:power_1Em5_odd_58}
\setlength{\tabcolsep}{4pt}
\renewcommand{\arraystretch}{1.15}
\begin{tabular}{cc||ccc|ccc}
\toprule
 &  & \multicolumn{3}{c}{$\mathcal{O}_{5}$} & \multicolumn{3}{c}{$\mathcal{O}_{8}$} \\
$\ell$ & $n$ & $p$ & $C$ & $\sigma/\mu$ & $p$ & $C$ & $\sigma/\mu$ \\
\midrule
2 & 0 & $2.0000\pm0.0000\,[2]$ & $\makebox[4em][r]{\ensuremath{119.93}}\,\vert\,\makebox[4em][l]{\ensuremath{477.18}}$ & $5.48\times10^{-5}$ & $2.0000\pm0.0000\,[2]$ & $\makebox[4em][r]{\ensuremath{7.49}}\,\vert\,\makebox[4em][l]{\ensuremath{29.82}}$ & $1.37\times10^{-5}$ \\
2 & 1 & $2.0000\pm0.0001\,[2]$ & $\makebox[4em][r]{\ensuremath{-35.58}}\,\vert\,\makebox[4em][l]{\ensuremath{596.37}}$ & $1.47\times10^{-4}$ & $2.0000\pm0.0000\,[2]$ & $\makebox[4em][r]{\ensuremath{-2.23}}\,\vert\,\makebox[4em][l]{\ensuremath{37.26}}$ & $3.68\times10^{-5}$ \\
2 & 2 & $2.0000\pm0.0001\,[2]$ & $\makebox[4em][r]{\ensuremath{-687.38}}\,\vert\,\makebox[4em][l]{\ensuremath{737.82}}$ & $1.97\times10^{-4}$ & $2.0000\pm0.0000\,[2]$ & $\makebox[4em][r]{\ensuremath{-42.96}}\,\vert\,\makebox[4em][l]{\ensuremath{46.07}}$ & $4.93\times10^{-5}$ \\
2 & 3 & $2.0001\pm0.0002\,[2]$ & $\makebox[4em][r]{\ensuremath{-2798.98}}\,\vert\,\makebox[4em][l]{\ensuremath{923.16}}$ & $2.52\times10^{-4}$ & $2.0000\pm0.0001\,[2]$ & $\makebox[4em][r]{\ensuremath{-174.80}}\,\vert\,\makebox[4em][l]{\ensuremath{57.61}}$ & $6.30\times10^{-5}$ \\
3 & 0 & $2.0000\pm0.0000\,[2]$ & $\makebox[4em][r]{\ensuremath{156.71}}\,\vert\,\makebox[4em][l]{\ensuremath{482.55}}$ & $3.72\times10^{-5}$ & $2.0000\pm0.0000\,[2]$ & $\makebox[4em][r]{\ensuremath{9.79}}\,\vert\,\makebox[4em][l]{\ensuremath{30.16}}$ & $9.28\times10^{-6}$ \\
3 & 1 & $2.0000\pm0.0000\,[2]$ & $\makebox[4em][r]{\ensuremath{136.30}}\,\vert\,\makebox[4em][l]{\ensuremath{598.69}}$ & $8.08\times10^{-5}$ & $2.0000\pm0.0000\,[2]$ & $\makebox[4em][r]{\ensuremath{8.51}}\,\vert\,\makebox[4em][l]{\ensuremath{37.41}}$ & $2.02\times10^{-5}$ \\
3 & 2 & $2.0000\pm0.0001\,[2]$ & $\makebox[4em][r]{\ensuremath{26.59}}\,\vert\,\makebox[4em][l]{\ensuremath{757.72}}$ & $1.88\times10^{-4}$ & $2.0000\pm0.0000\,[2]$ & $\makebox[4em][r]{\ensuremath{1.64}}\,\vert\,\makebox[4em][l]{\ensuremath{47.34}}$ & $4.71\times10^{-5}$ \\
3 & 3 & $2.0001\pm0.0002\,[2]$ & $\makebox[4em][r]{\ensuremath{-331.30}}\,\vert\,\makebox[4em][l]{\ensuremath{980.32}}$ & $2.91\times10^{-4}$ & $2.0000\pm0.0000\,[2]$ & $\makebox[4em][r]{\ensuremath{-20.74}}\,\vert\,\makebox[4em][l]{\ensuremath{61.21}}$ & $7.26\times10^{-5}$ \\
4 & 0 & $2.0000\pm0.0000\,[2]$ & $\makebox[4em][r]{\ensuremath{165.88}}\,\vert\,\makebox[4em][l]{\ensuremath{463.33}}$ & $6.36\times10^{-5}$ & $2.0000\pm0.0000\,[2]$ & $\makebox[4em][r]{\ensuremath{10.36}}\,\vert\,\makebox[4em][l]{\ensuremath{28.96}}$ & $1.59\times10^{-5}$ \\
4 & 1 & $2.0000\pm0.0000\,[2]$ & $\makebox[4em][r]{\ensuremath{168.83}}\,\vert\,\makebox[4em][l]{\ensuremath{556.57}}$ & $5.40\times10^{-5}$ & $2.0000\pm0.0000\,[2]$ & $\makebox[4em][r]{\ensuremath{10.55}}\,\vert\,\makebox[4em][l]{\ensuremath{34.79}}$ & $1.35\times10^{-5}$ \\
4 & 2 & $2.0000\pm0.0001\,[2]$ & $\makebox[4em][r]{\ensuremath{152.10}}\,\vert\,\makebox[4em][l]{\ensuremath{677.80}}$ & $1.17\times10^{-4}$ & $2.0000\pm0.0000\,[2]$ & $\makebox[4em][r]{\ensuremath{9.49}}\,\vert\,\makebox[4em][l]{\ensuremath{42.36}}$ & $2.92\times10^{-5}$ \\
4 & 3 & $2.0000\pm0.0001\,[2]$ & $\makebox[4em][r]{\ensuremath{68.03}}\,\vert\,\makebox[4em][l]{\ensuremath{850.98}}$ & $2.25\times10^{-4}$ & $2.0000\pm0.0000\,[2]$ & $\makebox[4em][r]{\ensuremath{4.22}}\,\vert\,\makebox[4em][l]{\ensuremath{53.16}}$ & $5.62\times10^{-5}$ \\
\bottomrule
\end{tabular}
\end{table*}
\begin{table*}[t]
\centering
\caption{Power-scaling summary for $\zeta_{\max}=10^{-5}$ and $\mathrm{even}$ parity.
Each $p$ entry is $p_{\rm eff}\pm\sigma_p$ with the chosen discrete $p_\star$ shown as $[p_\star]$.
$C$ lists the fitted coefficient components $\Re(C)\,|\,\Im(C)$ and $\sigma/\mu$ is the flatness ratio.}
\label{tab:power_1Em5_even_58}
\setlength{\tabcolsep}{4pt}
\renewcommand{\arraystretch}{1.15}
\begin{tabular}{cc||ccc|ccc}
\toprule
 &  & \multicolumn{3}{c}{$\mathcal{O}_{5}$} & \multicolumn{3}{c}{$\mathcal{O}_{8}$} \\
$\ell$ & $n$ & $p$ & $C$ & $\sigma/\mu$ & $p$ & $C$ & $\sigma/\mu$ \\
\midrule
2 & 0 & $2.0000\pm0.0000\,[2]$ & $\makebox[4em][r]{\ensuremath{124.59}}\,\vert\,\makebox[4em][l]{\ensuremath{497.35}}$ & $5.52\times10^{-5}$ & $2.0000\pm0.0000\,[2]$ & $\makebox[4em][r]{\ensuremath{7.78}}\,\vert\,\makebox[4em][l]{\ensuremath{31.08}}$ & $1.38\times10^{-5}$ \\
2 & 1 & $2.0000\pm0.0001\,[2]$ & $\makebox[4em][r]{\ensuremath{-43.13}}\,\vert\,\makebox[4em][l]{\ensuremath{599.43}}$ & $1.54\times10^{-4}$ & $2.0000\pm0.0000\,[2]$ & $\makebox[4em][r]{\ensuremath{-2.70}}\,\vert\,\makebox[4em][l]{\ensuremath{37.45}}$ & $3.84\times10^{-5}$ \\
2 & 2 & $2.0000\pm0.0001\,[2]$ & $\makebox[4em][r]{\ensuremath{-711.25}}\,\vert\,\makebox[4em][l]{\ensuremath{713.70}}$ & $1.98\times10^{-4}$ & $2.0000\pm0.0000\,[2]$ & $\makebox[4em][r]{\ensuremath{-44.44}}\,\vert\,\makebox[4em][l]{\ensuremath{44.57}}$ & $4.96\times10^{-5}$ \\
2 & 3 & $2.0001\pm0.0002\,[2]$ & $\makebox[4em][r]{\ensuremath{-2785.66}}\,\vert\,\makebox[4em][l]{\ensuremath{835.26}}$ & $2.58\times10^{-4}$ & $2.0000\pm0.0001\,[2]$ & $\makebox[4em][r]{\ensuremath{-173.96}}\,\vert\,\makebox[4em][l]{\ensuremath{52.13}}$ & $6.44\times10^{-5}$ \\
3 & 0 & $2.0000\pm0.0000\,[2]$ & $\makebox[4em][r]{\ensuremath{158.35}}\,\vert\,\makebox[4em][l]{\ensuremath{488.58}}$ & $3.79\times10^{-5}$ & $2.0000\pm0.0000\,[2]$ & $\makebox[4em][r]{\ensuremath{9.89}}\,\vert\,\makebox[4em][l]{\ensuremath{30.54}}$ & $9.47\times10^{-6}$ \\
3 & 1 & $2.0000\pm0.0000\,[2]$ & $\makebox[4em][r]{\ensuremath{136.42}}\,\vert\,\makebox[4em][l]{\ensuremath{603.43}}$ & $8.23\times10^{-5}$ & $2.0000\pm0.0000\,[2]$ & $\makebox[4em][r]{\ensuremath{8.52}}\,\vert\,\makebox[4em][l]{\ensuremath{37.71}}$ & $2.06\times10^{-5}$ \\
3 & 2 & $2.0000\pm0.0001\,[2]$ & $\makebox[4em][r]{\ensuremath{21.70}}\,\vert\,\makebox[4em][l]{\ensuremath{760.38}}$ & $1.90\times10^{-4}$ & $2.0000\pm0.0000\,[2]$ & $\makebox[4em][r]{\ensuremath{1.34}}\,\vert\,\makebox[4em][l]{\ensuremath{47.50}}$ & $4.76\times10^{-5}$ \\
3 & 3 & $2.0001\pm0.0002\,[2]$ & $\makebox[4em][r]{\ensuremath{-347.86}}\,\vert\,\makebox[4em][l]{\ensuremath{974.99}}$ & $2.87\times10^{-4}$ & $2.0000\pm0.0000\,[2]$ & $\makebox[4em][r]{\ensuremath{-21.77}}\,\vert\,\makebox[4em][l]{\ensuremath{60.88}}$ & $7.17\times10^{-5}$ \\
4 & 0 & $2.0000\pm0.0000\,[2]$ & $\makebox[4em][r]{\ensuremath{166.53}}\,\vert\,\makebox[4em][l]{\ensuremath{465.36}}$ & $6.44\times10^{-5}$ & $2.0000\pm0.0000\,[2]$ & $\makebox[4em][r]{\ensuremath{10.41}}\,\vert\,\makebox[4em][l]{\ensuremath{29.09}}$ & $1.61\times10^{-5}$ \\
4 & 1 & $2.0000\pm0.0000\,[2]$ & $\makebox[4em][r]{\ensuremath{169.30}}\,\vert\,\makebox[4em][l]{\ensuremath{558.58}}$ & $5.43\times10^{-5}$ & $2.0000\pm0.0000\,[2]$ & $\makebox[4em][r]{\ensuremath{10.57}}\,\vert\,\makebox[4em][l]{\ensuremath{34.91}}$ & $1.36\times10^{-5}$ \\
4 & 2 & $2.0000\pm0.0001\,[2]$ & $\makebox[4em][r]{\ensuremath{151.70}}\,\vert\,\makebox[4em][l]{\ensuremath{680.12}}$ & $1.18\times10^{-4}$ & $2.0000\pm0.0000\,[2]$ & $\makebox[4em][r]{\ensuremath{9.47}}\,\vert\,\makebox[4em][l]{\ensuremath{42.50}}$ & $2.94\times10^{-5}$ \\
4 & 3 & $2.0000\pm0.0001\,[2]$ & $\makebox[4em][r]{\ensuremath{64.93}}\,\vert\,\makebox[4em][l]{\ensuremath{852.71}}$ & $2.25\times10^{-4}$ & $2.0000\pm0.0000\,[2]$ & $\makebox[4em][r]{\ensuremath{4.03}}\,\vert\,\makebox[4em][l]{\ensuremath{53.27}}$ & $5.63\times10^{-5}$ \\
\bottomrule
\end{tabular}
\end{table*}

\end{document}